\documentclass[aps,pre,twocolumn,amsmath,amsfonts,floatfix,superscriptaddress,nofootinbib]{revtex4-1}

\usepackage{mathrsfs} \usepackage{amssymb} \usepackage{graphicx,color} \usepackage{bbm} \usepackage{multirow}
\usepackage{bm} \usepackage{mathrsfs} \usepackage{commath} \usepackage{stmaryrd} \usepackage{color} \usepackage{stmaryrd}
\usepackage{tikz}\usetikzlibrary{shapes,arrows,positioning,calc} \usepackage{comment}

\let\a=\alpha \let\b=\beta \let\g=\gamma \let\d=\delta
\let\e=\epsilon \let\z=\zeta \let\h=\eta \let\k=\kappa
\let\l=\lambda \let\m=\mu \let\n=\nu \let\x=\xi \let\p=\pi
\let\s=\sigma \let\t=\tau \let\f=\varphi 
   \let\G=\Gamma
\let\D=\Delta \let\Th=\Theta  
\let\Si=\Sigma   
\let\ee=\varepsilon \let\r=\rho \let\th=\theta \let\io=\infty
\let\om=\omega
\def\ie{{\textit{i.e.} }}\def\eg{{\textit{e.g.} }}

\def\MM{{\cal M}} \def\VV{{\cal V}}
\def\CC{{\cal C}}\def\FF{{\cal F}} 
 
\def\RR{{\cal R}}  \def\OO{{\cal O}}
\def\DD{{\cal D}}\def\AA{{\cal A}}\def\GG{{\cal G}} \def\SS{{\cal S}}

  \def\erf{\text{erf}}

\def\Dl{\D_{\rm liq}}\def\dd{\mathrm{d}}
\def\id{\mathbbm{1}}\def\omh{\bm{\bar{\om}}}
\def\sd{\bm{\D}}\def\mh{\bm{\bar{\m}}}

\def\restriction#1#2{\mathchoice
              {\setbox1\hbox{${\displaystyle #1}_{\scriptstyle #2}$}
              \restrictionaux{#1}{#2}}
              {\setbox1\hbox{${\textstyle #1}_{\scriptstyle #2}$}
              \restrictionaux{#1}{#2}}
              {\setbox1\hbox{${\scriptstyle #1}_{\scriptscriptstyle #2}$}
              \restrictionaux{#1}{#2}}
              {\setbox1\hbox{${\scriptscriptstyle #1}_{\scriptscriptstyle #2}$}
              \restrictionaux{#1}{#2}}}
\def\restrictionaux#1#2{{#1\,\smash{\vrule height .8\ht1 depth .85\dp1}}_{\,#2}}

\def\to{\rightarrow} \def\la{\left\langle} \def\ra{\right\rangle}
\def\sdet{\text{sdet}} \def\str{\text{str}}
\def\RRR{\mathbb{R}}
\newcommand{\afunc}[1]{\operatorname{\mathsf{#1}}}\def\DE{\afunc{D}}
\newcommand{\beq}{\begin{equation}} \newcommand{\eeq}{\end{equation}}
\newcommand{\wh}{\widehat} \newcommand{\wt}{\widetilde}

\def\SIcite{the Appendix}
\def\FE{\afunc{F}}
\def\de{\mathrm d}
\def\redv{\bar V}

\begin{document}

\title{
Solution of the  dynamics of liquids in the large-dimensional limit
} 

\author{Thibaud Maimbourg}
\affiliation{LPT,
\'Ecole Normale Sup\'erieure, UMR 8549 CNRS, 24 rue Lhomond, 75005 Paris, France}

\author{Jorge Kurchan}
\affiliation{LPS,
\'Ecole Normale Sup\'erieure, UMR 8550 CNRS, 24 rue Lhomond, 75005 Paris, France}

\author{Francesco Zamponi}
\affiliation{LPT,
\'Ecole Normale Sup\'erieure, UMR 8549 CNRS, 24 rue Lhomond, 75005 Paris, France}

\begin{abstract}
We obtain analytic expressions for the time correlation functions of a liquid of spherical particles, exact in the limit of high dimensions $d$.
The derivation is long but straightforward: a dynamic virial expansion for which only the first two terms survive,
 followed by a change to generalized spherical coordinates in the dynamic variables leading to saddle-point evaluation of
 integrals  for large $d$. The problem is thus mapped onto a one-dimensional  diffusion in a perturbed harmonic potential with colored noise. 
At high density, an ergodicity-breaking glass transition is found. 
In this regime, our results agree with thermodynamics, 
consistently with the general Random First Order Transition scenario.
The glass transition density is higher than the best known lower bound for hard sphere packings in large $d$. 
Because our calculation is, if not rigorous, elementary, an improvement in the bound for sphere packings in large dimensions is at hand.
\end{abstract}

\maketitle

\paragraph*{Introduction --}
The physics of liquids and glasses belongs to the group of fields that are victims of the lack of a small parameter.
Many  approximations  have been proposed over the years, but they suffer from the uncertainty
about  what is the  limit in which they are supposed to become exact. This has been true both for equilibrium
and for dynamic properties. From the point of view of dynamics, an extreme case  is that of  Mode-Coupling Theory (MCT)~\cite{Go09,Go98,RC05}:
 it may be introduced by an (uncontrolled) resummation of an infinite subset of diagrams. 
 The Mode-Coupling approximation yields Mode-Coupling dynamics:
 the phenomenology depends on the approximation itself~\cite{KS91},  somewhat like  
 a harmonic approximation is expected to predict oscillations.

An often used remedy to the absence of a small parameter~\cite{1N} is to promote the system to
$d$ dimensions, solve the large $d$ limit, and (eventually) expand around.
This strategy has been used with success for  liquids~\cite{WRF87,EF88,FP99}, strongly coupled electrons~\cite{GKKR96}, 
atomic physics~\cite{atomic},
gauge field theory \cite{DPS79}, and most recently the thermodynamics of amorphous systems \cite{PZ10,nature}.
In this  paper we extend this procedure to the dynamics of liquids made of spherical particles. We restrict
ourselves to equilibrium, although the extension to the glassy off-equilibrium ``aging'' regime is at hand.
It has been a long-standing question whether  MCT becomes exact in infinite dimension~\cite{KW87,KS91,SS10,IM10,JvW11,CIPZ11}, 
and the present computation gives an answer.

\paragraph*{Statement of the main result --}
We shall consider a system of $N$ identical particles, interacting via a spherical potential $V(r)$ 
of typical interaction length $\s$ in $d$ dimensions, 
and obtain a solution
for the equilibrium time correlations of the resulting liquid, that becomes exact in the limit $d \rightarrow \infty$.
We need to confine the particles in a finite volume ${\cal V}$. It is very convenient to do this in such a way that the ``box'' does not break 
rotational and translational invariances, which are crucial in our developments. A practical way to do this is to 
 consider particles living on points $x_i$ ($i=1,...,N$) on the $d$-dimensional surface of a hypersphere with $x_i \cdot x_i = \sum_\mu [x_i^\mu]^2=  R^2 \equiv \s^2 \Delta_{\rm liq}/(2d) \gg \sigma^2$, ($\mu=1,\cdots,d+1)$. The thermodynamic
  limit $R\to\io$ with constant density $\r=N/\VV$, in which the flat space is recovered, will be taken
  {\it before} $d\to\io$.
Rotations and translations in $d$ dimensions -- with dimensions $d(d-1)/2$ and $d$, respectively -- are encoded in the rotations in $d+1$ dimensions,
with dimension $d(d+1)/2$.
We consider a Langevin dynamics 
\beq\label{eq:HSLang}
m \ddot x_i(t) + \g \dot x_i(t) = -\nu_i(t) x_i(t) - \nabla_{x_i}H+ \x_i(t) \ ,
\eeq
where $\xi_i$ is a white noise with $\langle \x^\mu_i(t) \x^{\mu'}_j(t') \rangle = 2 T \g \d_{ij} \d_{\mu\mu'} \d(t-t')$, 
$H = \sum_{i < j} V(x_i - x_j)$ and $T=1/\b$ is the temperature.
  Here, and in what follows, $\langle \bullet \rangle$ denotes average over noise $\x_i$ and/or initial conditions. 
 The $\nu_i$ are  Lagrange multipliers,  imposing the spherical constraints
 $x_i \cdot \dot x_i=0$.
 For $d\to\io$, they do not fluctuate and their value $\nu_i \sim \nu =  d T p / R^2$ is  
 proportional to the equilibrium reduced pressure $p = \b P/\r$~\cite{hansen}.
 We shall in what follows treat the overdamped case $m=0$, but the inertial term may be reinstalled at any stage
 (and in that case the thermal bath may be disconnected setting  $\gamma=0$, to recover a purely Newtonian dynamics).

We define the adimensional scaled correlation and response functions
(see~Fig.1 in~\SIcite):
\beq  \label{eq:corrdef}
\begin{split}
C(t,t')&\sim  C_i(t,t')  =\frac{2 d}{\s^2}  x_i(t) \cdot x_i(t') \ ,  \\
R(t,t') &\sim  R_i(t,t')= \frac{2 d}{\s^2} \sum_{\mu}  \frac{\delta  x_i^\mu(t) }{\delta h_i^\mu(t')} \ ,\\
\Delta(t,t')&\sim\frac{ d}{\s^2}  |x_i(t) -x_i(t')|^2 =\D_{\rm liq}-C(t,t') \ , \\
\end{split}\eeq
where $h_i$ is an external field conjugated to $x_i$, and
we have also introduced the single particle mean square displacement $\D$~\cite{hansen}.
As we shall see, these quantities have a finite and non-fluctuating limit for $d\to\io$.
In the following and in the Appendix, we shall derive exact equations for the correlations~\eqref{eq:corrdef}.
For simplicity, we will restrict to equilibrium where $C(t,t') = C(t-t')$, and $R(t) = -\b \th(t) \dot C(t)$; here $\th(t)$
is the Heaviside step function.

Our main result is that correlation functions are obtained in terms of a memory kernel $M$ through the following equations:
\beq
\label{eq:HSLang1}
\begin{split}
&\wh\g \dot C(t) = - \frac{T}{\Delta_{\rm liq}}  C(t)- \beta \int_{0}^t \de u \; M(t-u) \dot C(u) \ , \\
&\wh\g \dot\Delta(t)  = T -\b\int_{0}^t \de u \; M(t-u) \dot \Delta(u) \ ,
\end{split}\eeq
the second relation being valid for $\Delta \ll \Delta_{\rm liq}$.

To give the expression of the memory kernel $M$,
let us define a scaled density $\wh\f = \r \VV_d(\s) / d$ where $\VV_d(\s)$ is the volume of a $d$-dimensional sphere of radius $\s$,
a scaled $\wh\g = \s^2 \g/(2d^2)$,
a scaled potential $\redv(y) = V(\s (1+y/d) )$, and
the inter-particle force $F(y)=- \redv'(y)$. 
$M$ is then obtained in terms of a one-dimensional effective
dynamics with a colored noise $\z$,
\begin{equation}\label{eq:HSfinal1}
\begin{split}
\wh\g \dot y(t) 
&= - w'(y(t)) - \b \int_{0}^t \de u \,   M(t-u) \dot y(u) + \z(t)  \ ,\\
  &\langle \z(t) \z(t') \rangle  = 2 \wh\g T \d(t-t') +   M(t-t') \ ,
\end{split}\end{equation}
from the self-consistent equation:
\beq\label{eq:HSfinal2}
M(t-t')  = \frac{\wh\f}{2}  \int \de y_0 \, e^{-\b w(y_0)} 
 \langle F(t) F(t') \rangle_{M,y_0} \ .
\eeq
The average in~\eqref{eq:HSfinal2} is over the process~\eqref{eq:HSfinal1}
with $y(t=0)=y_0$.
The total effective potential 
has the form (see Fig.4 in~\SIcite):
\begin{equation}
w(y_0) =  \redv(y_0) - T { y_0} + \frac{T y_0^2}{2\Dl} \ .
\end{equation} 
The quadratic part of the potential plays the role of the confining  ``box''.  In fact, it is negligible for finite times and large $\Delta_{\rm liq}$, 
and the probability distribution is exponential in $y$ near $y\sim 0$  
(expressing the growth of entropy as a function of distance along the $d$$+$$1$-dimensional sphere), 
the region relevant for those times. Reassuringly, box
details are irrelevant at short times.  


Finally, Eq.~(\ref{eq:HSfinal2}) has a microscopic counterpart (cf. Appendix);
in fact, the function $M(t)$ is, in the large $d$ limit, the autocorrelation of the
 inter-particle forces $F_{ij}(t) = - \nabla V(x_i(t) - x_j(t))$,
\begin{equation}
M(t) = \frac{1}{2dN} \sum_{i\neq j} \langle F_{ij}(t) \cdot F_{ij}(0) \rangle \ ,
\end{equation} 
a result that provides a physical interpretation of $M(t)$.

\paragraph*{Relation with MCT --}
If $M$ were a simple {\em function} of $C$, $M = \FE(C)$,
then Eq.~(\ref{eq:HSLang1}) would be in the {\em schematic MCT} form.
As is well known, schematic MCT is obtained as the exact dynamics of a system of 
spherical spins $\sum_i s_i^2=N$ with $p$-spin random interactions~\cite{Cu02,Bouchaud,CC05},
for which
\beq \label{eq:Mpspin}
M = \FE(C) = \frac{p}2 C^{p-1} \ .
\eeq
However, 
as soon as one considers non-spherical variables, e.g. soft-spins with a potential $V(s)= a (s^2-1)^2$,
one obtains an equation like (\ref{eq:HSfinal1}) with this $V(s)$~\cite{Bouchaud,SZ82,CK94}:
\beq\label{eq:softspin}
\begin{split}
\dot s(t) 
&= - V'(s)- \b \int_{0}^t \de u \,  M(t-u) \dot s(u) + \z(t)  \ .
\end{split}\eeq
Here again, Eq.~\eqref{eq:Mpspin} holds and the system is closed by
$C(t-t') = \langle s(t) s(t') \rangle$.
Within the liquid phase, 
this more general form of dynamic equation has essentially the same phenomenology as
schematic MCT.
Our system of equations belongs to this more general class, and they thus show exactly the same
MCT phenomenology for what concerns {\it universal} quantities that are independent of details
of the memory kernel (e.g. the dynamical scaling forms
and the relations between critical exponents)~\cite{ABB09,JMR14,RC05}.

However, important {\it quantitative} differences are observed with respect to applying the MCT approximation 
to the intermediate scattering function, which leads to the ``standard'' formulation of MCT for liquids~\cite{Go98,Go09}. 
Standard MCT has the same qualitative structure as schematic MCT, 
but also provides quantitative results for the self and collective scattering functions
 in all dimensions, in particular in $d=3$~\cite{KA95a,KA95b}; its 
$d \to \infty$ limit was discussed in Refs.~\cite{IM10,SS10}. 
Our result in $d\to\io$ is formulated in terms of $\Delta(t,t')$, and most of the other natural observables
are functionals of $\Delta(t,t')$.
For example, for $q\sigma/d^{3/2}\ll 1$,
we have for the self intermediate scattering function
(cf. Appendix):
\beq\label{eq:Phiq}
\phi_q^{\rm s}(t,t') = \exp\left[ - \frac{q^2\sigma^2}{2d^2}\Delta(t, t') \right] \ ,
\eeq 
in contrast to the non-Gaussianity in $q$ one finds within MCT close to the plateau~\cite{IM10,SS10}.
One could then write our equations
in terms of $\phi_q^{\rm s}$. The result, however, is different from standard MCT and in particular
our kernel $M$ is not an analytic function of~$\phi_q^{\rm s}$.

Because our equations fall in the same universality class as schematic MCT, but provide different quantitative
results with respect to standard MCT in $d\to\io$,
it remains a matter of taste if one wishes to call them with the same name 
or, more generally, ``dynamic Random First Order Transition (RFOT)''~\cite{KW87}.

\paragraph*{Sketch of the derivation --}
Let us outline the main steps in the derivation; more details are given in~\SIcite.
In order to construct the high-dimensional limit, a virial expansion is a reliable method. 
Following \cite{MK11}, we exploit the well-known analogy between trajectories and polymers. 
The dynamics are generated by a sum of trajectories in $d$-dimensional space,
in our case, the surface of a $d$$+$$1$-dimensional sphere. 
To each trajectory is associated an Onsager-Machlup probability weight which is the exponential of an action. 
The sum over all trajectories of this quantity is analogous to a partition function and is thus used to generate
 averages over the Langevin process~\eqref{eq:HSLang}.
The action is expressed in terms of trajectories of the $x_i$ and auxiliary ``response'' variables $\hat x_i$
(Martin-Siggia-Rose-De Dominicis-Janssen generating path integral)~\cite{Cu02,CC05}.
The result reads, in the It$\mathrm{\hat{o}}$ convention:
\begin{equation}\label{eq:action}
\begin{split}
&Z_N=\int \prod_{i=1}^N {\rm D}x_i{\rm D}\hat{x}_i \,e^{-\sum_{i=1}^N \Phi[x_i,\hat{x}_i]-\sum_{i<j}^{1,N}W[x_i-x_j,\hat{x}_i-\hat{x}_j] } \ , \\
 &\Phi[x,\hat{x}]=\gamma \int \mathrm{d}t \,\left(T\hat{x}^2 + i\dot{x}\cdot \hat{x} +  \nu \hat x\cdot  x\right)  \ , \\
&W[x_1-x_2,\hat{x}_1-\hat{x}_2]=i\int \mathrm{d}t\,(\hat{x}_1-\hat{x}_2)\cdot \nabla V(|x_1-x_2|) \ .
\end{split}\eeq
Following standard liquid theory~\cite{hansen,MK11},
we introduce the density function {\em for trajectories}
\begin{equation}
 \rho[x,\hat{x}]\equiv{\left\langle \frac1N\sum_{i=1}^N {\mathbf \delta}[x-x_i]{\mathbf \delta}[\hat{x}-\hat{x}_i]\right\rangle} \ ,
\label{rho}
\end{equation}
where $\delta[x]$ is the {\em functional} Dirac $\delta$ (a product of deltas over all times) and we construct a virial (Mayer)
expansion as a power series in $\r[x,\hat x]$.
One can show that all terms
involving a product of more than two density fields are subleading
for $d\to\io$~\cite{FP99}:
they are exponentially suppressed because of the requirement that the three trajectories overlap
(see Fig.1 in~\SIcite),
which is exponentially unlikely in large $d$.
Truncating the virial expansion accordingly, we get (cf. Appendix):
\beq
\label{eq:lnXixxtMT}
\begin{split}
\SS&\equiv \frac{\ln Z_N}N =-\int {\rm D}\r[x,\hat{x}] (\Phi[x,\hat{x}]+\mathrm{ln}\rho[x,\hat{x}])  \\
& + \frac{N}{2}\int {\rm D}\r[x_1,\hat{x}_1] {\rm D}\r[x_2,\hat{x}_2] f[x_1-x_2,\hat{x}_1-\hat{x}_2] \ ,
\end{split}\eeq
where $ {\rm D}\r[x,\hat{x}] = {\rm D}[x,\hat{x}] \rho[x,\hat{x}]$ and
$f[x_1-x_2,\hat{x}_1-\hat{x}_2] = e^{-W[x_1-x_2,\hat{x}_1-\hat{x}_2] }-1$. The physical
$\rho[x,\hat{x}]$ is determined by
$\delta \SS/\delta \rho[x,\hat{x}]=0$ and the normalization $\int {\rm D}\r[x,\hat{x}] =1$.
The first term in Eq.~\eqref{eq:lnXixxtMT} is an ideal gas contribution and the second accounts for interactions.

Following the thermodynamic treatment~\cite{KPZ12}, 
we may now argue that due to rotational invariance on the hypersphere, $\rho[ x(t), \hat x(t)] =\rho[C(t,t'),R(t,t'),D(t,t')]$ where
$R(t,t') \equiv (2d/\s^2)  x(t) \cdot \hat x(t')$ and $D(t,t') \equiv (2d/\s^2) \hat x(t) \cdot \hat x(t')$.
We can thus make a change of variables in the functional integration over $x(t),\hat x(t)$
to $Q(t,t') \equiv \{ C(t,t'),R(t,t'),D(t,t') \}$. 
The change of variables gives for density averages (cf. Appendix):
\beq\label{eq:sp1}
\int {\rm D}[x,\hat x]  \; \bullet \;  \rho \rightarrow \int {\rm D}Q \; \bullet \; e^{d \; \text{str}  \ln Q + d \Omega(Q)}  \ ,
\eeq 
where $e^{{d\; \text{str}} \ln Q}$ is the Jacobian of the transformation (cf. Appendix) and we defined $\rho(Q)= e^{d \Omega(Q)}$. 
The appearance of the dimension in the exponent leads to a narrowing of fluctuations of correlations, and saddle-point
evaluation becomes exact (cf. Appendix). In this way we can compute the ideal gas term in Eq.~\eqref{eq:lnXixxtMT}.
For the interaction term, that involves two $\rho$ functions, we need the variables corresponding to 
$(x_1,\hat x_1)$, $(x_2,\hat x_2)$ and also $\omega=|x_1-x_2|^2$, $\hat \omega= (x_1-x_2)\cdot (\hat x_1-\hat x_2)$.
The Jacobian may be calculated with the same methods (cf. Appendix), and the crucial result is that at the saddle-point 
level $Q_1 = Q_2 = Q$ with $Q$ determined by the same saddle-point as in Eq.~\eqref{eq:sp1}, 
while the remaining integration over $\omega(t),\hat \omega(t)$ is effectively one-dimensional. Changing variables with
$\omega(t) = \s^2 (1 + y(t)/d)$ leads to a finite integration over $y(t)$ that eventually gives Eq.~\eqref{eq:HSfinal1}.
Hence, the typical distance between two trajectories turns out to be $\sigma+ O(1/d)$. This scaling physically tells us that a particle vibrate inside a cage with $1/d$ amplitude and interacts with $O(d)$ neighbors.
Finally, equilibrium and causality at the saddle-point level imply $D(t,t')=0$ and the Fluctuation-Dissipation relation (FDT)
$R(t-t') = \b \th(t-t') \partial_{t'} C(t-t')$.
The saddle-point evaluation for the two-time variables is the mathematical
 justification of the above-mentioned fact that $C,R,D$ do not fluctuate.
 These saddle-point equations give Eq.~\eqref{eq:HSfinal2} and \eqref{eq:HSLang1} (cf. Appendix).
 
 In this paper we are treating an equilibrium situation. Within the liquid phase, this may be achieved by starting from
any configuration in the distant past. A more practical way, however, is to assume equilibrium at a convenient time $t_0$
(e.g. $t_0=0$). 
How does one deal with a non-Markovian equation of motion like (\ref{eq:HSfinal1})? The answer is simple: either  
one makes the memory kernel extend to the remote past, or, alternatively, one may assume equilibrium at $t_0$, in other words summing
all the past histories passing through $y(t_0)$ at $t_0$. It turns out \cite{Ha97} that this is implemented simply 
by cutting the memory at a lower limit $t_0$, as in Eq.~\eqref{eq:HSfinal1} (cf. Appendix). This completes the derivation of our basic dynamical equations.

\paragraph*{Dynamic transition and timescale separation --}
 We now apply the standard MCT methodology to locate the density or temperature at which
 a dynamic transition occurs with freezing in a cage, corresponding to the development of a plateau in $\D(t)$~\cite{Go09,Go98,RC05,CC05}. 

Consider the case when $M(t)$ falls from $M(0)$ to a plateau value $M_{\rm EA}$, and then, at much larger times, to zero.  
Concomitantly, $\Delta(t)$ grows to a plateau value $\Delta_{\rm EA}$, and then continues to grow at a slower (diffusive) pace. 
Denote the fast part $\delta M(t) = M(t) - M_{\rm EA}$.
 In the limit in which the plateau times are much larger than the microscopic times, and much smaller than the final relaxation times,
 the noise breaks into a fast variable $\delta \z(t)$ and a slow random variable $\bar \z$, as does the friction term. Their
 sum acts as an adiabatically slow  field $Y(t)$ at those times. We may thus split the equilibration in two steps~\cite{CK00}:
  $P\left(y | Y \right) $ and $P_{\rm slow}(Y)$.
 When $t - t' $ is in the plateau region, we may write the expectations:
\begin{equation}
 \langle A(t)B(t') \rangle =\int \de Y \, P_{\rm slow}(Y) \;   \langle A \rangle_{Y} \; \langle B \rangle_{Y}   \ ,
\end{equation}
 where $\langle \bullet \rangle_{Y} = \int \de Y \; \bullet \; P\left(y | Y\right)$, $P_{\rm eq}(y) \propto e^{-\b w_0(y)}$, and
 \beq\begin{split}
P\left(y | Y\right) &=  \frac{
 P_{\rm eq}(y) e^{- \frac{\beta^2M_{\rm EA}}2 \left( y-\frac{T}{M_{\rm EA}} Y\right)^2}
 }  {
 \int \de y' P_{\rm eq}(y') e^{- \frac{\beta^2M_{\rm EA}}2 \left( y'-\frac{T}{M_{\rm EA}} Y\right)^2}
 }    \ , \\
 P_{\rm slow}(Y) &= \frac{\int \de y \,
 P_{\rm eq}(y) e^{- \frac{\beta^2M_{\rm EA}}2 \left( y-\frac{T}{M_{\rm EA}} Y\right)^2}}
 { \sqrt{2 \pi M_{\rm EA}} } \ .
\end{split}\eeq
Obviously, $P_{\rm eq}(y) =\int dY  \;P\left(y|Y\right) P_{\rm slow}(Y)$.
We therefore obtain the self-consistent equation for $M_{\rm EA}$:
\beq
M_{\rm EA}  = \frac{ \wh\f}2 \int \de Y P_{\rm slow}(Y) \langle F \rangle_Y^2 \equiv \MM(M_{\rm EA})
\label{ddd} \ .
\eeq
 From Eq.~\eqref{eq:HSLang1} we obtain
\beq
\b^2 M_{\rm EA} = \frac1{\D_{\rm EA}} - \frac1{\Dl}  \sim  \frac1{\D_{\rm EA}} \label{aaa} \ .
\eeq
The dynamical transition point, at which the plateau becomes infinite, happens when
Eq.~(\ref{ddd}) first has a non-zero solution for $M_{\rm EA}=T^2/\Delta_{\rm EA}$. This point happens as a bifurcation, 
and may be quickly obtained by solving Eq.~(\ref{ddd}) together with $\MM'(M_{\rm EA})=0$ (cf. Appendix).

For hard spheres, the result is $\wh \f_{\rm d}= 4.807$. This result is fully consistent with the one based 
on thermodynamics~\cite{PZ10}, and in fact Eq.~\eqref{ddd} is exactly identical to
the one that can be derived using the replica method (cf. Appendix), consistently with the general RFOT 
scenario~\cite{KW87,KT88,KT89,KTW89,LW07,WL12,KT14}.
This is a particular instance of a general correspondence between thermodynamic and dynamic 
results that is verified by the infinite-$d$ solution, and
can be extended to critical MCT exponents~\cite{CFLPRR12}
and to correlation functions~\cite{FPRR11,PR12,FJPUZ13}.
In fact, expanding around $\D_{\rm EA}$~\cite{Go09,Go98,RC05}, one can show 
that $\D_{\rm EA} - \D(t)  \sim t^{-a}$ upon approaching the plateau, while 
$\D(t) - \D_{\rm EA} \sim t^b$ upon leaving the plateau. The exponents $a,b$
satisfy the famous relation~\cite{Go09,Go98,RC05,CFLPRR12,PR12}:
\beq
\frac{\G(1-a)^2}{\G(1-2a)} = \frac{\G(1+b)^2}{\G(1+2b)} = \l \ .
\eeq
For hard spheres, we obtain $\l = 0.707$ which implies
$a = 0.324$ and $b = 0.629$~\cite{KPUZ13}.
Finally, from Eq.~\eqref{eq:Phiq} 
one can show that the factorization property of MCT~\cite{Go98,Go09} 
still holds in $d\to\io$, namely that close to the plateau, 
$\phi_q^{\rm s}(t)-\phi_{q,\mathrm{EA}}^{\rm s} \sim - \frac{q^2\sigma^2}{2d^2} \phi_{q,\mathrm{EA}}^{\rm s} (\D(t) - \D_{\rm EA})$ 
factorizes in a function of $q$ and a function of $t$ (cf. Appendix).

\paragraph*{Diffusion, viscosity, Stokes-Einstein relation --}
At long times, in the liquid phase $\wh\f < \wh\f_{\rm d}$, the motion is diffusive and
$\D(t) \sim (2 d^2 D /\s^2) t$, where $D$ is the diffusion coefficient. 
Plugging this form in Eq.~\eqref{eq:HSLang1}, and recalling that $M(t)$ decays to zero
over a finite time,
we obtain an exact result for $D$:
\beq\label{eq:D}
\frac{2d^2}{\s^2} D =  \frac{T}{\wh\g + \b \int_0^\io \de t \, M(t)} \ .
\eeq
At low density $M(t) =0$ and we recover the bare diffusion coefficient $D = T/\g$.
Upon increasing density, $M(t)$ increases and $D$ decreases. For $\wh\f \to \wh\f_{\rm d}^-$,
where a finite plateau of $M(t)$ emerges, $\int_0^\io \de t\, M(t)$ diverges and the diffusion coefficient vanishes
as $D \sim (\wh\f_{\rm d} - \wh\f)^{\g}$
with the exponent $\g = 1/(2a) + 1/(2b) =2.34$, which is consistent with the numerical results of~\cite{CJPZ14}.

Within linear response theory, the shear viscosity $\h_S$ of the liquid is given by~\cite{hansen,Yo12}
\beq\begin{split}
\eta_S &= \frac{\b}\VV \int_0^\infty \de t \; \langle \sigma_{\mu\nu}(t) \sigma_{\mu\nu}(0) \rangle  \ , \\
\sigma_{\m\n}(t) &= \sum_{i<j} (x_i^\mu-x_j^\mu) \nabla^\nu V (x_i-x_j) \ ,
\end{split}\eeq
where $\mu \neq \nu$ are two arbitrary components of the stress tensor $\s_{\m\n}$. 
In~\SIcite~we show that
$\langle \sigma_{\m\n}(t) \sigma_{\m\n}(0) \rangle = d \, N \, M(t)$, and thus
\beq\label{eq:etaS}
\h_S = \b \r d \int_0^\io \de t \, M(t) \ .
\eeq
This relation shows that for $\wh\f \to \wh\f_{\rm d}^-$, $\h_S \sim 1/D \sim (\wh\f_{\rm d} - \wh\f)^{-\g}$, as it
is found in MCT.

Combining Eqs.~\eqref{eq:D} and \eqref{eq:etaS} we obtain a relation similar to the
Stokes-Einstein relation (SER)
\beq
D = \frac{T}{\g + \frac{2 d}{\r\s^2} \h_S} \approx \frac{T \r \s^2}{2d} \frac{1}{\h_S}  \ ,
\eeq
where the second expression holds close to $\wh\f_{\rm d}$. This relation is interesting 
because it predicts that the SER is not exactly satisfied in dense liquids: the quantity 
$D \h_S / T$, which is constant in SER, has instead a small variation proportional to $\r$. 
This is in agreement with results of~\cite[Fig.7b]{CCJPZ13}
that show a linear variation of $D\h_S$ with $\r$ in the dense regime for large enough dimension.

\paragraph*{Conclusions --}
In this work we obtained an exact solution of the dynamics of liquids in the limit of infinite spatial dimension.
The picture that emerges has a relevant ``caging'' lengthscale that is smaller than the particle radius by $O(1/d)$ 
(it is about $1/5$ for real colloids~\cite{WW02}). The physics of diffusion stems from interactions at that small scale, 
while all particle motion beyond that scale consists of uncorrelated 
steps of displacement, and memory of what happens at distances $\gg 1/d$ is lost. 
Diffusion coefficients and viscosity are thus decided at distances much smaller than the particle radius. 
This strongly suggests that the wavevectors $q$ that matter for this transition are not the ones associated with 
the first neighbor distance $1/\s$, but rather the much larger ones $\sim d/\s$ corresponding to the cage size.
Note that at the transition cages are correlated over large distances~\cite{KT88,FP00,BBBCEHLP05,BBMR06}.
Also, an expansion in the number of collisions~\cite{EF88} seems difficult to reconcile with our results, 
because we expect multiple collisions within a cage. 
 
Is the high-dimensional dynamics related to MCT? The answer is that the result is not the one obtained from the
usual procedure for building up a MCT equation, which for example gives a different scaling of 
$\wh\f_{\rm d}$ with dimension~\cite{SS10,IM10}. 
Instead, the system
we obtain is formally quite close to the  slightly more general case of soft-spin mean-field dynamics, Eq.~\eqref{eq:softspin}, 
because we have mapped the system into
a one-dimensional dynamics in the presence of a colored noise and friction, that have to be determined self-consistently
through Eqs.~\eqref{eq:HSfinal1} and \eqref{eq:HSfinal2}.

In the dense glassy regime we obtain predictions for the scaling
of the cage radius, of the dynamical transition density $\wh\f_{\rm d}$, and of the parameter $\l$, that differ from the ones of 
usual MCT~\cite{IM10,SS10,CIPZ11}.
Our results are fully consistent with those obtained from the thermodynamic approach~\cite{PZ10,KPUZ13,nature},
which proves the exactness of the RFOT scenario~\cite{KT88,KT89,KTW89,LW07,WL12,KT14} for statics and dynamics in $d\to\io$,
as conjectured in~\cite{KW87}. 
Our results are also in agreement with numerical simulations of hard spheres in large spatial dimension~\cite{CJPZ14,CCJPZ13}.

Interestingly, we find that an ergodic liquid phase of hard spheres 
exists for densities $\wh\f \leqslant \wh\f_{\rm d} = 4.807$. This implies that hard sphere packings exist
(at least) up to $\wh\f_{\rm d}$, and they can be constructed easily through a sufficiently slow compression of the liquid~\cite{LS90,MS06}.
Note that the value of $\wh\f_{\rm d}$ is larger than the best known lower bound for the existence of sphere packings, 
$\wh\f \geqslant 6/e$~\cite{Va11}, and that it took 20 years to improve the previous best lower bound $\wh\f \geqslant 2 $~\cite{Ba92}
by a small factor $3/e$. Our calculation is simple enough
that there is hope to transform it into a rigorous proof, along the lines of~\cite{BDG01}. This would result in an improved
constructive lower bound for sphere packings.

Future extensions of this work include the investigation of the effect of dissipation~\cite{BBK00,IB13,FC02,SF11}, 
the study of out-of-equilibrium aging dynamics~\cite{CK93},
and the study of non-perturbative processes in $1/d$ through an instantonic expansion. The thermodynamic partition function of {\it quantum}
systems is formally very similar to Eq.~\eqref{eq:action} and could also be studied along these lines.

\paragraph*{Acknowledgements --}
We warmly thank G.~Biroli, J.~P.~Bouchaud, P.~Charbonneau, M.~Fuchs, H.~Jacquin, Y.~Jin, T.~R.~Kirkpatrick, C.~Rainone, 
D.~Reichman, R.~Schilling, G.~Tarjus and P.~Urbani for very useful discussions.
We specially thank G.~Szamel for very useful comments and for pointing out an inconsistency in an earlier
version of this work.
T. M. acknowledges funding from a fondation CFM grant.


\clearpage

\begin{widetext}

\centerline{\bf \large APPENDIX}

\tableofcontents
\makeatletter
\let\toc@pre\relax
\let\toc@post\relax
\makeatother

\section{Formulation of the dynamics}
\subsection{Definition of the model}\label{sec:intro}

We consider an assembly of $N$ spheres of mass $m$ interacting through a repulsive finite-ranged radial pair potential $V(r)$. The interaction Hamiltonian is then $H=\sum_{i<j}V(x_i-x_j)$ with $x_i(t)$ the positions of the particles. $\s$ is the diameter of the spheres, we note $\bar V(\m)=V(\s(1+\m/d))$ and the rescaled force $F(\m)=-\bar V'(\m)$. In the following, we define the usual inverse temperature $\b=1/k_BT$ and take $k_B=1$.
In order to exploit efficiently simplifications given by the translational and rotational symmetries of the system, we constrain each particle to live on the surface of a $d+1$ dimensional hypersphere $\mathbb{S}^d(R)$ of radius $R$. A particle is thus represented by a point $x_i \in \RRR^{d+1}$ with $x_i^2 = R^2$. The volume of this space is $\VV = \Si_{d+1}(R) = \Omega_{d+1} R^d$. $\Si$ and $\Omega$ are respectively the surface and the solid angle. We also denote by 
\beq
\VV_d(R)=\frac{\Si_{d}(R)R}d=\frac{\p^{d/2}}{\G(1+d/2)}R^d
\eeq
the volume of a $d$-dimensional ball of radius $R$ (\ie the volume bounded by $\mathbb{S}^d(R)$ is $\VV_{d+1}(R)$). Adding this extra dimension, rotation and translation invariances in usual $d$ dimensional space are transposed to rotational invariance only on the sphere $\mathbb{S}^d(R)$. We recover in the limit $R\to\io$ the original definition in a $d$ dimensional periodic cubic volume. However, note that in subsections~\ref{sec:dynac} and~\ref{sec:virial}, we consider the original model in $d$ dimensional Euclidean space, so as to introduce the spherical model only when needed.
\subsection{The dynamical action}\label{sec:dynac}
A possible choice of the dynamics of the particles in the $d$-dimensional space is the following Langevin process:
\begin{equation}\label{eq:Langevin}
m\ddot{x}_i+\gamma \dot{x}_i=-\sum_{j\neq i} \nabla V(x_i-x_j)+{\xi}_i
\end{equation}
${\xi}_i(t)$ being a centered white Gaussian noise with $\langle {\xi}_i^\m(t){\xi}_j^\n(t') \rangle=2\gamma T \d_{ij}\d_{\m\n}\d(t-t')$. 
In the following, we will consider the overdamped\footnote{Either Newtonian or Brownian dynamics can be treated by changing the kernel associated to $\Phi$ in the following sections.} case where $m=0$. Given a set of initial conditions, the Martin-Siggia-Rose-De Dominicis-Janssen (MSRDDJ)~\cite{CC05,Cu02} generating path integral reads, with It$\mathrm{\hat{o}}$ convention:
\begin{equation}
Z_N=\int \prod_{i=1}^N {\rm D}x_i{\rm D}\hat{x}_i \,e^{-\AA[\{x_i,\hat{x}_i\}]}
\label{eq:Z}
\end{equation}
where the action is:
\begin{equation}\label{eq:action}
\begin{split}
\AA[\{x_i,\hat{x}_i\}]&=\sum_{i=1}^N \Phi[x_i,\hat{x}_i]+\sum_{i<j}^{1,N}W[x_i,\hat{x}_i,x_j,\hat{x}_j]\\
\Phi[x,\hat{x}]=\gamma\int \dd t \,\left( T\hat{x}^2 + i\hat{x}\cdot\dot{x} \right)\,,&\hskip15pt  W[x,\hat{x},y,\hat{y}]=i\int \dd t\,(\hat{x}-\hat{y})\cdot \nabla V(x-y)=W[x-y,\hat{x}-\hat{y}]
\end{split}
\end{equation}
The sum in $Z_N$ is done over all possible trajectories, with the measure given by ${\rm D}x_i=\prod_{n=1}^M\frac{\dd x_i^n}{(2\pi)^{\frac{d}{2}}}$ when discretizing the trajectory $x_i(t)$ in $M$ time steps. Time integrals are taken over an interval noted $[t_{\rm p},t_1]$ in section~\ref{sec:kernel} where the $\{x_i(t_{\rm p})\}$ are fixed (initial conditions, which would correspond to $n=0$ in the latter discretization) and the $\{x_i(t_1)\}$ are summed over (which would correspond to $n=M$). In section~\ref{sec:kernel} we will need averages of the type $\la x_i(t_0)x_i(t_1)\ra$ with $t_0\in[t_{\rm p},t_1]$, and it is not needed to consider times larger than $t_1$ in the action owing to causality, as they will give no contribution to the average by probability conservation (\ie for the same reason as $Z_N=1$). \\
The action $\AA$ is rotation invariant (for both position and response fields using the same global rotation) and translation invariant along positions\footnote{The invariances for response fields follow if one studies the system modulo `rigid' rotations and translations of the entire system.\label{ftn:resp}}. Because of the so-called kinetic term, it is not translation invariant along the response fields (which are already centered at the origin owing to the white noise), though the interaction term is.

\subsection{Derivation of the generating functional using a virial expansion}\label{sec:virial} 
\subsubsection{Dynamic virial}\label{sub:virial}
We define $\Xi=\sum_{N=0}^{+\infty}\frac{1}{N!}Z_N$ in order to use the Mayer expansion\footnote{This is why $W$ is cast in a symmetric form so that links in the product $\prod_{i<j}$ are not directed.} as in liquid theory~\cite{hansen}. This `grand canonical' form is handy as a generating functional, but note that we assume there is \textit{no} exchange of particles with a reservoir~; $\ln\Xi$ and $\ln Z_N$ are related by a Legendre transform and virial expansions are more fruitful with the former. We have:
\begin{equation}
\Xi=\sum_{N=0}^{+\infty}\frac{1}{N!} \prod_{i=1}^N \int {\rm D}[x_i,\hat{x}_i] \,z[x_i,\hat{x}_i] \prod_{i<j}\left(1+f[x_i,\hat{x}_i,x_j,\hat{x}_j]\right)
\end{equation}
with a generalized fugacity $z=e^{-\Phi}$ and a Mayer function $f=e^{-W}-1$. We Legendre transform $\mathrm{ln}\Xi$ with respect to $N\mathrm{ln}z$ since one has
\begin{equation}\label{eq:defrho}
 \frac{\delta \mathrm{ln} \Xi}{\delta (N\mathrm{ln}z[x,\hat{x}])}={\left\langle \frac1N\sum_{i=1}^N \delta(x-x_i)\delta(\hat{x}-\hat{x}_i)\right\rangle}_{\Xi} \equiv \rho[x,\hat{x}]
\end{equation}
where the mean is generated by the functional $\Xi$. Next, the usual Mayer expansion can be carried out in this dynamical case, and inverting the Legendre transform, $\ln\Xi$ can be written as an ideal gas contribution and the sum of all connected 1-irreducible Mayer diagrams\footnote{1-irreducible means that they do not disconnect upon removal of a node.} with $N\rho[x,\hat{x}]$ nodes and~$f[x-y,\hat{x}-\hat{y}]$ bonds~\cite{hansen}:
\begin{equation}
\ln\Xi =-N\int {\rm D}[x,\hat{x}] \,\rho[x,\hat{x}](\Phi[x,\hat{x}]+\mathrm{ln}\rho[x,\hat{x}])+\begin{tikzpicture}[baseline={([yshift=-.5ex]current bounding box.center)}] \draw[line width=1pt] (0,0) -- (1,0); \draw[fill=black] (0,0) circle (0.1) ;\draw[fill=black] (1,0) circle (0.1) ; \end{tikzpicture} + \begin{tikzpicture}[baseline={([yshift=-.5ex]current bounding box.center)}] \draw[line width=1pt] (0:0) -- (60:1) -- (0:1) -- (0:0);\draw[fill=black] (0:0) circle (0.1) ;\draw[fill=black] (60:1) circle (0.1) ;\draw[fill=black] (0:1) circle (0.1) ;\end{tikzpicture} + \begin{tikzpicture}[baseline={([yshift=-.5ex]current bounding box.center)}] \draw[line width=1pt] (0,0) -- (0,1) -- (1,1) -- (1,0) -- (0,0);\draw[fill=black] (0,0) circle (0.1) ;\draw[fill=black] (1,0) circle (0.1) ;\draw[fill=black] (0,1) circle (0.1) ;\draw[fill=black] (1,1) circle (0.1) ;\end{tikzpicture} + \begin{tikzpicture}[baseline={([yshift=-.5ex]current bounding box.center)}] \draw[line width=1pt] (0,0) -- (0,1) -- (1,1) -- (1,0) -- (0,0) -- (1,1) ;\draw[fill=black] (0,0) circle (0.1) ;\draw[fill=black] (1,0) circle (0.1) ;\draw[fill=black] (0,1) circle (0.1) ;\draw[fill=black] (1,1) circle (0.1) ;\end{tikzpicture} + \begin{tikzpicture}[baseline={([yshift=-.5ex]current bounding box.center)}] \draw[line width=1pt] (0,0) -- (0,1) -- (1,1) -- (1,0) -- (0,0) -- (1,1) ; \draw[line width=1pt] (1,0) -- (0,1) ; \draw[fill=black] (0,0) circle (0.1) ;\draw[fill=black] (1,0) circle (0.1) ;\draw[fill=black] (0,1) circle (0.1) ;\draw[fill=black] (1,1) circle (0.1) ;\end{tikzpicture} +  \dots 
\end{equation}
In infinite dimension, the Mayer expansion reduces to its first term. However, this is strictly true if we assume that we are in a regime 
where the trajectories have the time to wander away only a finite fraction of the box. Because we expect (and confirm) that all interesting dynamics (namely, the $\b$ relaxation with the formation of a plateau in correlations and the onset of a relaxation towards equilibrium - $\a$ relaxation -) occur on such scales, where the fluctuation around the initial position is of amplitude $ O(1/d)$ (a scaling consistent with the statics~\cite{PZ10,KPZ12,KPUZ13,nature}). We will show that one even gets the diffusive behaviour which is already decided at the $1/d$ scale (see figure 1(a)).\\ 
\begin{figure}
 \begin{tabular}{cc}
  \includegraphics[width=9cm]{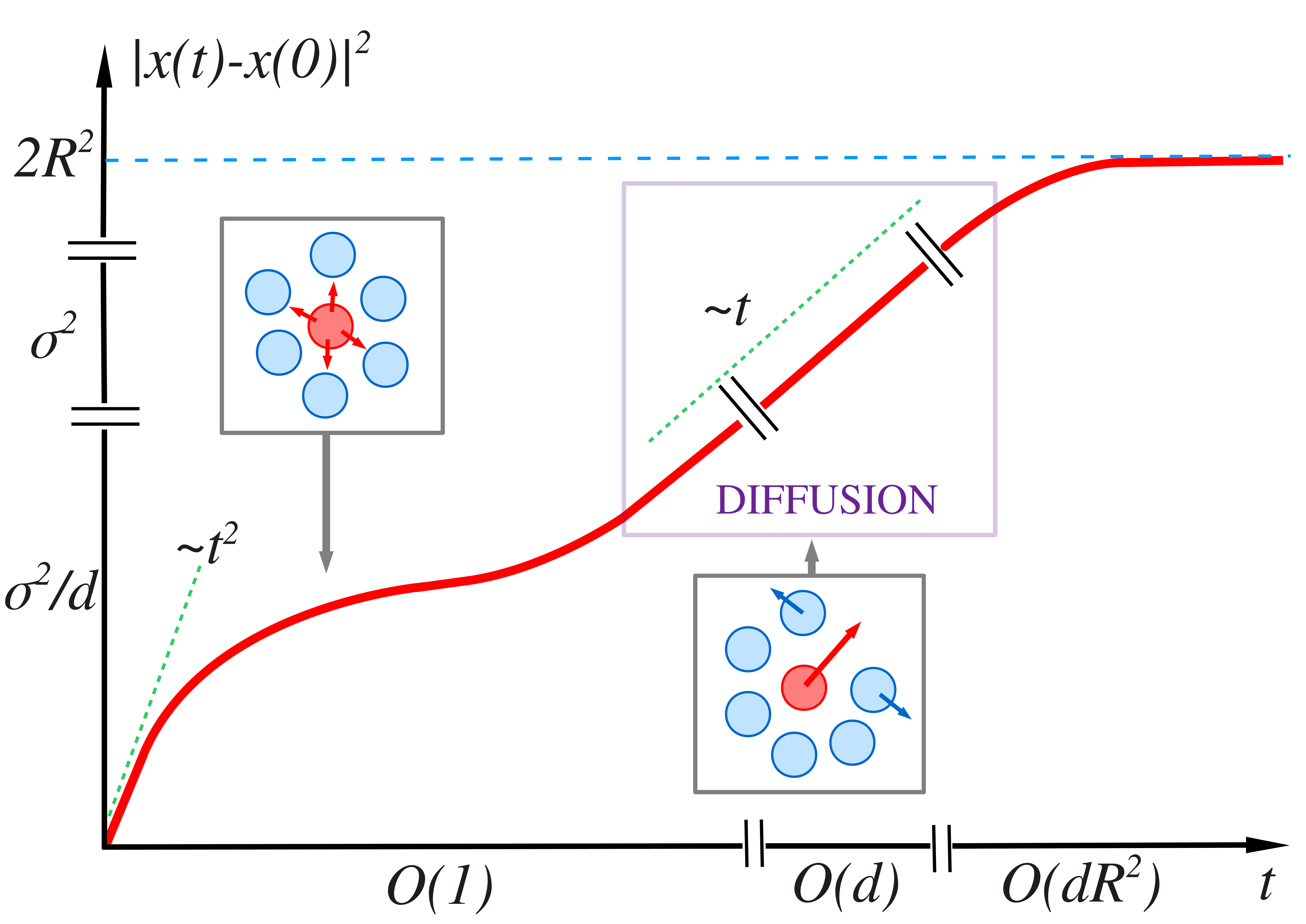} &
  \includegraphics[width=9cm]{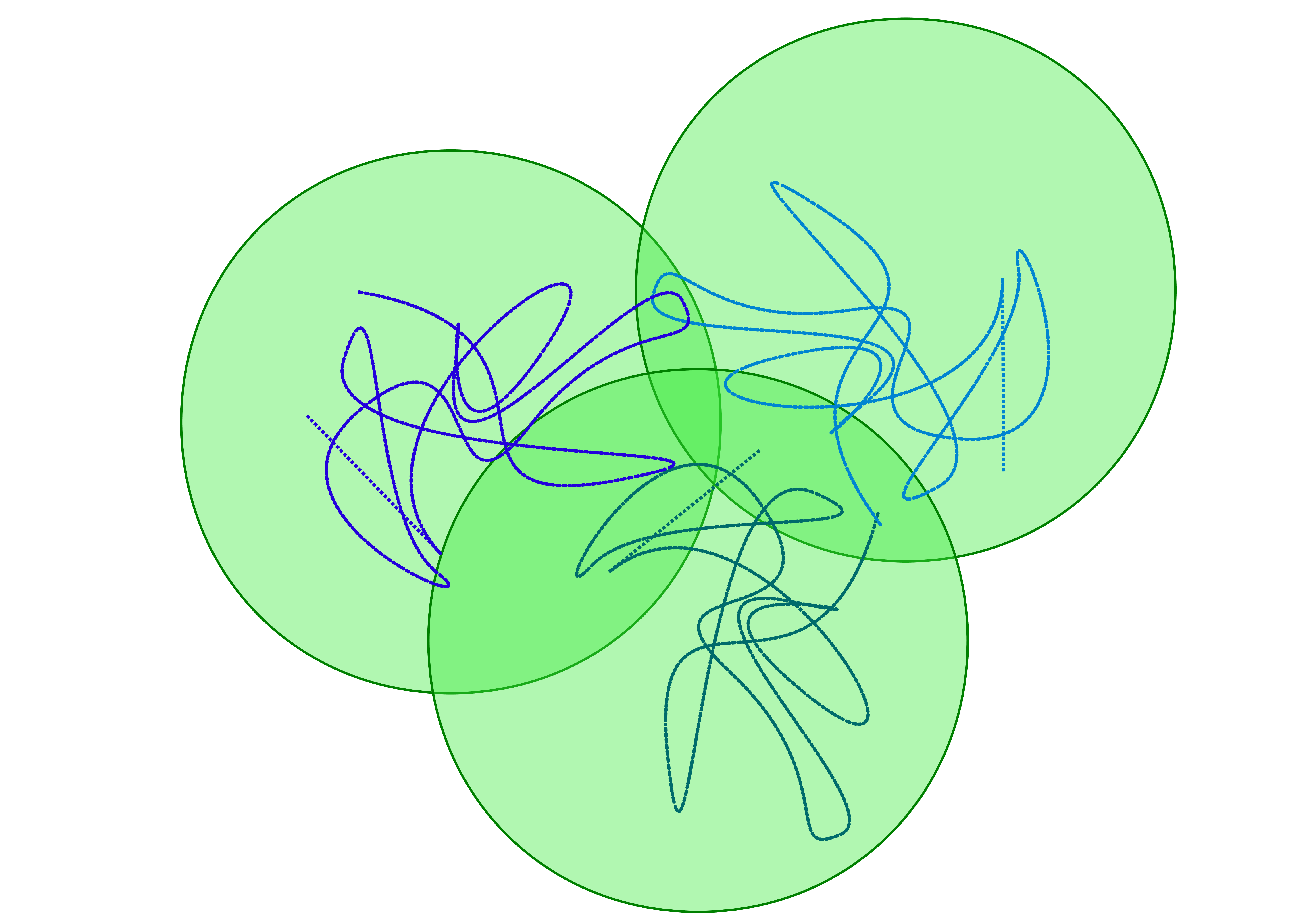} \\
  (a) & (b)
 \end{tabular}\label{fig:virialtraj}
\caption{(a) The different dynamical regimes, described by equation~\eqref{eq:MCTD}: most of the dynamics is determined at the cage size scaling $1/d$. We assume that the diffusive regime found already at this scale extends trivially at longer times. Concerning the slope at very short times, the mean-square displacement is $\propto t^2$ if inertia is not neglected; for purely overdamped motions it is $\propto t$. Finally, the particles feel the `box' for distances of order $R$, hence a saturation at this scale. (b) The dynamic triangle diagram with three trajectories.}
\end{figure}
To justify this truncation, let us consider for example the second term of this expansion (triangle diagram):
\begin{equation}\label{eq:tri}
 \begin{tikzpicture}[baseline={([yshift=-.5ex]current bounding box.center)}] \draw[line width=1pt] (0:0) -- (60:1) -- (0:1) -- (0:0);\draw[fill=black] (0:0) circle (0.1) ;\draw[fill=black] (60:1) circle (0.1) ;\draw[fill=black] (0:1) circle (0.1) ;\end{tikzpicture} =\frac{N^3}{3!}\int {\rm D}[x,\hat{x}]{\rm D}[y,\hat{y}]{\rm D}[z,\hat{z}]\,\rho[x,\hat{x}]\rho[y,\hat{y}]\rho[z,\hat{z}]f[x-y,\hat{x}-\hat{y}]f[y-z,\hat{y}-\hat{z}]f[z-x,\hat{z}-\hat{x}]
\end{equation}
For a finite-support potential $V$, 
$\nabla V(x-y)\propto \Theta(\s-\abs{x-y})$ (Heaviside's step function\footnote{In the main text it is noted $\th$ but this notation will be used for Grassmann variables from subsection~\ref{sub:susy} on.}), hence 
\begin{equation}
 f[x-y,\hat{x}-\hat{y}]=\left\{ \begin{array}{ll} 0& \mathrm{if}~\forall t,~\abs{x(t)-y(t)} > \s \\ \tau[x-y,\hat{x}-\hat{y}]& \mathrm{if}~\exists t,~\abs{x(t)-y(t)} < \s \end{array}\right. =[\Theta(\abs{x-y}-\s)-1]\tau[x-y,\hat{x}-\hat{y}] 
\end{equation}
Essentially, $\tau$ is just a numerical value depending on the details of $V$ (with $\abs{\tau}\leqslant 2$) and the physical content of $f$ lies in its finite support, \ie
\begin{equation}\label{eq:thetatraj}
f[x-y,\hat{x}-\hat{y}]\simeq \Theta(\abs{x-y}-\s)-1=-1+\prod_{n=1}^M\Theta(\abs{x^n-y^n}-\s)
\end{equation}
For finite times, a trajectory stays in a bounded region of space, represented as a ball of diameter the typical size $\propto\s/d$ of the trajectory in figure 1(b). The Mayer functions require that each couple of trajectories get closer than $\s$ at some time. To each set of three trajectories one can associate a corresponding static diagram with three overlapping balls (figure~1(b)). One can see this static diagram as the equivalence class of all dynamic diagrams with trajectories contained inside these balls. Actually there are lots of trajectories contained by these bounding balls that do not contribute because they do not get close enough. Then the sum over trajectories of the value of the integrand is at most of the same order of the static diagram value due to the normalization $\int {\rm D}[x,\hat{x}]\, \rho[x,\hat{x}]=1$ which accounts for the huge number of equivalent dynamic diagrams compared to the static one.
The same program applies to all terms in the expansion. We conclude, from~\cite{WRF87}, that the first term dominates the series\footnote{The critical packing fraction later found in~\eqref{eq:phid} is of order $ O(d/2^d)$ where it is known that the virial series does not converge anymore; nevertheless, Frisch \& Percus~\cite{FP99} have shown that in high $d$ the truncation used here is valid largely above this bound.} in the $d \to \infty$ limit. A more careful but maybe less intuitive argument is given in~\ref{app:virial}. Note that we did note specialize the discussion to hard spheres but to any finite-support interaction potential.
The result is then\footnote{In the thermodynamic limit $\mathrm{ln}\Xi/N=\mathrm{ln}Z_N/N$ (as if \textit{formally} the chemical potential was zero in the static partition functions analogy). That is why for simplicity we used $\mathrm{ln}Z_N/N$ in equation~(13) of the main text.}
\begin{equation}\label{eq:lnXixxt}
\boxed{
\SS\equiv\frac{\mathrm{ln}\Xi}N =-\int {\rm D}[x,\hat{x}] \,\rho[x,\hat{x}](\Phi[x,\hat{x}]+\mathrm{ln}\rho[x,\hat{x}])+\frac{N}{2}\int {\rm D}[x,\hat{x}] {\rm D}[y,\hat{y}] \,\rho[x,\hat{x}]\rho[y,\hat{y}]f[x-y,\hat{x}-\hat{y}]
}
\end{equation}
with $\boxed{\delta \SS/\delta \rho[x,\hat{x}]=0}$ from the Legendre transform and the normalization $\boxed{\int {\rm D}[x,\hat{x}]\, \rho[x,\hat{x}]=1}$. We neglected purely additive constants irrelevant for the dynamics.

\subsubsection{The Mayer expansion in infinite dimension}\label{app:virial}

We discuss here the truncation of the Mayer expansion when $d\to\io$ in a more pedestrian way. As in~\ref{sub:virial}, we focus on the triangle diagram in~\eqref{eq:tri} with the Mayer function given by~\eqref{eq:thetatraj}.\\
Consider two typical trajectories $x(t)$ and $y(t)$ which we expect from the static case to dominate the dynamics at large $d$, and let us focus on the positions $x^1$ and $y^1$ (any other couple would do). These typical trajectories scale as follows: the other positions of the trajectory $\{x^n\}_{n\neq1}$ (respectively $\{y^n\}_{n\neq1}$) fluctuate around $x^1$ (respectively $y^1$) over a neighborhood of size $1/d$, see figures~1(a),(b). Besides, the typical distance between the two trajectories is the diameter $\s$ (plus $ O(1/d)$ fluctuations). Then essentially $\prod_{n=1}^M\Theta(\abs{x^n-y^n}-\s)\simeq \Theta(|x^1-y^1|-\s)$ except in the neighborhood $\abs{x^1-y^1} \in [\s(1-\frac{\mu}{d}),\s(1+\frac{\mu}{d})]$ where these fluctuations of order $1/d$ matter ($\mu$ is of order 1). However this neighborhood has to be accounted for since its volume $\VV_d(\s)[(1+\frac{\mu}{d})^d-(1-\frac{\mu}{d})^d]\underset{d \to \infty}{\sim} \VV_d(\s)(e^{\mu}-e^{-\mu})$, is comparable to the volume $\VV_d(\s)e^{-\mu}$ of $\abs{x^1-y^1} \in [0,\s(1-\frac{\mu}{d})]$ where the contribution of $f[x-y,\hat x-\hat y]$~\eqref{eq:thetatraj} is also non zero 
(for $\abs{x^1-y^1} > \s(1+\frac{\mu}{d})$, $f[x-y,\hat{x}-\hat{y}]$ is zero).\\
Hence $f[x-y,\hat{x}-\hat{y}]\simeq f^*(x^1-y^1)$ with $f^*(x)$ being -1 for $\abs{x} \in [0,\s(1-\frac{\mu}{d})]$, of order 1 for $\abs{x} \in [\s(1-\frac{\mu}{d}),\s(1+\frac{\mu}{d})]$ and 0 further. What counts for the truncation done in~\cite{WRF87} is that $\int_{[0,\a]^d} f^* \underset{d \to \infty}{\propto} \VV_d(\s)$ with $\a$ around\footnote{More generally bounded by a finite and independent of $d$ quantity (for example here it is bounded by $2\s$ for large enough $d$).} $\s$ and $f^*$ is zero elsewhere. This is the case here since
\begin{equation}
\int_{[0,\s(1+\frac{\mu}{d})]^d} f^*\underset{d \to \infty}{\sim}-\VV_d(\s)e^{-\mu}+\underbrace{\int_{[\s(1-\frac{\mu}{d}),\s(1+\frac{\mu}{d})]^d} f^*}_{\propto \VV_d(\s)(e^{\mu}-e^{-\mu})}
\end{equation}
Thus we will replace $f[x-y,\hat{x}-\hat{y}]\simeq f_{\rm HS}(x^1-y^1)=-1+\Theta(\abs{x^1-y^1}-\s)$, where $f_{\rm HS}$ is the usual static Mayer function of hard spheres, since it does not change the scalings given by $f^*$ up to an irrelevant proportional constant. The term in equation~\eqref{eq:tri} reduces to
\begin{equation}
 \begin{tikzpicture}[baseline={([yshift=-.5ex]current bounding box.center)}] \draw[line width=1pt] (0:0) -- (60:1) -- (0:1) -- (0:0);\draw[fill=black] (0:0) circle (0.1) ;\draw[fill=black] (60:1) circle (0.1) ;\draw[fill=black] (0:1) circle (0.1) ;\end{tikzpicture}\simeq\frac{N^3}{3!} \int {\rm D}[x,\hat{x}]{\rm D}[y,\hat{y}]{\rm D}[z,\hat{z}]\,\rho[x,\hat{x}]\rho[y,\hat{y}]\rho[z,\hat{z}]f_{\rm HS}(x^1-y^1)f_{\rm HS}(y^1-z^1)f_{\rm HS}(z^1-x^1)
\end{equation}
We can integrate on all variables except $x^1$, $y^1$ and $z^1$ because 
\begin{equation}
 \int \frac{{\rm d}x^2}{(2\pi)^{\frac{d}{2}}} \ldots \frac{{\rm d}x^M}{(2\pi)^{\frac{d}{2}}} {\rm D}\hat{x}\,\rho(x^1,\ldots, x^M, \hat{x})=\rho(x^1)={\rm constant}=\frac{(2\pi)^{\frac{d}{2}}}{\VV}
\end{equation}
since $\rho(x^1)={\left\langle \frac1N\sum_{i=1}^N \d(x^1-x_i^1)\right\rangle}_{\Xi}$ must be a translation invariant function of $x^1$ \ie a constant. $\VV$ is the volume of the liquid. So the second term in the expansion is actually approximated for large dimension by
\begin{equation}
 \begin{tikzpicture}[baseline={([yshift=-.5ex]current bounding box.center)}] \draw[line width=1pt] (0:0) -- (60:1) -- (0:1) -- (0:0);\draw[fill=black] (0:0) circle (0.1) ;\draw[fill=black] (60:1) circle (0.1) ;\draw[fill=black] (0:1) circle (0.1) ;\end{tikzpicture}\simeq\frac{\r^3}{3!}\int \dd x\dd y\dd z\, f_{\rm HS}(x-y)f_{\rm HS}(y-z)f_{\rm HS}(z-x)
\end{equation}
where we defined the average particle density $\r=N/\VV$. We recognize the same term as in the usual static Mayer expansion of hard spheres and can conclude as in~\ref{sub:virial} from~\cite{WRF87}.

\subsection{Spherical setup}\label{sec:sphere}

 From now on, we constrain the particles to live on the surface of a sphere of radius $R$ embedded in $d+1$ dimensional space, $\mathbb{S}^d(R)$, see section~\ref{sec:intro}.
The field $x(t)$ is promoted to 
$x\,:\,\RRR\longrightarrow\RRR^{d+1}$ and $\forall t,~x(t)^2=R^2$ must be verified. Concerning the response field $\hat{x}(t)$, we will rather constrain it to be orthogonal to the position field: $\forall t,~x(t)\cdot\hat x (t)=0$, thus living at each time in the hyperplane tangential to the sphere at $x(t)$, cf. figure~2. This way we ensure that we will recover rotation and translation invariances for position fields and only rotation invariance for response fields once $R\to\io$ is taken\footnote{Another way to see it is that the spherical constraint implies $\dot{x}_i\cdot x_i=0$ and only tangential fields $\hat x_i$ are needed to exponentiate the Langevin equation in the MSRDDJ path integral.}.\\
There are many possible ways to enforce these constraints, which are eventually equivalent:
\begin{enumerate}
\item One strategy is to use Lagrange multipliers via a term $-\nu_i(t)x_i(t)$ in the right-hand side of~\eqref{eq:Langevin}, promoted to $d+1$ dimensions, which ensures that the trajectory does not get out of the sphere due to interactions or thermal noise~\cite{CC05}. It is the one adopted in the main text for concision. 
 \newcounter{enumTemp}
 \setcounter{enumTemp}{\theenumi}
\end{enumerate}  
In $d\to\io$, the value of $\n_i\sim\n$ is non-fluctuating and is obtained by discretizing~\eqref{eq:Langevin} as follows (in the It\^o sense):
\beq
x_i(t+\dd t) = x_i(t) -\frac1\g \nu_i(t) x_i(t) \dd t - \frac1\g \nabla_{x_i}H \dd t + \frac1\g\h_i \ ,
\hskip30pt
\langle \h_i^\m \h^\n_j \rangle = 2 \g T  \dd t \d_{ij} \d_{\m\n} 
\eeq
At order $\dd t$, using that for large $d$ one has $B \cdot \h_i \to 0$ (for any vector $B$ uncorrelated with $\h_i$ in the It\^o sense) and $\h_i \cdot \h_i \sim 2 d T \g \dd t$ due to the central limit theorem, we impose the constraint
\beq
R^2 = x_i(t+\dd t) \cdot x_i(t+\dd t) = x_i(t)\cdot x_i(t) - \frac{2 \dd t}\g x_i(t) \cdot \left[
\n_i(t) x_i(t) +\nabla_{x_i}H  \right] +\frac{2d T \dd t}{\g} 
\eeq
and therefore
\beq\label{eq:HSmu}
\n_i(t) = - \frac1{R^2} x_i \cdot  \nabla_{x_i} H + \frac{d \, T}{R^2} 
\eeq
We have a general relation~\cite[Eq.(2.2.10)]{hansen} for the reduced pressure $p$: 
\beq\label{eq:pdef}
p \equiv \frac{\b P}{\r} = 1 - \frac{\b}{d \, N} \la \sum_i x_i\cdot\nabla_{x_i}H \ra
\eeq
For $d\to\io$ the fluctuations vanish because we average over $d$ dimensions, we thus have
\beq
\frac 1N\sum_i x_i\cdot\nabla_{x_i}H\sim \frac 1N\la \sum_i x_i\cdot\nabla_{x_i}H \ra = d T (1-p)
\eeq
and plugging this in Eq.~\eqref{eq:HSmu} we obtain 
that all $\nu_i(t)$ are equal and constant in time, given by
\beq
\boxed{\nu_i(t) \sim \nu =  \frac{d \, T}{R^2} p}
\eeq
From the static entropy in $d\to\io$~\cite{WRF87,FP99,PZ10,MK11}, $p = 1 + d\wh\f/2$ in the liquid phase, where $\wh\f$ is the rescaled packing fraction $\wh\varphi=2^d\varphi/d=\r \VV_d(\s)/d$ and the number density $\r=N/V$.\\
This choice results in an additional term $\int \dd t\,\n \hat x\cdot x$ in the definition of $\Phi$~\eqref{eq:action}, and the summation of paths in~\eqref{eq:lnXixxt} is over the sphere $\mathbb{S}^d(R)$ for positions and the tangential hyperplane. In practice, we will integrate on the whole $d+1$ dimensional space and enforce the constraint through Dirac deltas. Note that, with $x$ (resp. $\hat x$) representing the position (resp. response) field at some time and $E=\textrm{Span}(x)$, 
\begin{equation}
 \begin{split}
\int_{\RRR^{d+1}} \dd x\, \d(x^2 - R^2)=\Omega_{d+1} \int_0^\io \dd r \, r^d& \d(r^2 - R^2)=\Omega_{d+1} \frac{R^d}{2R}=\frac{\VV}{2R}= \frac{1}{2R} \int_\VV \dd x \\
\int_{\RRR^{d+1}} \dd \hat x\,\d(2x\cdot\hat x)=\int_{E \times E^{\perp}} \dd \hat x_{\sslash}\dd \hat x_{\perp}\,\d(2\abs{x}\hat x_{\sslash})&=\int_{E \times E^{\perp}} \dd \hat x_{\sslash}\dd \hat x_{\perp}\,\d(2R\hat x_{\sslash})=\frac{1}{2R}\int_{E^{\perp}}\dd \hat x_{\perp}
\end{split}
\end{equation}
These choices rescale the path integral measures with respect to the ones on $\RRR^d$, which does not affect the dynamics. 
In the thermodynamic limit of infinite radius $R$ (with $\r=N/\VV$ fixed), we recover the original $d$-dimensional space. 

\begin{enumerate}
\setcounter{enumi}{\theenumTemp}
\item In the previous method, we would exponentiate the Dirac delta functions, giving additional terms in the exponent $\int\dd t\,\m x\cdot\hat x$ and $\int\dd t\, \hat\m(x^2-R^2)$. We see that the Lagrange multiplier would just shift $\m$. Hence, another technique is directly (and somewhat physically blindly) to promote the path integrals in~\eqref{eq:lnXixxt} to $d+1$ dimensions and use Dirac deltas to constrain the $x$, $\hat{x}$ fields. This is what we are going to follow with fields $\n$, $\hat{\n}$ but note that \textit{it is not exactly the Lagrange multiplier}, even if it plays a similar role.
\item We might as well add a soft constraint $A\sum_i(x_i^2-R^2)^2$ in the Hamiltonian. It would add a single-particle term $2A\int\dd t \, i\hat x\cdot x(x^2-R^2)$ to $\Phi$. Then we can write (still at a given time)
\beq
e^{-2Ai\hat x\cdot x(x^2-R^2)}\propto\int_{\RRR^{d+1}}\dd\hat{\n}\,\d\left(\frac{\hat{\n}}{2A}-\hat{x}\cdot x\right)e^{-i\hat{\n}(x^2-R^2)}\propto\int_{\RRR^{d+1}}\dd\n\dd\hat\n\,e^{i\n\hat{\n}/A-2i\n x\cdot\hat{x}-i\hat{\n}(x^2-R^2)}
\eeq
and in the hard limit $A\to\io$ this is the same as enforcing the constraints through $\d(x^2 - R^2)$, $\d(2x\cdot\hat x)$.
\end{enumerate}

\subsection{Superspace notation}\label{sub:susy}
\subsubsection{Translation of the dynamics into superfield language}
As a compact way to write dynamical equations, we will use superspace notation~\cite{K92,K03}. At any step, one can unfold this notation to recover the standard dynamical variables. $\{{\theta}_i,\bar{{\theta}}_i\}$ are Grassmann variables\footnote{Let us emphasize here that this is only a compact notation without any physical content (for our present purpose), but we note by way of excuse that the computation is prohibitively complicated proceeding otherwise.}. Let us define $\tilde x=i\hat x$ for convenience. We encode the position and response fields $[x,\hat x]$ in a superfield\footnote{We do not need to resort to purely fermionic components (\ie linear in $\th_1$ and $\bar\th_1$) in the superfield $\bm x$ since we interpreted the Langevin equation in the It\^o sense, implying the MSRDDJ Jacobian is 1. Any other discretization scheme can be equivalently considered by adding two such ghosts.} $\bm{x}(a)=x(t)+\bar{{\theta}}_1{\theta}_1 \tilde{x}(t)$, where arguments are denoted by $a=({\theta}_1,\bar{{\theta}}_1,t)$. The Mayer function and the kinetic part can be explicitly written:
\begin{equation}\label{eq:fphi}
\begin{split}
f(\bm{x})&=e^{-\int \dd a \,V(\bm{x})}-1 ~~~~\textrm{with}~~\int \dd a\,=\int \dd {\theta}_1 \dd  \bar{\theta}_1 \dd t\\
\Phi(\bm{x})&=\g \int \dd a\, \frac{\partial \bm{x}}{\partial {\theta}_1}\cdot \left(T\frac{\partial \bm{x}}{\partial \bar{\theta}_1}-{\theta}_1 \frac{\partial \bm{x}}{\partial t}\right)
\end{split}
\end{equation}
$\bm x(a)^2=x(t)^2+2\bar{\th}_1\th_1x(t)\cdot \tilde x(t)$ implies for the constraints that $\d(x(t)^2-R^2)\d(2x(t)\cdot \hat{x}(t))=\d(\bm x(a)^2-R^2)$.\\
The measure $\mathrm{D}[x,\hat{x}]$ is replaced by $\mathrm{D}\bm{x}=\mathrm{D}[x,\hat{x}]$ where integration over $\hat x$ is on the `imaginary axis' $i\RRR^{d+1}$. Then the action can be written in the form:
\begin{equation}\label{eq:SSspace}
\boxed{
\SS =-\int \mathrm{D}\bm{x}\, \d(\bm x(a)^2-R^2)\rho(\bm{x})(\mathrm{ln}\rho(\bm{x})+\Phi(\bm{x}))+\frac{N}{2}\int \mathrm{D}\bm{x}\mathrm{D}\bm{y} \,\d(\bm x(a)^2-R^2) \d(\bm y (b)^2-R^2) \rho(\bm{x})\rho(\bm{y})f(\bm{x}-\bm{y})
}
\end{equation}
still with $\boxed{\d\SS/\delta \rho(\bm{x})=0}$ and $\boxed{\int \mathrm{D}\bm{x}\, \d(\bm x(a)^2-R^2) \rho(\bm{x})=1}$.

\subsubsection{Definitions about superfields} \label{app}

Here we provide some useful generalization of usual field-theoretic tools and other identities in superfield language, used in the following sections. Proofs are not given here, but are easily obtained using the definitions and by direct computations.
\begin{enumerate}
\item \underline{\textit{Superfields:}} We will consider (one-component) superfields and define them as $\bm x(a)=x(t)+ \bar{{\theta}}_1{\theta}_1\hat x(t)$.\\We will also use operator (two components) superfields analogous to the replica case, such as $\bm{q}(a,b)=\bm{x}(a)\cdot \bm{x}(b)$. Similarly, an operator superfield $\bm r$ can be cast in the canonical expression with Grassmann variables and real scalar fields:
 \begin{equation}
 \label{eq:notq}
 \bm{r}(a,b)=r_1(t,t')+\bar{{\theta}}_1{\theta}_1\hat{r}_1(t,t')+\bar{{\theta}}_2{\theta}_2\hat{r}_2(t,t')+\bar{{\theta}}_1{\theta}_1\bar{{\theta}}_2{\theta}_2 r_2(t,t')
 \end{equation}
\item \underline{\textit{Dirac deltas:}} For superfields, they are simply functionally defined as $\delta(\bm{x}(a))=\delta(x(t))\delta(\hat{x}(t))$. For operator superfields, they are simply functionally defined as a product of the functional deltas of their components appearing in notation~\eqref{eq:notq}. If the superfield is symmetric we need to introduce deltas only on the independent part, which is the case for $\bm q$.
\item \underline{\textit{Path integral measure:}} We clarify here the path integral measure for future needs. For a general superfield $\bm r$, the path integral measure is defined as $\mathrm{D}\bm r=\mathrm{D}[r_1,\hat{r}_1,\hat{r}_2,r_2]$. For symmetric superfields such as $\bm q$, we will only sum on the symmetric components of it, and call it $\mathrm{D}^{s}\bm q=\mathrm{D}^{s}[q_1,q_2]\mathrm{D}[\hat{q}_1,\hat{q}_2]$ with $\mathrm{D}^{s}q_1=\mathrm{D}q_1\prod_{t>t'}\d(q_1(t,t')-q_1(t',t))$ in a discretized point of view. Therefore $\bm q(a,b)$ with `$a>b$' (loosely speaking) will appear in the path integral, the previous Dirac deltas imposing them to be $\bm q(b,a)$, so that we can use all components and introduce $\bm q$ as a symmetric superfield in the path integral as it should be from its definition.
 \item \underline{\textit{Product of superfields:}} We define the product of two superfields $\bm r=\bm p\bm q$ as the generalization of the operator product
 \begin{equation}
  \bm r(a,b)=\int \dd c\, \bm p(a,c)\bm q(c,b)
 \end{equation}
\item \underline{\textit{Identity:}} The identity for superfields reads in components
 \begin{equation}
  \bm{1}(a,b)=\bar{{\theta}}_1{\theta}_1\delta(t-t')+\bar{{\theta}_2{\theta}_2\d(t-t')}
 \end{equation}
\item \underline{\textit{Inverse:}} The definition of the inverse $\bm r^{-1}$ in the superfield sense reads, through the relation $\bm r^{-1}\bm r=\bm r\bm r^{-1}=\bm 1$,
\begin{equation}
 \bm r^{-1}=\b+\bar{{\theta}}_1{\theta}_1\b\a+\bar{{\theta}}_2{\theta}_2\g\b+\bar{{\theta}}_1{\theta}_1\bar{{\theta}}_2{\theta}_2 (\g\b\a+r_1^{-1})
 \end{equation}
 with $\b=(r_2-\hat{r}_1r_1^{-1}\hat{r}_2)^{-1}$, $\g=-r_1^{-1}\hat{r}_2 $ and $\a=-\hat{r}_1r_1^{-1}$. 
\item\underline{\textit{Determinant:}} We define a `superdeterminant' for superfields as
 \begin{equation}\label{eq:sdet}
 \sdet(\bm r)=\det(r_1)\det(r_2-\hat{r}_1r_1^{-1}\hat{r}_2)
 \end{equation}
\item \underline{\textit{Trace:}} The trace is defined by $\str(\bm r)=\int \dd a \,\bm r(a,a)=\int \dd t\, (\hat{r}_1+\hat{r}_2)(t,t)$. 
\item\underline{\textit{Integral representation of Dirac deltas for superfields:}} 
an integral representation is expressed as
 \begin{equation}
 \begin{split}
   \d(\bm q)&=\int \mathrm{D}\bm p\, e^{ i\int \dd a\dd b\,\bm p(a,b)\bm q(b,a)}=\int \mathrm{D}\bm p\, e^{ i\str(\bm p\bm q)} ~~~~ \rm for~a~superfield~operator\\
   \d(\bm x)&=\int \mathrm{D}\bm y\, e^{ i\int \dd a\,\bm y(a)\cdot\bm x(a)} ~~~~ \textrm{similarly~for~a~one-component~superfield}
 \end{split}
 \end{equation}
 For a symmetric superfield such as $\bm q(a,b)=\bm x(a) \cdot \bm x(b)$, we only need to introduce the independent part of it, that is, taking the additional superfield $\bm p$ to be also symmetric, we only have to sum in the exponential over half of $\int \dd a\dd b\,\bm p(a,b)\bm q(a,b)$. Rescaling $\bm p$ with the factor $\frac{1}{2}$, the formula above is unchanged\footnote{All numerical constants, such as this one coming from the rescaling, will be omitted as we will eventually calculate all proportionality constants in another way.} provided that the measure is understood as $\mathrm{D}^{\mathrm{s}} \bm p$ since $\bm p$ is symmetric. 
\item\underline{\textit{Derivation with respect to a superfield:}} Defining the derivative with respect to a superfield in a way similar to operators, using the convention for functional derivation:
\beq
\frac{\d \bm r (a,b)}{\d \bm r (c,d)}=\bm 1(a,c)\bm 1(b,d)
\eeq
where $\bm r$ is a superfield, and $\bm 1$ is the equivalent of the Dirac delta for superfields. Similarly for a symmetric superfield $\bm q$ we get 
 \beq
 \frac{\d \bm q (a,b)}{\d \bm q (c,d)}=\bm 1(a,c)\bm 1(b,d)+\bm 1(a,d)\bm 1(b,c)
 \eeq
 which gives the factors 2. We will thus use the formulas 
 \beq
 \frac{\d \ln\sdet\bm q}{\d \bm q}=2\bm q^{-1}\;,\hskip15pt
 \frac{\d \str(\bm p\bm q)}{\d \bm q}=2\bm p\;,\hskip15pt
 \frac{\d}{\d\bm q}\str(\bm{q}^{-1}\bm p)=-2\bm{q}^{-1}\bm{p}\bm{q}^{-1}
 \eeq
\item\underline{\textit{Gaussian integration on superfields:}} A direct calculation with components shows that the Gaussian integral can be cast in the familiar form:
 \begin{equation}
\int \mathrm{D}\bm x\, e^{-\frac{1}{2}\int \dd a\dd b\,\bm q(a,b)\bm x(a)\bm x(b)+\int \dd a\,\bm h(a)\bm x(a)}= \frac{e^{\frac{1}{2}\int \dd a\dd b\,\bm q^{-1}(a,b)\bm h(a)\bm h(b)}}{\sqrt{\sdet(\bm q)}}
 \end{equation}
 $\bm h$ and $\bm x$ being here scalar fields for the sake of clarity (\ie $d=1$). $\bm q$ is assumed symmetric, \ie $q_1$ and $q_2$ are symmetric operators and $\hat{q}_1=\hat{q}_2^\mathrm{T}$. The Gaussian integrations require that $q_1$ (respectively $q_2$) and $q_2-\hat{q}_1q_1^{-1}\hat{q}_2$ (respectively $q_1-\hat{q}_2q_2^{-1}\hat{q}_1$) are positive definite.
\item\underline{\textit{Gaussian moments:}} Moments of a Gaussian random variable in notation SUSY enjoy similar properties than discrete Gaussian random variables. As an example, we can compute the following variance:
 \beq
\sqrt{\sdet(\bm q)}\int \mathrm{D}\bm x \,\bm x(a)\bm x (b) e^{-\frac12\int \dd a\dd b\,\bm q(a,b)\bm x(a)\bm x(b)}=\bm q^{-1}(a,b)
 \eeq
\item\underline{\textit{Other useful identities:}} The following relations are useful for the derivation of the interaction term, for any superfields $\bm A$ and $\bm B$ such that the series converge:
 \beq\label{eq:powerseries}
 \begin{split}
 \ln\sdet(\bm A + \bm B)&=\ln\sdet\bm A+\str\sum_{n\geqslant1}\frac{(-1)^{n-1}}{n}({\bm A}^{-1}\bm B)^n=\ln\sdet\bm A+\str\sum_{n\geqslant1}\frac{(-1)^{n-1}}{n}(\bm B{\bm A}^{-1})^n\\
 (\bm A+\bm B)^{-1}&=\left(\sum_{n\geqslant0}(-1)^n({\bm A}^{-1}\bm B)^n\right)\bm A^{-1}=\bm A^{-1}\sum_{n\geqslant0}(-1)^n(\bm B{\bm A}^{-1})^n
 \end{split}
 \eeq
These formulas are just generalizations of power series to superfields. The radius of convergence of the series used is 1.
\item\underline{\textit{The projector $\bm P$:}} We define a superfield $\bm P(a,b)=1$ and $\l$ a scalar. We get the following important formulas:
\beq\label{eq:lndetinv}
 \begin{split}
  \ln\sdet(\l\bm P + \bm B)&=\ln\sdet(\bm B)+\ln\left[1+\l\str(\bm P\bm B^{-1})\right]\\
  (\l\bm P+\bm B)^{-1}&=\bm B^{-1}\left[\bm 1 -\frac{\l}{1+\l\str(\bm P\bm B^{-1})}\bm P \bm B^{-1}\right]
 \end{split}
\eeq
 Both formulas require $\abs{\str(\bm P\bm B^{-1})}<1/\abs{\l}$.
 
\end{enumerate}

\section{Translational and rotational invariances}\label{sec:inv}
We now take into account rotational and translational invariances and take the limit $d\to\io$. In some cases the order of the two limits is irrelevant, but when relevant, we should take the $R\to\io$ limit first. In other words, we should consider for example that $R/d$ is a large quantity.

\subsection{`Functional spherical coordinates': invariances using the mean-square displacement}\label{sub:invar}

As emphasized in section~\ref{sec:intro}, the aim of the introduction of $\mathbb{S}^d(R)$ is to take into account both translation and rotation invariances on Euclidean $d$ dimensional space by only rotation invariance on a sphere of a $d+1$ dimensional space, which is actually easier to handle in the viewpoint of the dynamics. Indeed, the MSRDDJ action $\AA$ in~\eqref{eq:action}, now promoted to $d+1$ dimensional fields, and the constraints in~\eqref{eq:SSspace} are invariant by the same rotation $\RR$ for both fields \ie $(x,\hat{x})\rightarrow (\RR x,\RR \hat{x})$, which is transposed to superfields as a global rotation\footnote{See subsection~\ref{sec:sphere} and footnote~\ref{ftn:resp}.} $\bm{x}\rightarrow \RR \bm{x}$. \\
Now considering expression~\eqref{eq:SSspace}, we define a superfield $\bm{q}(a,b)=\bm{x}(a)\cdot \bm{x}(b)$. We assume that the $d+1$ dimensional liquid is invariant by rotation \ie $\rho(\bm{x})=\rho(\{ \bm{x}(a)\cdot \bm{x}(b)\}_{a,b})=\rho(\bm{q})$. This way we will remove all irrelevant variables and be able to use a saddle point method.\\
Eventually, as regards the $R\to\io$ limit, it is more convenient to consider the mean-square displacement (MSD) $\bm{D}(a,b)=(\bm x(a)-\bm x(b))^2$ since it is a finite quantity as long as the difference $t-t'$ is finite, at equilibrium. Before the dynamical transition is met, $\bm{D}$ is of order $R^2$ when $t-t'\to\io$. We thus expect an artificial second plateau of order $R^2$ due to finite size effects, that will be removed when $R\to\io$, giving back diffusion at long times (see~figure 1(a)). One can check explicitly that in this limit the original $d$-dimensional MSD is recovered and that it is translation and rotation invariant for the position and rotation invariant for the response field, see~figure 2.
\begin{figure}\label{fig:sphere}
 \centering
 \includegraphics[width=7cm,trim={32cm 13cm 21cm 15cm},clip]{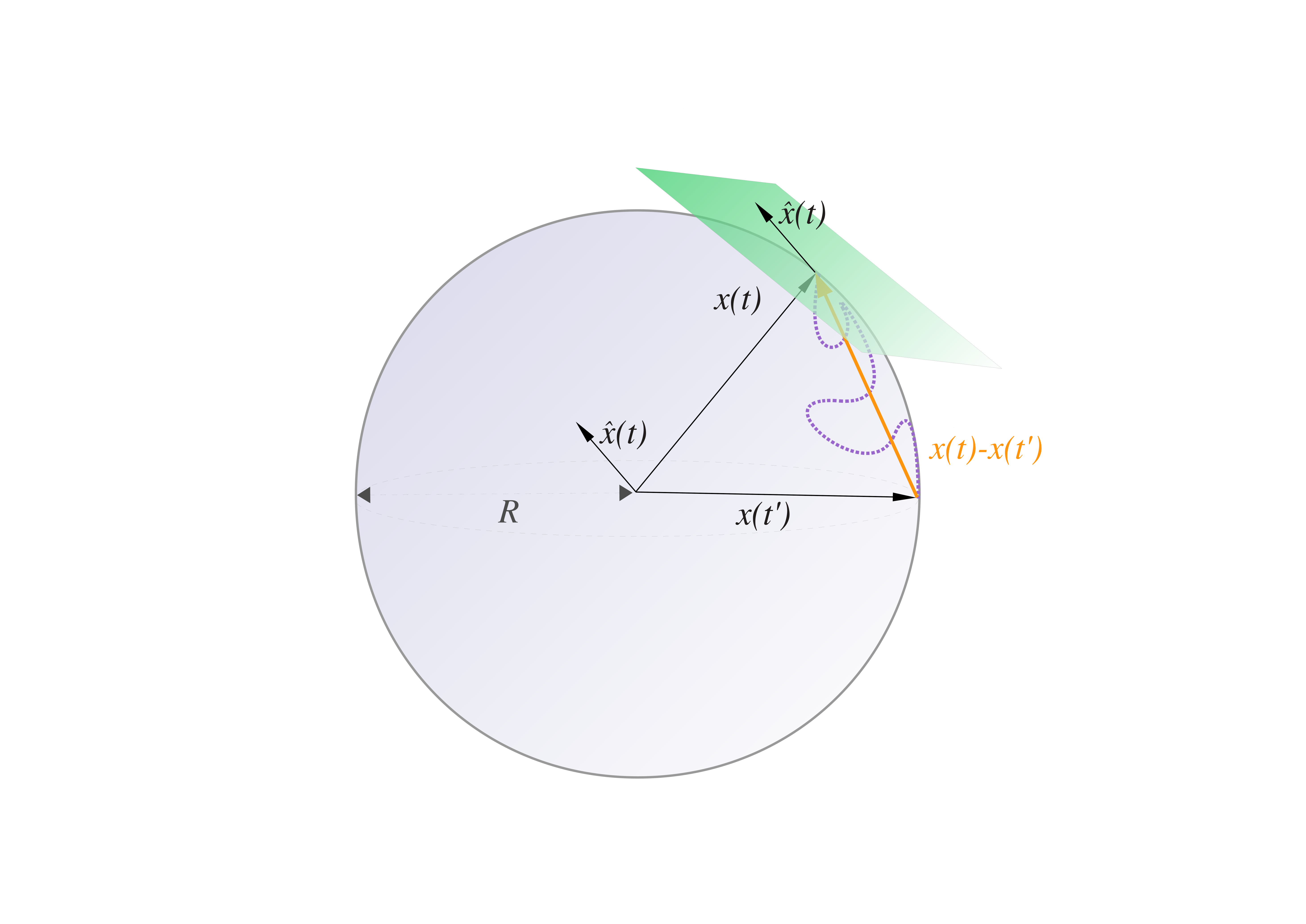}
 \caption{Notations on the sphere for $x$, $\hat x$. The vector $x(t)-x(t')$ is along a chord and when $R\to\io$, it lives in the same $d$-dimensional space as $\hat x$.}
\end{figure}

\subsection{Scalings in the infinite $d$ limit}\label{sub:scalings}
First, from now on we measure distances in units of the diameter $\s$, \ie we take $\s=1$.\\
To use the saddle point method, we need to specify how the quantities defined here scale with dimension. From the statics in~\cite{PZ10}, if we define $u=x-X$ where $X$ denotes the translational degree of freedom of a particle so that $u$ characterizes motion around an average position, $u$ scales as $1/d$. As $\Phi$ is translation invariant along $x$,
\begin{equation}
\Phi[x,\tilde{x}]=\Phi[u,\tilde{u}\equiv\tilde{x}]=\gamma \int \dd t \left(-T\tilde{u}^2 + \tilde{u}\cdot\dot{u} \right)
\end{equation}
$\dot{u} \sim 1/d$ implies that, for the action to describe the Langevin process above (\eqref{eq:Langevin}), we need the two terms in the integral to scale identically in the limit $d \to \infty$. Thus $\tilde{u}\sim 1/d$ as well. This implies that $\boxed{\bm{D}=\bm\D/d}$ where $\bm\D$ is of order 1. Hence, for convenience, noticing that $\bm{D}=2R^2\bm P-2\bm q$, we define the rescaled quantities
\beq
\boxed{\Dl=2dR^2}\,,\hskip20pt \boxed{\bm Q=\Dl\bm P-\sd=2d\bm q}
\eeq
$\Dl$ corresponds to a typical MSD between particles (in the liquid phase) on the sphere $\mathbb{S}^d(R)$. $\bm Q(a,a)=\Dl$ is the spherical constraint.\\
We then choose $\boxed{\g = 2\wh{\g} d^2}$ in order to write $\boxed{\Phi=d\wh{\Phi}}$ with $\wh{\g}$ and $\wh{\Phi}$ of order 1 so that $\Phi(\bm x)$ scales like $\mathrm{ln}\rho(\bm x)$, otherwise the infinite $d$ limit would not be well defined. \\
Indeed, from the statics of the liquid phase~\cite{WRF87,FP99,PZ10,MK11,KPZ12}, we will assume that $\rho$ is exponential in $d$, that is
\begin{equation}
 \boxed{\rho(\bm q)=\Lambda(\bm Q)e^{d\Omega(\bm Q)}} ~~~~ \textrm{\ie} ~~~~ \mathrm{ln}\rho(\bm q)\underset{d \to \infty}{\sim} d\Omega(\bm Q)
\end{equation}
with $\Omega$ of order 1 and $\Lambda$ a subdominant factor (\ie non-exponential in $d$).\\
Concerning the potential $V(r)$, $r$ will typically scale here as $r=\s(1+\frac{\m}{d})$ where $\m$ is of order 1. Besides the two terms in the action $\AA$ in~\eqref{eq:action} must have the same scalings to describe the underlying Langevin process~\eqref{eq:Langevin}. As $\AA$ can be cast in a form similar to~\eqref{eq:lnXixxt} 
using $\rho[x, \tilde{x}]$, thus similar to an ideal gas term plus an interaction term with `propagator' $W$, the following derivation of the infinite $d$ limit will indicate that such interaction terms have the same scalings as the ideal gas term only if the `propagators' $W$ (or $f$) scales as $\wh{\Phi}$ \ie 1 here. Thus $V$, which scales like $W$ (cf.~\eqref{eq:fphi}), must be of order 1. For example, the soft harmonic spheres potential is $V_{\rm SS}(r)=\k (1-r)^2\Theta(1-r)=\k (\m/d)^2\Theta(-\m)$. The amplitude must be $\k=\hat{\k}d^2$ with $\hat{\k}$ of order 1.

\subsection{Ideal gas term}\label{sub:norm1}

We write equation~\eqref{eq:SSspace} as $\SS=\SS_{\mathrm{IG}}+\SS_{\mathrm{int}}$ and focus on the ideal gas term $\SS_{\mathrm{IG}}$.
Exploiting the invariances, we apply the program mentioned in~\ref{sub:invar}:
\beq\label{eq:chig}
\begin{split}
 \SS_{\mathrm{IG}}&\propto\int \mathrm{D}\bm{x}\, \d(\bm x(a)^2-R^2)\rho(\bm{x})(\mathrm{ln}\rho(\bm{x})+\Phi(\bm{x}))\\
 &=\int \mathrm{D}\bm{x}\mathrm{D}^{\mathrm{s}}\bm{q}\, \d(\bm q(a,b)-\bm x(a)\cdot\bm x(b))\d(\bm q(a,a)-R^2)\rho(\bm{q})(\mathrm{ln}\rho(\bm{q})+\Phi(\bm{q}))\\
 &=\int \mathrm{D}\bm{x}\mathrm{D}^{\mathrm{s}}\bm{q} \mathrm{D}^{\mathrm{s}}\bm{q'}\,\d(\bm q(a,a)-R^2)e^{i\str(\bm{qq'})-i\int\dd a\dd b\,\bm{q'}(a,b)\bm x(a)\cdot\bm x(b)}\rho(\bm{q})(\mathrm{ln}\rho(\bm{q})+\Phi(\bm{q}))\\
 &=\int \mathrm{D}^{\mathrm{s}}\bm{q} \mathrm{D}^{\mathrm{s}}\bm{q'}\,\d(\bm q(a,a)-R^2)e^{i\str(\bm{qq'})-\frac{d+1}{2}\ln\sdet(2i\bm{q'})}\rho(\bm{q})(\mathrm{ln}\rho(\bm{q})+\Phi(\bm{q}))
\end{split}
\eeq
The explicit expression of $\Phi(\bm{q})$ will be computed in~\ref{sub:formphi}. The integral over $\bm{q'}$ is evaluated through a saddle point method for $d\to\io$. Using the tools from subsection~\ref{app}, the saddle-point equation is $2i\bm q=(d+1){\restriction{\bm{q'}^{-1}}{{\rm sp}}}$. We introduce the rescaled variable $\bm Q$ and neglect subdominant terms which will be calculated in subsection~\ref{sub:norm1}. For convenience with respect to the saddle-point equation in section~\ref{sec:sp}, we exponentiate the resulting spherical constraint $\bm Q(a,a)-\Dl$,  
\beq
 \SS_{\mathrm{IG}}\propto \int \mathrm{D}[\bm{Q},\bm\n] e^{\frac{d}{2}\ln\sdet\bm{Q}-\frac{d}{2}\int\dd a\, \bm\n(a)\left(\bm Q(a,a)-\Dl\right)}\rho(\bm{Q})(\mathrm{ln}\rho(\bm{Q})+\Phi(\bm{Q}))
\eeq
In the limit $d\to\io$ we apply a saddle point method, thanks to the scalings provided in~\ref{sub:scalings}. We can get rid of proportionality constants using the normalization of the density:
\beq
 \SS_{\mathrm{IG}}=\frac{\SS_{\mathrm{IG}}}{\int \mathrm{D}\bm{x}\, \d(\bm x(a)^2-R^2) \rho(\bm{x})}=-\frac{\int \mathrm{D}[\bm{Q},\bm\n]\CC(\bm Q)e^{\frac{d}{2}\G(\bm Q,\bm\n)}(\mathrm{ln}\rho(\bm{Q})+\Phi(\bm{Q}))}{\int \mathrm{D}[\bm{Q},\bm\n]\CC(\bm Q)e^{\frac{d}{2}\G(\bm Q,\bm\n)}}
\eeq
where $\CC(\bm Q)$ accounts for forgotten subdominant contributions to the integral, and
\beq\label{eq:G}
\boxed{\G(\bm Q,\bm\n)=\ln\sdet \bm Q-\int\dd a\, \bm\n(a)\left(\bm Q(a,a)-\Dl\right)+2\Omega(\bm Q)}
\eeq
is the saddle point function. In $d\to\io$ we maximize $\G$, in particular
\beq
\restriction{\frac{\d\G}{\d\bm Q}}{\rm sp}=\bm 0~~~~~~\textrm{and}~~~~~~\restriction{\frac{\d\G}{\d\bm \n}}{\rm sp}=\bm 0 \Leftrightarrow \bm Q(a,a)=\Dl
\eeq
giving the result
\beq
\SS_{\mathrm{IG}}\underset{d\to\io}{=}-d[\Omega({\bm{Q}}^{{\rm sp}})+\wh{\Phi}({\bm{Q}}^{{\rm sp}})]
\eeq
This can be expressed explicitly using once again the normalization condition
\beq
\int \mathrm{D}[\bm{Q},\bm\n]\CC(\bm Q)e^{\frac{d}{2}\G(\bm Q,\bm\n)}=1
\eeq
Evaluating it for $d\to\io$ and taking the logarithm of the resulting equation at dominant order $ O(d)$ provides $\boxed{\G({\bm{Q}}^{{\rm sp}},{\bm{\n}}^{{\rm sp}})=0}$ up to irrelevant additive constants\footnote{These constants were hidden in $\CC({\bm{Q}}^{{\rm sp}})$ and do not depend upon ${\bm{Q}}^{{\rm sp}}$, therefore having no influence on the dynamics.}, from which we get
\beq
\boxed{\SS_{\mathrm{IG}}\underset{d\to\io}{=}\frac d2\ln\sdet\bm Q^{\rm sp}-\frac{d}{2}\int\dd a\, \bm{\bm{\n}}^{{\rm sp}}(a)\left({\bm{Q}}^{{\rm sp}}(a,a)-\Dl\right)-d\wh\Phi({\bm{Q}}^{{\rm sp}})}
\eeq

\subsection{Interaction term}\label{sub:int}
\subsubsection{Changes of variables to `functional spherical coordinates'}

Now focusing on $\SS_{\mathrm{int}}$, we introduce the superfields $\bm q(a,b)=\bm x(a)\cdot\bm x(b)$, $\bm p(a,b)=\bm y(a)\cdot\bm y(b)$ (through a symmetric measure $\mathrm{D}^{\mathrm{s}}$) and the interaction superfield $\bm \om(a)=(\bm x(a)-\bm y(a))^2$, a one component variable since the Mayer function $f[x-y,\hat x - \hat{y}]$ in~\eqref{eq:fphi} needs only scalar products between its variables at equal times.
\beq\label{eq:chvarint}
\begin{split}
 \SS_{\mathrm{int}}&\propto\int \mathrm{D}[\bm{x},\bm{y}] \d(\bm x(a)^2-R^2) \d(\bm y (b)^2-R^2) \rho(\bm{x})\rho(\bm{y})f(\bm{x}-\bm{y})\\
 &=\int\mathrm{D}[\bm{x},\bm{y},\bm{q},\bm{q'},\bm{p},\bm{p'},\bm{\om},\bm{\om'}]\d(\bm q(a,a)-R^2)\d(\bm p(a,a)-R^2)e^{i\str(\bm{qq'}+\bm{pp'})}\rho(\bm{q})\rho(\bm{p})f(\sqrt{\bm{\om}})\\
 &~~~~~~~~~~~~~~~\times e^{-i\int\dd a\dd b\,[\bm{q'}(a,b)\bm x(a)\cdot\bm x(b)+\bm{p'}(a,b)\bm y(a)\cdot\bm y(b)]-i\int\dd a\,\bm{\om'}(a)[(\bm x(a)-\bm y(a))^2-\bm{\om}(a)]}\\
 &=\int\mathrm{D}[\bm{q},\bm{q'},\bm{p},\bm{p'},\bm{\om},\bm{\om'}]\d(\bm q(a,a)-R^2)\d(\bm p(a,a)-R^2)\rho(\bm{q})\rho(\bm{p})f(\sqrt{\bm{\om}})\\
 &~~~~~~~~~~~~~~~\times\exp\left(i\str(\bm{qq'}+\bm{pp'})+i\int\dd a\,\bm{\om'}(a)\bm{\om}(a)-\frac{d+1}{2}\ln\det\begin{pmatrix}\bm{q'}+\omh&-\omh\\-\omh&\bm{p'}+\omh\end{pmatrix}\right)
\end{split}
\eeq
The last line is obtained by Gaussian integration on superfields $\bm x$ and $\bm y$, introducing the superfield\\$\omh(a,b)=\bm{\om'}(a)\bm 1(a,b)$, and $\det$ means the determinant of the $2\times2$ block matrix consisting of superfields. We now change variables to exploit the $\bm x \leftrightarrow \bm y$ symmetry in~\eqref{eq:chvarint},
\beq
\bm{q_{\pm}}=\frac{\bm q\pm\bm p}{2}~~~~~~\textrm{and}~~~~~~\bm{q'_{\pm}}=\frac{\bm{q'}\pm\bm{p'}}{2}
\eeq
whose Jacobian is subdominant. The $2\times 2$ block matrix can be transformed through a (Weyl) rotation which does not affect the determinant: 
\beq
\frac1{\sqrt2} \begin{pmatrix}\id & -\id\\\id&\id\end{pmatrix}\begin{pmatrix}\bm{q'}+\omh&-\omh\\-\omh&\bm{p'}+\omh\end{pmatrix}\frac1{\sqrt2}\begin{pmatrix}\id & \id\\-\id&\id\end{pmatrix}=\begin{pmatrix}\bm{q'_+}+2\omh&\bm{q'_-}\\\bm{q'_-}&\bm{q'_+}\end{pmatrix}
\eeq
In $d\to\io$, owing to the $\bm x \leftrightarrow \bm y$ or $\bm q \leftrightarrow\bm p$ (respectively $\bm{q'} \leftrightarrow\bm{p'}$) symmetry in~\eqref{eq:chvarint}, the saddle-point value of $\bm{q_-}$ (respectively $\bm{q'_-}$) is zero. Dropping the $+$ index for the two other superfields\footnote{The symmetry also implies that the saddle-point value of $\bm{q_+}$ is the same than $\bm q$ or equivalently $\bm p$ (same for the primes).}, we get
\beq
 \SS_{\mathrm{int}}\propto\int\mathrm{D}[\bm{q},\bm{q'},\bm{\om},\bm{\om'}]\d(\bm{q}(a,a)-R^2)e^{2i\str(\bm{qq'})+i\int\dd a\,\bm{\om'}(a)\bm{\om}(a)-\frac{d+1}{2}\ln\sdet\bm{q'}-\frac{d+1}{2}\ln\sdet(\bm{q'}+2\omh)}\rho(\bm{q})^2f(\sqrt{\bm{\om}})
\eeq
To simplify the last term in the exponential, we make the following change of variables:
\beq\label{eq:chqq}
\bm q=R^2\bm P-\frac12\bm{D}~~~~~~\textrm{and}~~~~~~ \bm{q'}=\frac{d+1}{2i}\left(R^2\bm P-\frac12\bm{D'}\right)^{-1}
\eeq
The Jacobian is subdominant ; all such subdominant terms will be calculated in subsection~\ref{sub:norm2}. The aim is to recover the same saddle point function $\G$ at $ O(d)$ in the exponential in $\SS_{\mathrm{int}}$ as in the ideal gas term. Let us focus on the last term in the exponential, neglecting irrelevant constants:
\beq
\begin{split}
 \ln\sdet(\bm{q'}+2\omh) &=-\ln\sdet\left(R^2\bm P-\frac{\bm{D'}}2\right) + \ln\sdet\left(\bm1-\frac{2i}{d+1}\omh\bm{D'}\right)+\ln\left(1+\frac{2R^2}{d+1}\str\left[\bm P\left(\frac{\omh^{-1}}{2i}-\frac{\bm{D'}}{d+1}\right)^{-1}\right] \right)\\
 &=-\ln\sdet\left(R^2\bm P-\frac{\bm{D'}}2\right)+ \ln\sdet\left(\bm1-\frac{2i}{d+1}\omh\bm{D'}\right)+\ln\str\left[\bm P\left(\frac{\omh^{-1}}{2i}-\frac{\bm{D'}}{d+1}\right)^{-1}\right]
\end{split}
\eeq
We used~\eqref{eq:lndetinv} and expanded using the limit $R\to\io$ before $d\to\io$ in the last line as emphasized in the introduction to this section, which is valid if $\str\left[\bm P\left(\frac{\omh^{-1}}{2i}-\frac{\bm{D'}}{d+1}\right)^{-1}\right]$ is not zero.\\
We expect the same scalings for $\bm{D}$ and $\bm{D'}$, so let us assume $\bm{D'}=\bm{\D'}/d$ and $\bm{\om'}=d\bm{\m'}$, hence $\omh\bm{D'}=\mh\bm{\D'}= O(d^0)$. Using~\eqref{eq:powerseries}, we can use expansions:
\beq
\begin{split}
\ln\sdet\left(\bm1-\frac{2i}{d+1}\omh\bm{D'}\right)&\underset{d\to\io}{=}-\frac{2i}{d+1}\str(\mh\bm{\D'})+ O\left(\frac1{d^2}\right)\\
 \ln\str\left[\bm P\left(\frac{\omh^{-1}}{2i}-\frac{\bm{D'}}{d+1}\right)^{-1}\right]&\underset{d\to\io}{=}\ln\str(\bm P\mh)+\frac{2i}{d+1}\frac{\str(\bm P\mh\bm{\D'}\mh)}{\str(\bm P\mh)}+ O\left(\frac1{d^2}\right)
\end{split}
\eeq
Summarizing,
\beq
\begin{split}
 \SS_{\mathrm{int}}\propto\int\mathrm{D}[\bm{D}&,\bm{D'}]\d(\bm{D}(a,a))e^{2i\str\left[\left(R^2\bm P-\frac12\bm{D}\right)\frac{d+1}{2i}\left(R^2\bm P-\frac12\bm{D'}\right)^{-1}\right]-(d+1)\ln\sdet\left(R^2\bm P-\frac{\bm{D'}}2\right)}\rho(\bm{D})^2\\
 &~~~~~~~~~~\times \int\mathrm{D}[\bm{\om},\bm{\m'}]e^{id\int\dd a\,\bm{\m'}(a)\bm{\om}(a)+i\str(\mh\bm{\D'})-\frac{d+1}{2}\ln\str(\bm P\mh)-i\str(\bm P\mh\bm{\D'}\mh)/\str(\bm P\mh)}f(\sqrt{\bm{\om}})
\end{split}
 \eeq
At order $ O(d)$ in the exponential in $\SS_{\mathrm{int}}$, we get (twice) the same terms that we had in~\eqref{eq:chig}.  ; therefore in the last line, all dependence in $\bm{D'}$ is of $ O(d^0)$. 
We now go back to the original variables $\bm q$ and $\bm{q'}$ by making again the change of variables~\eqref{eq:chqq}. The terms in the exponent that must be kept for the saddle point in $d$ becomes
\beq
 2i\str(\bm{qq'})-(d+1)\ln\sdet(2i\bm{q'})+2\ln\rho(\bm{q})
\eeq
which is exactly twice what we had for $\SS_{\mathrm{IG}}$. Hence, $2i\bm q=(d+1){\restriction{\bm{q'}}{{\rm sp}}}^{-1}$ \ie $\restriction{\bm{D'}}{{\rm sp}}=\bm{D}$. \\Then the term $\str(\mh\bm{\D})=\int\dd a\,\bm{\m'}(a)\sd(a,a)=0$ because of the constraint. We also note that
\beq
\bm{\om}(a)=(\bm x(a)-\bm y(a))^2=(x(t)-y(t))^2+2\bar{\th}_1\th_1(x(t)-y(t))\cdot(\tilde x(t)-\tilde y(t))
\eeq
As we expect $(x(t)-y(t))^2=\s^2(1+ O(1/d))$, 
 we define $\bm{\om}(a)=1+2\bm{\m}(a)/d$. We set $\l=\int\dd a\,\bm{\m'}(a)=\str(\bm P \mh)$ with a Dirac delta and exponentiate it with a conjugated $\l'$ as usual. The interaction term\footnote{For now we put all subexponential dependence in $\FF$ ; explicit expressions will be given in~\ref{sub:norm2}.} now reads, once again exponentiating the constraint $\d(\bm Q(a,a)-\Dl)$ through a superfield $\bm\n$,
\beq
\begin{split}
  \SS_{\mathrm{int}}&\propto\int\mathrm{D}[\bm{Q},\bm\n]e^{d\G(\bm Q,\bm\n)}\FF(\bm Q),~\textrm{with}~\FF~\textrm{defined by:}\\
  \FF(\bm Q)&\propto\int\mathrm{D}[\bm{\m},\bm{\m'}]\dd \l\dd \l'\,\exp\left(i\l\l'+id\l-\frac{d+1}{2}\ln\l+2i\int\dd a\,\bm{\m'}(a)(\bm{\m}(a)-\l'/2)\right)\\
  &\hskip90pt \times \exp\left(-i\l\Dl+i\str(\bm P\mh\bm{Q}\mh)/\l\right)f\left(\sqrt{1+\frac{2\bm{\m}}{d}}\right)
\end{split}
  \eeq
We used the simplification $\str(\bm P\mh\bm P\mh)=\l^2$. The integral over $\l$ can be performed in the infinite $d$ limit: the saddle-point equation gives $\restriction{\l}{{\rm sp}}=1/2i$. Performing the following steps:
\begin{enumerate}
 \item In the Mayer function $f$, expand the square root $\sqrt{1+2\bm{\m}(a)/d}$ in the limit $d\to\io$
 \item Rescale $\bm{\m}(a)-\l'/2\longrightarrow\bm{\m}(a)$, still calling the new superfield $\bm{\m}$
 \item Rescale $\l'/2\longrightarrow\l$, dropping the prime for convenience
 \item Perform the Gaussian integration on $\bm{\m'}$
\end{enumerate}
we obtain
\beq
 \FF(\bm Q)\propto \frac{e^{-\Dl/2}}{\sqrt{\sdet\bm Q}}\int\dd \l\mathrm{D}\bm{\m}\,e^{\l-\frac12\int\dd a\dd b\,\bm{\m}(a){\bm Q}^{-1}(a,b)\bm{\m}(b)}f\left(1+\frac{\bm{\m}+\l}{d}\right)
\eeq

\subsubsection{Normalization}\label{sub:norm2}

Note that all the nonexponential in $d$ dependences overlooked during the procedures of the different changes of variables does not depend upon the choice of the Mayer function $f$. Here we benefit from this to give the explicit expression of $\SS_{\mathrm{int}}$.\\
In the MSRDDJ action $\AA$~\eqref{eq:action} we sum on times belonging to an interval $[t_{\rm p},t_1]$, where initial conditions are fixed at $t_{\rm p}$. $t_1$ labels the final state, and if we sum on all positions at $t_1$, we have $Z_N=1$ in~\eqref{eq:Z}. Let us pick $s \in ]t_{\rm p},t_1[$ and define a test function $f_0[x,\hat{x}]=\Th(1-\abs{x(s)})$. Note that the choice of the test function is not completely arbitrary: as seen in~\ref{sec:virial}, it should satisfy the properties of the
true Mayer function $f$ that we used to derive $\SS$, as it must reject all trajectories that do not get close at some time. 
Making a choice that does not respect these properties would lead to absurd results. We obtain, using first the expression of $\SS_{\mathrm{int}}$ in~\eqref{eq:lnXixxt} and setting $y=u+x$ and $\hat{y}=\hat{u}$:
\beq
\begin{split}
\SS_{\mathrm{int}}[f_0]&= \frac{N}{2}\int {\rm D}[x,\hat{x}] {\rm D}[u,\hat{u}] \rho[x,\hat{x}]\rho[u+x,\hat{u}]\Th(1-\abs{u(s)})=\frac{N}{2}\int\frac{\dd  x^{n_s}}{(2\p)^{\frac{d}{2}}}\frac{\dd  u^{n_s}}{(2\p)^{\frac{d}{2}}}\r(x^{n_s})\r(x^{n_s}+u^{n_s})\Th(1-\abs{u^{n_s}})\\
&=\frac{N}{2}\left(\frac{(2\p)^{\frac{d}{2}}}{\VV} \right)^2\frac{\VV}{(2\p)^{\frac{d}{2}}}\int\frac{\dd  u^{n_s}}{(2\p)^{\frac{d}{2}}}\Th(1-\abs{u^{n_s}})=\frac{\r \VV_d(1)}{2}
\end{split}
\eeq
As in~\ref{app:virial}, we discretized the trajectories and used that translation invariance and the normalization $\int\mathrm{D}[x,\hat x]\r[x,\hat x]=1$ imply $\int\mathrm{D}\hat x \prod_{n\neq n_s}^{1,M}\frac{\dd x^n}{(2\p)^{\frac{d}{2}}}\r[\{x^1,\cdots,x^M\},\hat{x}]=\textrm{constant}=(2\p)^{\frac{d}{2}}/\VV$. $n_s$ labels the time $s$, \ie $s=t_{\rm p}+n_s(t_1-t_{\rm p})/M$. We define $\CC'(\bm Q,\Dl)$ accounting for all the overlooked terms. We have, taking the saddle point over $\bm{Q}$,
\beq
\begin{split}
 \frac{\SS_{\mathrm{int}}[f]}{\SS_{\mathrm{int}}[f_0]}&=\frac{\SS_{\mathrm{int}}[f]}{\r \VV_d(1)/2}\\
 &=\frac{e^{-\Dl/2}\int\dd \l\mathrm{D}[\bm{\bm Q},\bm{\n},\bm{\m}]\CC'(\bm Q,\Dl)e^{d\G(\bm Q,\bm\n)}e^{\l-\frac12\int\dd a\dd b\,\bm{\m}(a){\bm Q}^{-1}(a,b)\bm{\m}(b)}f\left(1+\frac{\bm{\m}+\l}{d}\right)}{e^{-\Dl/2}\int\dd \l\mathrm{D}[\bm{\bm Q},\bm{\n},\bm{\m}]\CC'(\bm Q,\Dl)e^{d\G(\bm Q,\bm\n)}e^{\l-\frac12\int\dd a\dd b\,\bm{\m}(a){\bm Q}^{-1}(a,b)\bm{\m}(b)}f_0\left(1+\frac{\bm{\m}+\l}{d}\right)}\\
 &=\frac{1}{\CC''({\bm Q}^{{\rm sp}},\Dl)}\int\mathrm{D}\bm{\m}\dd \l\, e^{\l-\frac12\int\dd a\dd b\,\bm{\m}(a)\restriction{{\bm{Q}}^{-1}}{{\rm sp}}(a,b)\bm{\m}(b)}f\left(1+\frac{\bm{\m}+\l}{d}\right)
\end{split}
\eeq
$\CC''$ is given by 
\beq
\begin{split}
 \CC''({\bm Q}^{{\rm sp}},\Dl)&=\int\mathrm{D}\bm{\m}\dd \l\, e^{\l-\frac12\int\dd a\dd b\,\bm{\m}(a)\restriction{{\bm{Q}}^{-1}}{{\rm sp}}(a,b)\bm{\m}(b)}\th\left(-\frac{\m(s)+\l}{d}\right)\\
 &=\int\mathrm{D}\bm{\m}\,e^{\int\dd a\,\bm{\m}(a)\bm g(a)-\frac12\int\dd a\dd b\,\bm{\m}(a)\restriction{{\bm{Q}}^{-1}}{{\rm sp}}(a,b)\bm{\m}(b)}=e^{\Dl/2}\sqrt{\sdet\bm Q^{\rm sp}}
\end{split}
\eeq
where we introduced the superfield $\bm g(a)=-\bar{\th}_1\th_1\d(t-s)$ to integrate the Gaussian. We also used the constraint ${\bm{Q}}^{\rm sp}(a,a)=\Dl$. Now we can conclude:
\beq\label{eq:defFF}
\boxed{
\begin{split}
\SS_{\mathrm{int}}&\underset{d\to\io}{=}\frac{\r \VV_d(1)}2\FF({\bm{Q}}^{{\rm sp}})\\
\textrm{where}\hskip15pt\FF(\bm{Q})=\frac{e^{-\Dl/2}}{\sqrt{\sdet\bm Q}}&\int\mathrm{D}\bm{\m}\dd \l\, e^{\l-\frac12\int\dd a\dd b\,\bm{\m}(a){\bm Q}^{-1}(a,b)\bm{\m}(b)}f\left(1+\frac{\bm{\m}+\l}{d}\right)
\end{split}
}
\eeq

\subsection{Final result in the limit $d\to\io$}

Collecting the results from the last two subsections, and using~\eqref{eq:lndetinv}, we obtain the final result in the infinite dimension limit:
\beq\label{eq:finSS}
\boxed{\SS\underset{d\to\io}{=}\restriction{\frac d2\ln\sdet\bm Q-\frac{d}{2}\int \dd a \, \bm\n(a)\left(\bm Q(a,a)-\Dl\right)-d\wh\Phi(\bm{Q})+\frac{d\wh\varphi}2\FF(\bm{Q})}{{\rm sp}}}
\eeq
up to irrelevant additive constants (cf.~\ref{sub:norm1}), with $\FF$ defined in~\eqref{eq:defFF}. 

\section{Saddle-point equation}\label{sec:sp}

\subsection{Explicit form of $\Phi(\bm Q)$}\label{sub:formphi}

We now make explicit the $\bm Q$ dependence of $\Phi$ using $\wh\Phi$ justified in subsection~\ref{sub:scalings}. From~\eqref{eq:action}, it is Gaussian in $x$ and $\tilde{x}$,
\begin{equation}\label{eq:Phi}
\begin{split}
 \wh\Phi(\bm{x})=\frac1d\Phi(\bm{x})=\wh{\g}d\int & \dd t \left(\tilde{x}\cdot\dot{x}-T\tilde{x}^2 \right)=2d\left[\int \dd t\dd t'\, \tilde{x}(t)k(t,t')\tilde{x}(t')+2\int \dd t\dd t'\, \tilde{x}(t)\hat{k}(t,t')x(t')\right]\\
& \boxed{\textrm{with} ~ k(t,t')=-\wh{\g} T \d(t-t') ~ \textrm{and} ~ \hat{k}(t,t')=\frac{\wh{\g}}{2} \frac{\partial}{\partial t}\d(t-t')}
\end{split}
\end{equation}
The kernel $k$ is symmetric while $\hat{k}$ is antisymmetric. Let us define a symmetric superfield:
\begin{equation}
 \boxed{\bm k(a,b)=k(t,t')-\bar{\th}_1\th_1\hat{k}(t,t')+\bar{\th}_2\th_2\hat{k}(t,t')}
\end{equation}
One can check that $\boxed{\wh\Phi(\bm Q)=\str(\bm{k}\bm Q)}$ gives back expression~\eqref{eq:Phi}.

\subsection{Saddle-point equation for the dynamic correlations}

From subsection~\ref{sec:sphere}, the probability density of trajectories $\r$ is given by the saddle-point equation $\d\SS/\d\r(\bm x)=\d\SS/\d\r(\bm Q)=0$. In the infinite $d$ limit, $\SS$ depends on $\r$, or equivalently on its logarithm $\Omega$, only through its saddle-point value $\Omega(\bm{Q}^{\rm sp})$. From the relation $\G(\bm{Q}^{\rm sp},\bm{\n}^{\rm sp})=0$ derived in~\ref{sub:norm1} thanks to the normalization of $\r$, $\Omega(\bm{Q}^{\rm sp})$ is explicitly determined by the saddle-point values $\bm{Q}^{\rm sp}$ and $\bm{\n}^{\rm sp}$. Hence the saddle point condition is equivalent\footnote{The condition $\restriction{\d\SS/\d\bm{\n}}{{\rm sp}}=0$ is once again the spherical constraint ${\bm Q}^{\rm sp}(a,a)=\Dl$.} to $\restriction{\d\SS/\d\bm{Q}}{{\rm sp}}=0$ in~\eqref{eq:finSS}, and is, as a consequence, equivalent to the saddle point condition used in the virial terms for $d \to\infty$,
\beq
\restriction{\frac{\d\SS}{\d\bm Q}}{{\rm sp}}=0 \Leftrightarrow \restriction{-{\bm Q}^{-1}-\bm\n\bm 1+2\bm k+\frac{\wh\varphi}{2}\frac{\d\FF}{\d\bm Q}}{{\rm sp}}=0
\eeq
The derivative of $\FF$ is
\beq
\begin{split}
\hspace{-0.3cm} \frac{\d\FF}{\d\bm Q(a,b)}&=\FF(\bm\D){\bm Q}^{-1}(a,b)-\int\dd a'\dd b'\,{\bm Q}^{-1}(a,a') \int\dd \l\, e^{\l-\Dl/2}\int \DD\bm\m\,\bm{\m}(a')\bm{\m}(b')f\left(1+\frac{\bm{\m}+\l}{d}\right){\bm Q}^{-1}(b',b)\\
 &\textrm{with the Gaussian measure}~~\boxed{\int \DD\bm\m\,\bullet=\frac{1}{\sqrt{\sdet\bm Q}}\int\mathrm{D}\bm{\m}\,\bullet e^{-\frac12\int\dd a\dd b\,\bm{\m}(a){\bm Q}^{-1}(a,b)\bm{\m}(b)}}
\end{split}
\eeq
Hence the saddle-point equation, $\forall(a,b)$,
\beq\label{eq:sp}
\boxed{\restriction{\left(1-\frac{\wh\varphi\FF(\bm Q)}{2}\right){\bm Q}^{-1}(a,b)-2\bm k(a,b)+\restriction{\frac{\wh\varphi}{2}{\bm Q}^{-1}\int\dd \l\, e^{\l-\Dl/2}\int \DD\bm\m\,\bm{\m}\bm{\m}\,f\left(1+\frac{\bm{\m}+\l}{d}\right){\bm Q}^{-1}}{(a,b)}-\bm\n(a)\bm 1(a,b)=0}{{\rm sp}}}
\eeq
Together with the spherical constraint ${\bm Q}^{\rm sp}=\Dl$ which shall provide ${\bm \n}^{\rm sp}$, this determines ${\bm Q}^{\rm sp}$.

\subsection{Simplification of the saddle-point equation}

\subsubsection{Exploiting Ward-Takahashi-like identities}
Here we drop the labels 'sp' for convenience. Generically, derivatives of $\SS$ are needed for example to compute the Mode-Coupling Theory (MCT) exponents. They can be simplified using Ward-Takahashi-like identities. From the definition of the Mayer function $f$~\eqref{eq:fphi}, a quantity like $\la \bullet f\ra_{\bm{\m}}$ is a difference between two averages, one with potential $V$ and the other without. Let us focus on the non Gaussian part, with potential:
\beq\label{eq:ward}
\int \DD\bm\m\,e^{-\int\dd a\,\bar V(\bm\m(a)+\l)}\equiv \la 1 \ra_V
\eeq
Usually one shifts $\bm\m$ by $\bm\epsilon$ in some well chosen average $\la  O \ra_V$ and demands the cancellation of all non-zero orders in $\bm\epsilon$, but a more compact way to get moments of $\bm Q\bm\m$ is to write, $\forall(a,b)$,
\beq\begin{split}\label{eq:simpli}
\boxed{\restriction{{\bm Q}^{-1}\la\bm\m\bm\m\ra_V{\bm Q}^{-1}}{(a,b)}}=&\frac{1}{\sqrt{\sdet\bm Q}}\int \mathrm{D}\bm\m \, e^{- \int \dd c \, \bar V(\m(c) + \l )} \left[ \frac{\d^2}{\d \bm\m(a) \d \bm\m(b)} + {\bm Q}^{-1}(a,b) \right]
e^{  -\frac{1}2 \int \dd c \dd e \, \bm\m(c) {\bm Q}^{-1}(c,e) \bm\m(e)} \\
=&{\bm Q}^{-1}(a,b) + \int \DD \bm\m \,
\frac{\d^2}{\d \bm\m(a) \d \bm\m(b)} 
e^{- \int \dd  c \, \bar V(\bm\m(c) + \l )} \\
=& \boxed{{\bm Q}^{-1}(a,b) + \la F(\bm\m(a)+\l)F(\bm\m(b)+\l)\ra_V +\la F'(\bm\m(a)+\l)\ra_V \bm 1(a,b)}
 \end{split}\eeq
where we integrated by parts twice. This method can be easily generalized to higher moments.

\subsubsection{The value of $\FF(\bm{Q})$ at the saddle point}\label{subsub:FF}

The measure in~\eqref{eq:ward} can be interpreted as an average over a Langevin process with potential $V$. Provided we sum over all possible trajectories of $\bm\m$, equation~\eqref{eq:ward} is actually the normalization of probability $\la 1\ra_V=1$.\\
Similarly $\FF(\bm{Q})$ can be interpreted as a difference between averages over two dynamical processes, one with potential $V$ and the other free,
\beq\label{eq:Fval}
\begin{split}
\FF(\bm{Q})&=\int\dd \l\, e^{\l-\Dl/2}\int\DD \bm\m\, f\left(1+\frac{\bm{\m}+\l}{d}\right)=\int\dd \l\, e^{\l-\Dl/2}\left[\int\DD \bm\m\,e^{-\int\dd a\,\bar V(\bm\m(a)+\l)}-\int\DD \bm\m\,1\right]\\
&=\int\dd \l\, e^{\l-\Dl/2}\left[\la1\ra_V-\la1\ra_0\right]\Leftrightarrow\boxed{\FF(\bm{Q})=0}
\end{split}
\eeq
Hence $\FF$ is zero for all acceptable dynamical propagators (positive definite, as we expect $\bm Q$ is, at least at its saddle-point value dominating the dynamics) due to normalization.

\subsubsection{Definition of the memory kernel $\bm M$}

From equations~\eqref{eq:sp}, \eqref{eq:simpli} and \eqref{eq:Fval}, the saddle-point equation is simplified as:
\beq\label{eq:spM}
\boxed{{\bm Q}^{-1}(a,b)=2\bm k(a,b)-\bm M(a,b)+(\bm\nu(a)+ \bm{\d\n}(a))\bm 1(a,b)}
\eeq
where
\beq\label{eq:SPM}
\boxed{
\begin{split}
  \bm M(a,b)&=\frac{\wh\varphi}{2}\int\dd \l\,e^{\l-\Dl/2}\la F(\bm\m(a)+\l)F(\bm\m(b)+\l)\ra_V\\
  \bm{\d\n}(a)&=-\frac{\wh\varphi}{2}\int\dd \l\,e^{\l-\Dl/2}\la F'(\bm\m(a)+\l)\ra_V
\end{split}
}
\eeq
In our study of the dynamics of the system, $\bm M$ will play the role of the analog of the MCT kernel.

\section{Equation for the dynamic correlations}\label{sec:kernel}
\subsection{Equilibrium hypothesis}
In this subsection and the next ones we will unfold the SUSY notation to get rid of it, coming back to the standard dynamical variables.\\
We focus on the equilibrium dynamics of the system, assuming that time translation invariance (TTI) as well as causality hold, and consequently fluctuation dissipation theorem (FDT): we assume we start in the remote past (time $t_{\rm p}$, formally sent to $-\infty$), so that the system is at equilibrium when a finite $t_0$ is reached.\\
$\bm Q=2d\bm q$ (equivalently $\sd$) takes its equilibrium form and so does $\bm M$:
\beq\label{eq:qsp}
\begin{split}
 \bm Q(a,b)&=C(t-t')+\bar\th_1\th_1 R(t'-t)+\bar\th_2\th_2 R(t-t')\\
 \bm M(a,b)&=M(t-t')+\bar\th_1\th_1\wh M(t'-t)+\bar\th_2\th_2 \wh M(t-t')
\end{split}
\eeq
Along with TTI and~\eqref{eq:spM}, this implies that $\bm\nu(a)=\nu$ and $\bm{\d\n}(a)=\d\n$ are constant and real quantities. From this the inverse $\bm Q^{-1}$ reads: 
\beq\label{eq:qinv}
 \bm Q^{-1}(a,b)=\wt C(t-t')+\bar\th_1\th_1\wt R(t'-t)+\bar\th_2\th_2 \wt R(t-t')~~\textrm{with}~~\wt C=-R^{-\mathrm{T}}CR^{-1}~~\textrm{and}~~ \wt R=R^{-1}
\eeq
$C$ and $R$ satisfying fluctuation-dissipation theorem, one can check that $\wt C$ and $\wt R$ also do. 
Similarly, we can verify directly with~\eqref{eq:Phi} that $\bm k$ and $\bm 1$ verify the latter relation. We conclude, from~\eqref{eq:spM}, that $\bm M$ also satisfies FDT (which was suggested by~\eqref{eq:SPM}). 
\subsection{Mode-coupling form of the saddle-point equation and the effective stiffness}

We can cast the saddle-point equation into a mode-coupling form by multiplying~\eqref{eq:spM} by $\bm Q$ on the right. The scalar component of the equation obtained reads at $(t',t)$:
\beq
\begin{split}
 0=&-2(\hat{k} C+k {R}^\mathrm{T})-\wh M C-M{R}^\mathrm{T}+(\n+\d\n) C\\
 =&\wh\g\partial_{t'} C(t'-t)-2\wh\g TR(t-t')+(\n+\d\n) C(t-t')-\b\int_{-\infty}^{t'}\dd u\,\left(\partial_uM(t'-u)\right)C(u-t)\\
 &-\b\int_{-\infty}^{t}\dd u\, M(t'-u)\partial_uC(t-u)
\end{split}
\eeq
We can assume, for instance, that $t>t'$, and use FDT:
\beq
0=\wh\g\dot{C}(t-t')+(\n+\d\n) C(t-t')-\b\int_{t'}^t\dd u\,M(t'-u)\partial_uC(t-u)-\b\int_{-\infty}^{t'}\dd u\, \partial_u\left[M(t'-u)C(t-u)\right]\\
\eeq
Using the relaxation for long times and making the substitution $v=t+t'-u$,
\beq
\begin{split}
 &\wh\g\dot{C}(t-t')=-\left(\n+\d\n-\b M(0)\right) C(t-t')-\b\int_{t'}^t\dd v\,M(t-v)\dot{C}(v-t')\\
 &t'\to t^-~\textrm{gives}~~\n+\d\n-\b M(0)=-\frac{\wh\g \dot C(0)}{\Dl}
\end{split}
\eeq
$\dot C(0)=-TR(0^+)$ represents the immediate response to a perturbation. At such very short times the potential is not relevant, particles follow a free dynamics. $\dot C(0)$ can thus be obtained using~\eqref{eq:Langevin} $\g \dot{x}_i=-\n_ix_i+\xi_i$ with $\n_i=dT/R^2$ since, at equilibrium, TTI and the equipartition theorem holds. One can solve this Ornstein-Uhlenbeck process and compute $C(t)=\Dl\exp(-dTt/\g R^2)$, giving $-\wh\g \dot C(0)=T$, hence
\beq\label{eq:nu}
\boxed{\n+\d\n-\b M(0)=\frac{T}{\Dl}}
\eeq
We conclude that the mode-coupling-like equation for $C$ is, for $t>t'$:
\beq\label{eq:MCTC}
\boxed{\wh\g\dot{C}(t-t')=-\frac{T}{\Dl} C(t-t')-\b\int_{t'}^t\dd v\,M(t-v)\dot{C}(v-t')}
\eeq
or equivalently for the MSD $\D=d\DE=\Dl-C$ at $t>t'$:
\beq
\label{eq:MCTD}
\boxed{\wh\g\dot{\D}(t-t')=T-\frac{T}{\Dl} \D(t-t')-\b\int_{t'}^t\dd v\,M(t-v)\dot{\D}(v-t')}
\eeq

\subsection{Effective Langevin process}

The aim is to compute $\bm M$, the mode-coupling-like kernel, as a function of $\bm{Q}^{{\rm sp}}$ to solve the saddle-point equation for $\bm{Q}^{{\rm sp}}$, providing correlation and response of the system. To do this, we must calculate correlations of the force $F$ at two times $t_0$ and $t_1\geqslant t_0$.
To achieve this program, we will interpret, as mentioned in~\ref{subsub:FF}, the average defining $\bm M$ as two-point correlation functions of a Langevin dynamics with potential $V$. We drop the notation $V$ in averages since we will be refering only to this process from now on. To this end, let us unfold the MSRDDJ path integral in SUSY notation~\eqref{eq:ward} using the saddle-point equation~\eqref{eq:spM}, where $\bm\m(a)=\m(t)+\bar\th_1\th_1\tilde\m(t)$:
\beq
\begin{split}\label{eq:probMSR}
\la 1 \ra&\equiv\frac{1}{\sqrt{\sdet\bm Q}}\int \mathrm{D}\bm\m\,e^{-\frac{1}2 \int \dd a \dd b \, \bm\m(a) {\bm Q}^{-1}(a,b) \bm\m(b)-\int\dd a\,\bar V(\bm\m(a)+\l)}\\
&=\frac{1}{\sqrt{\sdet\bm Q}}\int \mathrm{D}\bm\m\,e^{-\frac{1}2 \int \dd a \dd b \, \bm\m(a)\left[(\bm\n+\bm{\d\n})\bm1+2\bm k-\bm M \right](a,b) \bm\m(b)-\int\dd a\,\bar V(\bm\m(a)+\l)}
\end{split}
\eeq
$\sqrt{\sdet\bm Q}$ plays the role of normalization\footnote{Note that $\sdet\bm Q=$ numerical constant at the saddle point level if the system is causal.}. The corresponding Langevin process with potential $V$, depending on $\l$, is:
\beq\label{eq:Leff}
\begin{split}
 \wh\g\dot\m(t)&=-(\n+\d\n)\m(t)+\int_{t_{\rm p}}^{t}\dd t'\,\wh M(t-t')\m(t')+F(\m(t)+\l)+\z(t)\\
 \textrm{with}&~~\la\z(t)\ra=0~~\textrm{and}~~\la\z(t)\z(t')\ra=2\wh\g T\d(t-t')+M(t-t')
\end{split}
\eeq
We used that $\wh M$ is causal and consider times $t\in[t_0, t_1]$. $M$ (and $\wh M$ by FDT) can be computed self-consistently with force correlation functions of this process through its definition~\eqref{eq:SPM}:
\beq\label{eq:defMC}
M(t-t')=\frac{\wh\varphi}{2}\int\dd \l\,e^{\l-\Dl/2}\la F(\m(t)+\l)F(\m(t')+\l)\ra
\eeq
Using FDT for $\bm M$, an integration by part with the fact that $t-t_{\rm p}>t_0-t_{\rm p}\gg \t_{\a}$ the relaxation time of the system, above which correlations vanish, and~\eqref{eq:nu}, we get the generalized Langevin equation equivalent to~\eqref{eq:Leff}:
\beq\label{eq:Langeff}
\wh\g\dot\m(t)=-\frac{T}{\Dl}\m(t)-\b\int_{t_{\rm p}}^{t}\dd t'\,M(t-t')\dot\m(t')+F(\m(t)+\l)+\z(t)
\eeq
Note that so far in this section we have used equilibrium properties for all observables. It is either because we considered them at a time $t\geqslant t_0$ where equilibrium is reached, or in the case of convolution products with $\bm M$ where the integral extends to the remote past, because we expect that $\wh M$ (respectively $M$) vanishes quickly (respectively vanishes on a finite time scale $\t_\a$). Then for finite times where they do not vanish, the system has equilibrated from the initial condition in the remote past $t_{\rm p}$, and we can consider their equilibrium properties.

\subsection{Memory kernel in equilibrium: resumming trajectories from the remote past}\label{sub:markov}

As emphasized before, dynamical two points functions at $(t_0,t_1)$ should be function of a single argument, the time difference $t_1-t_0$, as we assume that equilibrium is reached. But equilibrium properties hold \textit{only} if we consider times $t\geqslant t_0$. Therefore, we wish to use $t_0$ as our initial time, {\em but in the understanding that at this time the system is at equilibrium}. Similarly, we are not interested in what happens after $t_1$, as it should be irrelevant due to causality, as stressed in section~\ref{sec:dynac} (see also figure~3). However one expects the Langevin equation~\eqref{eq:Langeff} to describe a non Markovian process in which the memory kernel persists for a duration of the order of $\t_{\a}$. How are we going to ignore the times $t<t_0$ if the kernel $M$ extends to the remote past?\\
An equation like~\eqref{eq:Langeff} may be thought of as having originated in a system coupled linearly to a bath of harmonic oscillators, \`a la Zwanzig~\cite{Z73,Ha97}. Let us consider the Hamiltonian evolution of this system, described by coordinates denoted collectively by $\G=\{\m,p_{\a},q_{\a}\}$, according to the Hamiltonian\footnote{If we restore the inertia term in the dynamics, we would add to $\G$ one more coordinate $p_{\m}$, momentum of the particle of position $\m(t)$. Similarly, $H_{\G}$ would have one more term $p_{\m}^2/2m$.}:
\beq
H_{\rm tot}(\G)=\underbrace{\frac{T}{2\Dl}\m^2+\bar V(\m+\l)}_{H_{\rm eff}(\m)}+\underbrace{\sum_{\a}\left[\frac{p_{\a}^2}{2m_{\a}}+\frac{m_{\a}\om_{\a}^2}2\left(q_{\a}-\frac{c_{\a}}{m_{\a}\om_{\a}^2}\m\right)^2\right]}_{H_{\rm B}(\G)}
\eeq
$\{c_\alpha,\omega_\alpha,m_\alpha\}$ are chosen suitably to reproduce dissipation and noise terms in~\eqref{eq:Langeff}. As before, assume we start in the remote past $t_{\rm p}$ at a point in phase space $\G_{\rm p}$ and let us distinguish two times $t_0$ and $t_1$, such that $t_1>t_0\gg t_{\rm p}$. We can rewrite averages over the effective Langevin process as averages over the Markovian process $\G(t)$, by defining the transition probabilities, for fixed $\G_{\rm p}$ and $\G_1$,
\beq
 P_V(\G_1,t_1|\G_{\rm p},t_{\rm p})=\int_{\G_{\rm p},t_{\rm p}}^{\G_1,t_1} \mathrm{D}[\G]\,\r_V[\G(t)|\G_{\rm p},t_{\rm p}]
\eeq
where $\r_V$ is the transition probability density of the whole system \{effective~particle\}~$\cup$~\{bath\}. The marginal of $\r_V[\G(t)|\G_{\rm p},t_{\rm p}]$ over all trajectories of the bath degrees of freedom $\{p_{\a},q_{\a}\}$ is the marginal of $\r_V[\m,\tilde\m|\m_{\rm p},t_{\rm p}]$ over trajectories of the response field $\tilde\m$ of the MSRDDJ probability density in~\eqref{eq:probMSR}, namely in compact SUSY notation with fixed initial conditions at $t_{\rm p}$:
\beq
\mathrm{D}\bm\m\,e^{-\frac{1}2 \int_{t_{\rm p}}^{t_1}\int_{t_{\rm p}}^{t_1} \dd a \dd b \, \bm\m(a)\bm{Q}^{-1}(a,b) \bm\m(b)-\int_{t_{\rm p}}^{t_1}\dd a\,\bar V(\bm\m(a)+\l)}=\mathrm{D}[\m,\tilde\m]\r_V[\m,\tilde\m|\m_{\rm p},t_{\rm p}]
\eeq
We can now write, using Markovianity:
\beq
P_V(\G_1,t_1|\G_{\rm p},t_{\rm p})=\int\dd \G_0\,P_V(\G_1,t_1|\G_0,t_0)P_V(\G_0,t_0|\G_{\rm p},t_{\rm p})
\eeq
where\footnote{Constants are arbitrary here, they can be absorbed in the normalizations.} $\dd \G=\frac{\dd \m}{\sqrt{2\pi}}\prod_\a\dd p_\a\dd q_\a$. For $t_0-t_{\rm p}\gg\t_\a$ the system relaxes, \ie we may assume that $P_V(\G_0,t_0|\G_{\rm p},t_{\rm p})=P_V^{\rm eq}(\G_0)$, the equilibrium distribution in the phase space of the whole system:
\beq
 P_V^{\rm eq}(\G_0)=\frac{e^{-\b H_{\rm tot}(\G_0)}}{Z_VZ_{\rm B}}=P_V^{\rm eq}(\m_0)P_{\rm B}^{\rm eq}(\G_0)
\eeq
with the marginals describing respectively the effective particle 
\begin{eqnarray}
 P_V^{\rm eq}(\m_0)=\frac{e^{-\b H_{\rm eff}(\m_0)}}{Z_V}=\frac{e^{-\m_0^2/2\Dl-\b\bar V(\m_0+\l)}}{Z_V}
\end{eqnarray}
and the bath
\beq
 P_{\rm B}^{\rm eq}(\G_0)=\frac{e^{-\b H_{\rm B}(\G_0)}}{Z_{\rm B}}
 \hskip15pt
 \textrm{with}
 \hskip15pt
 Z_{\rm B}=\int\prod_\a\dd p_\a\dd q_\a\,e^{-\b H_{\rm B}(\G_0)}=\prod_\a\frac{2\pi}{\b\om_\a}
\eeq
We may now get rid of the bath degrees of freedom as in Zwanzig's calculation, here integrating them out:
\beq
 P_V(\G_1,t_1|\G_{\rm p},t_{\rm p})=\int\frac{\dd \m_0}{\sqrt{2\pi}}\,P_V^{\rm eq}(\m_0)\underbrace{\int\prod_\a\dd p_{\a}^0\dd q_{\a}^0\,P_{\rm B}^{\rm eq}(\G_0)\int_{\G_0,t_0}^{\G_1,t_1} \mathrm{D}[\G]\,\r_V[\G(t)|\G_0,t_0]}
\eeq
Following H\"anggi~\cite{Ha97}, the last term (underbraced) can be interpreted as the transition probability of a Langevin process, with full equilibrium at $t_0$ between the harmonic oscillators and the effective particle, given by $H_{\rm B}$, \ie not neglecting the coupling to the particle. This generating functional reads, using again MSRDDJ:
\beq
1=\int\dd \G_1\, P_V(\G_1,t_1|\G_{\rm p},t_{\rm p})=\int\frac{\dd \m_0}{\sqrt{2\pi}}\,P_V^{\rm eq}(\m_0)\int\mathrm{D}[\m,\tilde\m]\,\r_V[\m,\tilde\m|\m_0,t_0]
\eeq
where now the effective Langevin process ruling the dynamics is:
\beq\label{eq:Leff2}
\boxed{
\begin{split}
 \wh\g\dot\m(t)&=-\frac{T}{\Dl}\m(t)-\b\int_{t_0}^{t}\dd t'\,M(t-t')\dot\m(t')+F(\m(t)+\l)+\z(t)\\
 \textrm{with}&~~\la\z(t)\ra=0~~\textrm{and}~~\la\z(t)\z(t')\ra=2\wh\g T\d(t-t')+M(t-t')\\
 \m(t_0)&=\m_0~\textrm{is picked with the equilibrium measure}~P_V^{\rm eq}(\m_0)
\end{split}
}
\eeq
Notice the lower limit of the friction kernel and the fact that there is no extra term, called `initial slip' in~\cite{Ha97}, due to the somewhat unusual `conditional average' over the perturbed bath, given by $H_{\rm B}$.\\
\begin{figure}
 \includegraphics[height=6cm]{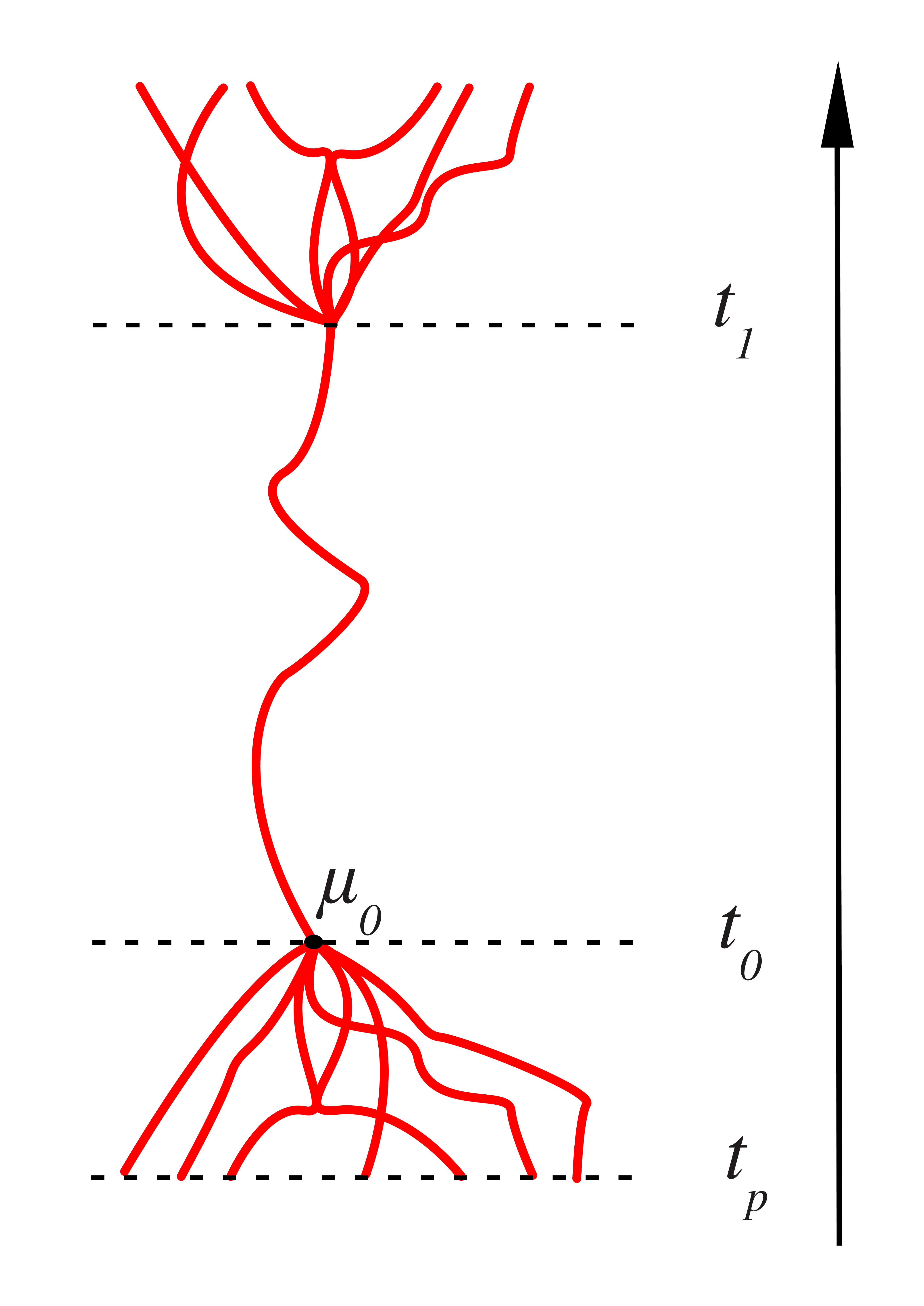}
 \caption{The procedure can be more intuitively interpreted as a resummation before $t_0$ of all possible trajectories starting from a fixed initial position at $t_{\rm p}$ (thanks to the relaxation towards equilibrium) and after $t_1$ (owing to causality), leaving us with finite times motions in the interval $[t_0,t_1]$.}
\end{figure}

\subsection{Relaxation at long times}

From now on we will be able to compute physically relevant observables, and with this in mind we will use standard MCT methodology~\cite{Go09,Go98}. At long time difference $t-t'\to\io$, equation~\eqref{eq:MCTC} gives 
\beq\label{eq:CMio}
0=-\frac{T}{\Dl} C(\io) - \b M(\io) [ C(\io) - R^2 ] 
\hskip5pt
\Rightarrow
\hskip5pt
C(\io) = \frac{ \Dl^2 \b^2M(\io)}{1 + \Dl \b^2M(\io)}
\hskip5pt
\Leftrightarrow
\hskip5pt
\b^2 M(\io) = \frac{1}{\Dl} \frac{C(\io)}{\Dl - C(\io)}
\eeq
In the long time limit, decorrelation occurs \textit{if we assume ergodicity} (which will be questioned at the dynamical transition, see~\ref{sub:dyntrans}), the average in~\eqref{eq:defMC} splits and by TTI we have to compute equilibrium averages:
\beq
M(\io) = \frac{\wh\varphi}{2}\int\dd \l\,e^{\l-\Dl/2}\la F(\m_0+\l)\ra^2
\eeq
Using the procedure of the last subsection, $\la F(\m_0+\l)\ra=\int\frac{\dd \m_0}{\sqrt{2\pi}}\,P_V^{\rm eq}(\m_0)F(\m_0+\l)$. Note that in equilibrium we can show through an integration by parts that, for any observable $ O(\m_0)$,
\beq\label{eq:IBP}
T \la \frac{\dd  O}{\dd \m_0} \ra = \frac{T}{\Dl}\la \m_0  O\ra -\la F(\m_0+\l) O\ra
\eeq
In particular by choosing $ O=1$ we obtain
\beq
\la  F(\m_0+\l)\ra = \frac{T}{\Dl} \la \m_0 \ra \
\eeq
For simplicity we focus here on hard spheres. For this potential the normalization reads:
\beq
Z_V=\int_{-\l}^{\infty}\frac{\dd \m_0}{\sqrt{2\pi}}\,e^{-\m_0^2/2\Dl}=\sqrt{\Dl}\,\Th_0(-\l / \sqrt{\Dl})~~\textrm{with}~~\Th_0(x)=\frac{1+\erf(-x/\sqrt{2})}{2}
\eeq
Given that
\beq
\la \m_0 \ra = \frac{\int_{-\l}^\io\frac{\dd \m_0}{\sqrt{2\p}}\,e^{-\m_0^2/2\Dl}\m_0}{\sqrt{\Dl}\,\Th_0(-\l / \sqrt{\Dl})}=\sqrt{\frac{\Dl}{2\p}}\frac{e^{-\l^2/2\Dl}}{\Th_0(-\l / \sqrt{\Dl})}
\eeq
after a short computation we find in the limit $R\to\io$:
\beq
\boxed{\b^2 M(\io) \underset{\Dl\to\io}{\sim}\frac{\wh\varphi}{4\sqrt\p}\frac{e^{-\Dl/4}}{\sqrt{\Dl}}}
\eeq
which shows with~\eqref{eq:CMio} that both $M(\io)$ and $C(\io)$ go exponentially to zero for $\Dl \to \io$, as one would expect.

\subsection{Getting rid of the sphere $\mathbb{S}^d(R)$: infinite radius limit}

Applying the method in~\ref{sub:markov}, the equation for $M$~\eqref{eq:defMC} reads
\beq
M(t-t')=\frac{\wh\varphi}{2}\int\dd \l\,e^{\l-\Dl/2}\int\frac{\dd \m_0}{\sqrt{2\pi}}\,P_V^{\rm eq}(\m_0)\la F(\m(t)+\l)F(\m(t')+\l)\ra
\eeq
where the average is computed over the Langevin process~\eqref{eq:Leff2}. Fixing $\l$, we make the change of variables\footnote{Note that $h$ is called $y$ in the main text.} $h(t)=\m(t)+\l$ centered at the wall. The normalization of $P_V^{\rm eq}$ is unchanged and
\beq
P_V^{\rm eq}(h_0)=\frac{e^{-(h_0-\l)^2/2\Dl-\b\bar V(h_0)}}{Z_V}
\eeq
We can simplify the expression of $M$ with a saddle point method\footnote{From the Langevin equation, the bracketed term $\la F(h(t))F(h(t'))\ra$ depends upon $\a$ but it cannot be exponential in $\Delta_{\rm liq}$ and as a consequence gives no contribution to the saddle-point equation.\label{ftn:subdominant}} for $\Dl\to\io$, setting $\a=\l/\Dl$:
\beq\label{eq:MCio}
\begin{split}
& M(t-t')=\frac{\wh\varphi}{2}\sqrt{\Dl}\int\frac{\dd h_0}{\sqrt{2\pi}}\dd \a\,e^{-\b\bar V(h_0)-\frac{h_0^2}{2\Dl}-\frac{\Dl}{2}(\a-1)^2+h_0\a}\la F(h(t))F(h(t'))\ra\\
 \Rightarrow& \boxed{M(t-t')\underset{\Dl\to\io}{\sim}\frac{\wh\varphi}{2}\int\dd h_0\,e^{-\b w(h_0)}\la F(h(t))F(h(t'))\ra\hskip15pt\textrm{with}\hskip15pt w(h)=\bar V(h)-Th+\underbrace{\frac{T}{2\Dl}h^2}}
\end{split}
\eeq
since $\a=1$ at the saddle point. We replaced the normalization by 
\beq
Z_V=\int\frac{\dd\m_0}{\sqrt{2\p}}\, e^{-\m_0^2/2\Dl-\b\bar V(\m_0+\a\D_{\rm liq})}\underset{\Dl\to\io}{\sim}\sqrt{\D_{\rm liq}}
\eeq
since the potential goes to zero at long distances. The Langevin equation is affected by the change of variables only with an additional term $T\a\simeq T$:
\beq\label{eq:Leff3}
\boxed{\wh\g\dot h(t)=T\underbrace{-\frac{T}{\Dl}h(t)}-\b\int_{t_0}^{t}\dd t'\,M(t-t')\dot h(t')+F(h(t))+\z(t)}
\eeq
Our problem is mapped onto a one-dimensional diffusion with colored noise (as usual in mean field~\cite{CC05}) and a harmonic effective potential $w(h)$ perturbed by the spheres' repulsion (cf. figure 4). The underbraced terms -the harmonic potential well- are negligible for finite times, but necessary to confine the system: they represent the `box'.
\begin{figure}[b]
 \includegraphics[width=8cm]{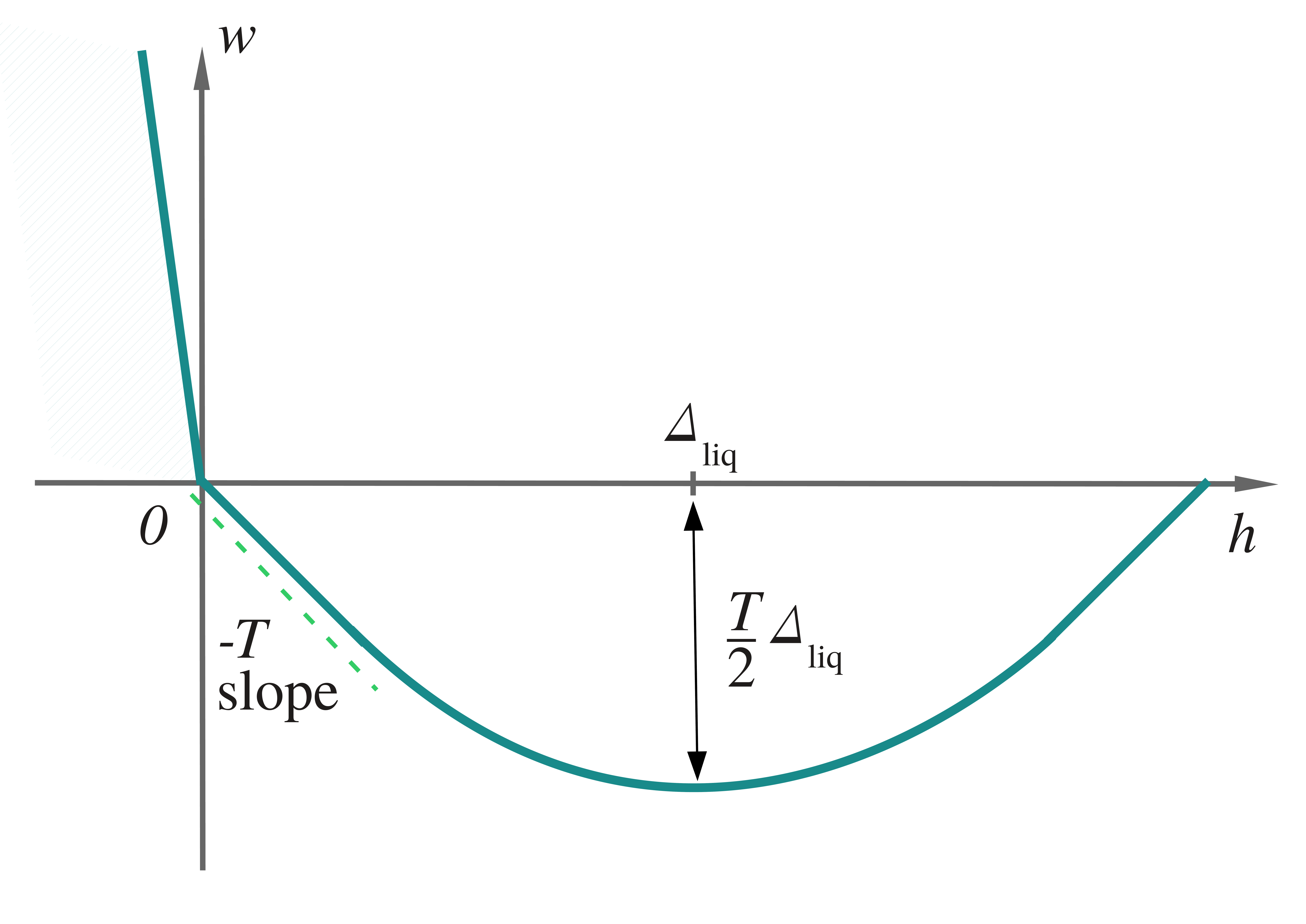}
 \caption{Effective potential landscape. If $V$ is hard, there is an infinite wall at $h=0$ prohibiting any motion in the $h<0$ half line. If the constraint is softer, as drawn here, motion is possible for $h<0$ but is rather unlikely.}
\end{figure}

\subsection{The `Lagrange multiplier'}
Plugging in~\eqref{eq:nu} the value at equal times\footnote{$M(0)$ diverges in the hard-sphere limit, which is natural given its interpretation as a force-force correlation.} $M(0)$, which can be computed at the equilibrium using TTI, we have
\beq
\n - \frac{T}\Dl = - \d\n + \b M(0)  =\frac{\wh\varphi}{2}\int_{-\io}^0\dd h_0\,e^{h_0-\b\bar V(h_0)}F'(h_0)+\b \frac{\wh\varphi}{2}\int_{-\io}^0\dd h_0\,e^{h_0-\b\bar V(h_0)}F(h_0)^2
\eeq
since $\bar V(h_0)=0$ for $h_0>0$. We will focus in the following on hard spheres and assume a regularization of the potential which is of class $C^1$, \eg soft spheres\footnote{It is actually valid for a linear regularization $\bar V_{\rm lin}(h)=-\k h\Th(-h)$ too. Indeed we get $\d\n=\frac{\wh\varphi}{2}\k$ and $\b M(0)=\frac{\wh\varphi}{2}(\k-T)+ O(1/\k)$.} $\bar V_{\rm SS}(h)=\k r^2\Th(-h)$. We have, with integration by parts, due to the continuity of $\bar V$ and $F$ in 0,
\beq
\begin{split}
 \n - \frac{T}\Dl &=\frac{\wh\varphi}{2}\left\{ \left[e^{h_0-\b\bar V(h_0)}F(h_0)\right]_{-\io}^0-\int_{-\io}^0\dd h_0\,e^{h_0-\b\bar V(h_0)}F(h_0)\left(1+\b F(h_0)\right) +\b \int_{-\io}^0\dd h_0\,e^{h_0-\b\bar V(h_0)}F(h_0)^2\right\}\\
 &=\frac{\wh\varphi}{2}\left\{ \left[e^{h_0-\b\bar V(h_0)}\left(F(h_0)-T\right)\right]_{-\io}^0+T\int_{-\io}^0\dd h_0\,e^{h_0-\b\bar V(h_0)}\right\}
\end{split}
\eeq
The last integral being zero for hard spheres, we conclude that the Lagrange multiplier\footnote{As noted in~\ref{sec:sphere}, $\n$ is not the actual Lagrange multiplier, but is related to it.} is (up to exponentially small corrections in $\Dl\to\io$ due to $M(0)$ and $\d\n$):
\beq
\boxed{\b\n=-\frac{\wh\varphi}{2}+\frac1\Dl }
\eeq

\section{Physical consequences of the dynamical equations}
\subsection{Plateau and dynamical transition}\label{sub:dyntrans}
\subsubsection{Metastable glassy states: plateau value}

We now look for a plateau in the dynamics: we assume a strong separation between a fast and a slow motion. We can split the correlations (and similarly the memory kernel) into a vibrational short-lived contribution and a slowly decaying function:
\beq\label{eq:CMp}
C(t-t')= C^f(t-t')+C^s(t-t')\,, \hskip15pt M(t-t')=M^f(t-t')+M^s(t-t')
\eeq
each decaying on timescales $\t_f\ll\t_s$, respectively. Let us look at intermediate times $\t_f\ll t-t'\ll \t_s$. The slowly varying functions are approximately constant at this scale, equal to the plateau value noted $C_{\rm EA}$ (respectively $M_{\rm EA}$). We have $C_{\rm EA}=\Dl-\D_{\rm EA}$ where $\D_{\rm EA}$ is the plateau of the MSD. Similarly to~\eqref{eq:CMio}, we get from~\eqref{eq:MCTC} the relation between $M_{\rm EA}$ and $\D_{\rm EA}$:
\beq\label{eq:Mp}
\b^2 M_{\rm EA} = \frac{1}{\Dl} \frac{C_{\rm EA}}{\Dl - C_{\rm EA}}=\frac{1}{\D_{\rm EA}}-\frac{1}{\Dl}
\eeq
From~\eqref{eq:CMp}, we can consider the Langevin noise as the sum of two independent centered Gaussian noises, a slowly varying one $\bar \z$ and a fast one $\z^f$. In this limit, the Langevin equation~\eqref{eq:Leff3} reads, using~\eqref{eq:Mp},
\beq\label{eq:Langplateau}
\begin{split}
  \wh\g\dot h(t)&=s-\frac{T}{\D_{\rm EA}}h(t)-\b\int_{t_0}^{t}\dd t'\,M^f(t-t')\dot h(t')+F(h(t))+\z^f(t)\\
  \textrm{with}\hskip5pt s&\equiv\bar\z+\b M_{\rm EA}h_0+T\,,\hskip5pt  \la\z^f(t)\z^f(t')\ra=2\wh\g T\d(t-t')+M^f(t-t')\,,\\
 &\la\bar\z(t)\bar\z(t')\ra=M^s(t-t')\,\hskip5pt\textrm{and for}~\t_f\ll t-t'\ll\t_s\,,\hskip5pt \la\bar\z^2\ra\simeq M_{\rm EA}
\end{split}
\eeq
Following~\cite[Sec. 4.]{CK00}, $s$ acts as a quasistatic field: for times $t-t'\ll\t_s$, $(\bar\z,h_0)$ or equivalently $s$ can be considered as quenched variables, picked with probability $P_{\rm slow}(s)$. For $t-t'\gg\t_f$, the process relaxes to an `equilibrium' state selected by $s$, which is the actual metastable glassy state, with probability\footnote{One has to be careful here with the limit $\Dl\to\io$: we cannot use directly equation~\eqref{eq:MCio}, this is why we have to compute $P_{\rm slow}$ on the sphere and take the infinite radius limit. In principle we would also do it for $P_1(h|s)$ but it is subdominant (as in footnote~\ref{ftn:subdominant}), so that we can compute it directly with~\eqref{eq:Langplateau}, where $\Dl\to\io$ has already been taken. Note that one recovers equation (16) of the main text by setting $Y\equiv s-T$.}
\beq
\begin{split}
P_1(h|s)&=\frac{e^{-\b H_1(h,s)}}{Z_1(s)}\hskip5pt\textrm{with}\hskip5pt H_1(h,s)=\frac{T}{2\D_{\rm EA}}h^2-sh+\bar V(h)\\
P_{\rm slow}(s)&=\int\frac{\dd h_0}{\sqrt{2\p}}\dd\bar\z\, P_V^{\rm eq}(h_0)\,\frac{e^{-\bar\z^2/2M_{\rm EA}}}{\sqrt{2\p M_{\rm EA}}}\,\d(s-\bar\z-\b M_{\rm EA}h_0-T)\\
&=\int\frac{\dd h_0}{\sqrt{2\p}}\,e^{-\b w(h_0)}\frac{e^{-\frac{\b^2M_{\rm EA}}{2}\left(h_0-\frac{T}{M_{\rm EA}}(s-T)\right)^2}}{\sqrt{2\p M_{\rm EA}}}\times\frac{e^{\Dl/2-\l +h_0(\a-1)-\Dl(\a-1)^2/2}}{Z_V}
\end{split}
\eeq
with $\a=\l/\Dl$. Taking $\Dl\to\io$ as in~\eqref{eq:MCio} provides the plateau value:
\beq
\sqrt{M_{\rm EA}}=\frac{\wh\varphi}{2}\int\dd s\, e^{-(s-T)^2/2M_{\rm EA}}Z_1(s)\la F(h)\ra_1^2
\eeq
As an example, we restrict ourselves to the hard-sphere case. Then, with an integration by parts,
\beq
\begin{split}
\la F(h)\ra_1 &=\int\frac{\dd h}{\sqrt{2\p}}\, P_1(h|s)F(h)=\frac{T}{Z_1(s)\sqrt{2\p}}\\
\textrm{with}
\hskip15pt
Z_1(s)&=\int\frac{\dd h}{\sqrt{2\p}}\,e^{\b sh-h^2/2\D_{\rm EA}-\b\bar V(h)}=\sqrt{\D_{\rm EA}}\;\Th_0\left(- s\b\sqrt{\D_{\rm EA}}\right)e^{\b^2s^2\D_{\rm EA}/2}
\end{split}
\eeq
setting $u=\sqrt{\D_{\rm EA}}(\b s-1)$ finally gives
\beq\label{eq:plateau}
\boxed{
\frac{1}{\sqrt{\D_{\rm EA}}}=\wh\varphi\int\frac{\dd u}{4\p}\,\frac{e^{-u^2/2-(u+\sqrt{\D_{\rm EA}})^2/2}}{\Th_0\left(-u-\sqrt{\D_{\rm EA}}\right)}
} 
\eeq

\subsubsection{Equivalence with the one-step replica-symmetry-breaking result}

In~\cite{PZ10,KPZ12}, the equation for the plateau value was obtained, maximizing the hard spheres' replicated entropy in $d\to\io$ in the glass phase where the 1RSB ansatz is stable. It reads
\beq
 \frac{1-m}{\hat A}=\wh\varphi\frac{\dd\GG_m}{\dd\hat A}(\hat A)
 \hskip15pt
 \textrm{with}
 \hskip15pt
 \GG_m(\hat A)=1-m\int\frac{\dd v}{\sqrt{2\p}}\,e^{-v^2/2}\Th_0\left(v-\sqrt{2\hat A}\right)^{m-1}
\eeq
where $2\hat A\equiv\D_{\rm EA}$. Here we are interested in the dynamical transition where the replica symmetric solution is valid (the liquid state is our `paramagnetic' state), which can be recovered in the limit $m\to1$ in Monasson's scheme. Then
\beq
\GG_m(\hat A)\underset{m\to1}{=}(1-m)\left[1+\int\frac{\dd v}{\sqrt{2\p}}\,e^{-v^2/2}\ln\Th_0\left(v-\sqrt{2\hat A}\right)\right]+ O((m-1)^2)
\eeq
Deriving with respect to $\hat A$, we get
\beq
\frac{1}{\sqrt{2\hat A}}=\wh\varphi\int\frac{\dd v}{4\p}\,\frac{e^{-v^2/2-\left(v-\sqrt{2\hat A}\right)^2/2}}{\Th_0\left(v-\sqrt{2\hat A}\right)}
\eeq
which is the same as~\eqref{eq:plateau} with $u=-v$.

\subsubsection{Dynamical transition}

Plotting the function $\wh\f$ versus $\D_{\rm EA}$ using~\eqref{eq:plateau} we get a minimum at the critical packing fraction
\beq\label{eq:phid}
\boxed{\varphi_{\dd}\simeq4.80678 \frac{d}{2^d}}
\eeq
where the plateau is $\D_{\rm EA}(\widehat\varphi_{\dd})\simeq 1.15336$. This packing fraction is larger than the best known lower bound for the existence of sphere packings $\wh\f \geqslant 6/e$~\cite{Va11}. As mentioned in the abstract, this means that a hard sphere system may be prepared in equilibrium up to these densities in times that do not scale with the size of the box. In other words, packings as good as this are easy to obtain, and we conclude that this would be a constructive improvement on the best bound known in high $d$. It would require the present derivation to be turned into a rigorous proof, which seems feasible along the lines of~\cite{BDG01} since our calculation, though a bit tedious, is quite elementary.
\begin{figure}[!h]
\centering
\includegraphics[width=7cm]{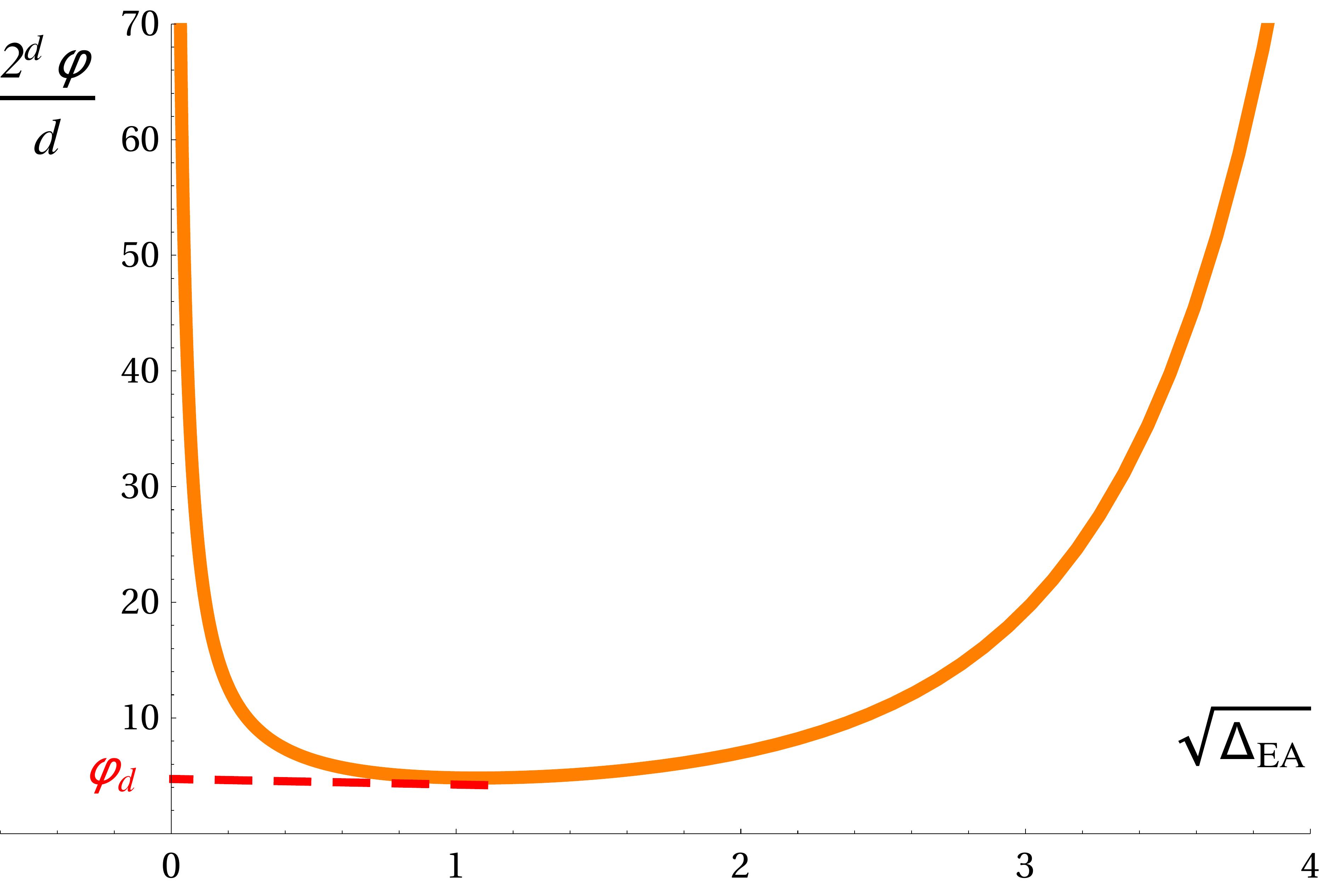}
\caption{Plateau value and critical packing fraction}
\end{figure}

\subsection{Diffusion at long times}

From~\eqref{eq:MCTD} we obtain an expression for the diffusion coefficient for times larger than the relaxation time but still $\D\ll\Dl$. In this regime, the mode-coupling-like equation for the MSD reduces to
\beq
\g\dot{\DE}(t-t')=2dT-2d^2\b\int_{t'}^t\dd v\,M(t-v)\dot{\DE}(v-t')
\eeq
where $\DE=\D/d$ is the non-rescaled MSD, that is, the MSD of the original system of particles. Using Laplace transform we have
\begin{equation}
 \wt\DE(p)=\frac{1}{p^2}\frac{2dT}{\g+2d^2\b\wt M(p)}\underset{p\to0}{\sim}\frac{1}{p^2}\frac{2dT}{\g+2d^2\b\wt M(0)}
\end{equation}
By definition $\wt M(0)=\int_0^\io\dd t\, M(t)$. A Tauberian theorem then gives the long-time diffusive behaviour of the MSD from the small $p$ behaviour of its Laplace transform~\cite{Feller}:
\begin{equation}
 \boxed{\DE(t)\sim 2dD t
\hskip20pt
\textrm{with}
\hskip20pt
D = \frac{T}{\g + 2d^2\b \int_0^\io M} }
\end{equation}

At low density $M\simeq0$ and we recover the usual diffusion coefficient $D = T/\g$ of the free dynamics.
Upon increasing density, $M$ increases and the diffusion coefficient decreases. At the dynamical transition,
$M$ displays a persistence of a plateau, the relaxation time $\t_\a\propto\int_0^\io M$ diverges and the diffusion coefficient vanishes.
One usually defines an exponent $\g$ such that, for $\varphi\to\varphi_{\dd}^-$,
\beq
\t_\a\sim\left(1-\frac{\varphi}{\varphi_{\dd}}\right)^{-\g}\hskip15pt\textrm{and}\hskip15pt D\sim\left(1-\frac{\varphi}{\varphi_{\dd}}\right)^{\g}
\eeq
$\g$ is one of the so-called MCT exponents~\cite{Go98,Go09}.

\subsection{Relation to the standard density formulation of MCT}
\subsubsection{Intermediate scattering functions}\label{subsub:phiq}

In its standard formulation, MCT provides equations for density correlators between time $t$ and the origin $\phi_q(t)=\la\r_q(t)^*\r_q\ra/\la |\r_q|^2\ra$, where $\la\bullet\ra$ is a canonical average over initial conditions, and $\r_q=\sum_{i=1}^Ne^{iq\cdot x_i}$ is the Fourier transform of the particle density~\cite{Go09,Go98}. This correlator thus reads
\beq
\phi_q(t)\propto \la\sum_{ij}e^{iq\cdot[x_i(t)-x_j(0)]}\ra\, ,\hskip15pt \phi_q(0)=1
\eeq
We can define, as in~\ref{sub:virial} and similarly to standard liquid theory~\cite{hansen}, the (non-averaged) local densities of trajectories
\beq\label{eq:rho2}
\tilde\r^{(1)}[x,\hat x]=\sum_i\d(x-x_i)\d(\hat x-\hat x_i)\,,\hskip15pt \tilde\r^{(2)} [x, y,\hat x,\hat y]=\sum_{i\neq j}\d(x-x_i)\d(\hat x-\hat x_i)\d(y-x_j)\d(\hat y-\hat x_j)
\eeq
so that the intermediate scattering functions can be written as
\beq
\phi_q(t)=\phi_q^{\rm s}(t)+\phi_q^{\rm d}(t)\propto\la \int\mathrm{D}[x,\hat x]\,\tilde\rho^{(1)}[x,\hat x]e^{iq\cdot[x(t)-x(0)]} +\int\mathrm{D}[x,y,\hat x,\hat y]\,\tilde\rho^{(2)}[x,y,\hat x,\hat y]e^{iq\cdot[x(t)-y(0)]} \ra 
\eeq
where the self ($i=j$) and distinct ($i\neq j$) parts are defined.
Both parts can then be expressed as a function of $\D(t)$ through the saddle point evaluation of the integrals. In the following for simplicity we
discuss only the self part.

\subsubsection{The self part in infinite dimension}

In $d\to\io$, the self part is simply expressed in terms of the MSD. Using rotation invariance of $\tilde\rho^{(1)}$ to average over $d$-dimensional random rotations $\RR$:
\beq
\begin{split}
\phi_q^{\rm s}(t)&\propto  \la \int\mathrm{D}[x,\hat x]\,\tilde\rho^{(1)}[x,\hat x]e^{iq\cdot[x(t)-x(0)]}\ra=\la\int\mathrm{D}[x,\hat x] \dd\RR\,\tilde\rho^{(1)}[\RR x,\RR\hat x]e^{iq\cdot\RR[x(t)-x(0)]}\ra\\
&\propto\la\int\mathrm{D}[x,\hat x]\,\tilde\rho^{(1)}[x,\hat x]\int_0^\p\dd\th\,\sin^{d-2}\theta\,e^{i|q||x(t)-x(0)|\cos\th}\ra
\end{split}
\eeq
where $\th$ denotes the angle between $q$ and $\RR[x(t)-x(0)]$ and we used hyperspherical coordinates\footnote{In $d$ dimensions, hyperspherical coordinates are defined by $x^1=r\cos\th_1$, $x^2=r\sin\th_1\cos\th_2$, ... , $x^d=r\sin\th_1\cdots\sin\th_{d-2}\cos\th_{d-1}$ with $\th_{d-1}\in[0,2\p[$, $\th_{\m\neq d-1}\in [0,\p]$. The measure is $\prod_{\m=1}^d\dd x^\m=r^{d-1}\sin^{d-2}\th_1\sin^{d-3}\th_2\cdots\sin\th_{d-2}\,\dd r \prod_{\m=1}^{d-1}\dd\th_\m$.}. Now we can proceed as in~\ref{sub:norm1} and express the dynamical variables in terms of the MSD:
\beq\label{eq:phiq2}
\phi_q^{\rm s}(t)\propto\la \int\mathrm{D}[\bm{Q},\bm\n] \,e^{\frac{d}{2}\ln\sdet\bm{Q}-\frac{d}{2}\int\dd a\, \bm\n(a)\left(\bm Q(a,a)-\Dl\right)}\tilde\rho^{(1)}(\bm{Q})\int_0^\p\dd\th\,\sin^{d-2}\theta\,e^{i|q|\sqrt{\D(t)/d}\cos\th}\ra
\eeq
The last integral can be evaluated through a saddle-point method in $d\to\infty$. Provided $|q|\sqrt{\D(t)}/d^{3/2}\ll1$, the $q$-dependent term is irrelevant for the saddle-point evaluation\footnote{If $q= O(d^{3/2})$, which, a priori, does not represent any physical distance (much less than the typical zone spanned by vibrations inside a cage $q= O(d)$), the saddle-point values of $\th$ are different, although also simple; however the $q$-dependent term now contributes to the saddle-point equation on $\sd$, changing its value and the formula~\eqref{eq:phiq} does not hold anymore.} on $\bm Q$. The saddle-point value of $\th$ is imposed at the equator $\p/2$,
\beq
\begin{split}
\int_0^\p\dd\th\,\sin^{d-2}\theta\,e^{i|q|\sqrt{\D(t)/d}\cos\th}& \underset{\th=\p/2+\e}{=}\int_0^\p\dd\e\,e^{(d-2)\ln\cos\e -i|q|\sqrt{\D(t)/d}\,\sin\e}\\
&\underset{d\to\io}{\sim}\int_\RRR \dd\e\, e^{-(d-2)\e^2/2-i|q|\sqrt{\D(t)/d}\,\e}\propto e^{-q^2\D(t)/2d^2}
\end{split}
\eeq
The remaining integral over $\bm Q$ and $\bm\n$ is dealt with as in~\ref{sub:norm1} and is normalized to 1, since the saddle-point is not affected by the last term in~\eqref{eq:phiq2} as long as $q^2\D(t)/d^{3}\ll1$. Together with the normalization $\phi_q^{\rm s}(0)=1$, we finally conclude for all wavevectors satisfying the latter condition,
\beq\label{eq:phiq}
\boxed{\phi_q^{\rm s}(t)\underset{d\to\io}{=}\exp\left(-\frac{q^2}{2d^2}\D(t)\right)}
\eeq
First, we note that the self correlator is Gaussian in $d\to\io$, in contrast to what is found in~\cite{SS10,IM10}. In these articles, the MCT equations for the plateau value (the so-called Debye-Waller factor or non-ergodic parameter, which reads here $\phi_{q,\textrm{EA}}^{\rm s}=e^{-q^2\D_{\rm EA}/2d^2}$) are solved numerically for the hard spheres system up to $d=800$, and its shape is found to be non-Gaussian.\\
This expression is also exact for any dimension both in the free-particle regime (lengths and time small compared to mean free path and collision time respectively) and hydrodynamic limit (lengths and time large compared to mean free path and collision time respectively)~\cite{hansen}.\\
\eqref{eq:phiq} implies, by substitution in~\eqref{eq:MCTD}, equations for the $\phi_q^{\rm s}$, with noticeable qualitative differences with respect to MCT equations (such as a non-local memory kernel $M$).  

\subsubsection{The factorization property}

A crucial outcome of MCT is the so-called factorization property~\cite{Go09,Go98}, which allows to get MCT scaling laws. It states that, in the $\b$-relaxation window (\ie close to the plateau), the difference between the value of the intermediate scattering functions and their value at the plateau can be factorized into a product of a function of the wavector \textit{only} and a function of time \textit{only}:
\beq
\d\phi_q^{\rm s}(t)\equiv\phi_q^{\rm s}(t)-\phi_{q,{\rm EA}}^{\rm s}\simeq H(q)G(t)
\eeq
This property is a stringent test of MCT in simulations~\cite{KA95a}. In $d\to\io$, the self intermediate scattering functions for all wavevectors are governed by a single quantity, the MSD. Close to the plateau, $\d\D(t)=\D(t)-\D_{\rm EA}$ is small and from equation~\eqref{eq:phiq},
\beq
\d\phi_q^{\rm s}(t)\simeq-\frac{q^2\phi_{q,{\rm EA}}^{\rm s}}{2d^2}\d\D(t)
\eeq 
We conclude that the factorization property holds in the infinite $d$ limit. Besides, equation~\eqref{eq:phiq} is more general since it provides all orders in $\d\D$ and is valid even far from the plateau.

\subsection{MCT exponents}

Starting from equations~\eqref{eq:MCTC},\eqref{eq:MCTD}, one can compute the different MCT exponents related to the approach to the plateau $\D_{EA} - \D(t)  \sim t^{-a}$ or the departure from it $\D(t) - \D_{EA} \sim t^b$, by expanding around the plateau value $\D_{EA}$~\cite{Go09,Go98,RC05}. We do not report here the full dynamical computation~\cite{CFLPRR12,FPRR11,PR12,FJPUZ13}. One finds that the MCT exponents are controlled by the exponent parameter $\l$ through the relations~\cite{Go09,Go98,RC05,CFLPRR12}:
\beq
\frac{\G(1-a)^2}{\G(1-2a)} = \frac{\G(1+b)^2}{\G(1+2b)} = \l\,, \hskip30pt \g=\frac{1}{2a}+\frac{1}{2b}
\eeq 
For hard spheres, we obtain $\l \simeq 0.70698$ which implies
$a \simeq 0.324016$, $b \simeq 0.629148$ and $\g\simeq 2.33786$~\cite{KPUZ13}. We emphasize that the value of $\g$ is consistent with numerical results obtained in~\cite{CJPZ14}.

\subsection{Connections with the microscopic model}
\subsubsection{Correlation, response and mean-square displacement}\label{sub:connMSD}

We wish to establish a connection between microscopic quantities and their counterpart at the saddle-point level in $d\to\io$. Let us look at the MSD, but a similar reasoning can be done for correlations and responses (which are related to it). Let us look at the microscopic MSD for a particle $i$
\beq
\la \left(x_i(t)-x_i(t')\right)^2 \ra_{Z_N}=\frac1N \la \sum_{i=1}^N\left(x_i(t)-x_i(t')\right)^2 \ra_{Z_N}=\restriction{\frac{\d \left(\ln Z_N[h]/N[h]\right)}{\d h(t,t')}}{h=0}
\eeq
since $Z_N[h=0]=1$, with $Z_N[h]$ (or $\Xi[h]$) obtained by adding to the dynamical action a coupling to a generating field $h(t,t')$,
\beq
\AA[\{x_i,\hat{x}_i\},h]=\AA[\{x_i,\hat{x}_i\}]-\sum_{i=1}^N\int\dd t\dd t'\left(x_i(t)-x_i(t')\right)^2h(t,t')
\eeq
This is a single-particle term, which amounts to shift the kinetic term in the following way:
\beq
\Phi[x,\hat x,h]=\Phi[x,\hat x]-\int\dd t\dd t'\left(x(t)-x(t')\right)^2h(t,t')
\eeq
Then we use the above-derived $d\to\io$ limit of $\ln Z_N/N\underset{N\to\io}{\sim}\ln\Xi/N=\SS$ with $\SS[h]=\SS_{\rm IG}[h]+\SS_{\rm int}$, where only the ideal gas term depends on $h$: $\SS_{\rm IG}=-\int {\rm D}[x,\hat{x}] \rho[x,\hat{x}](\Phi[x,\hat{x},h]+\mathrm{ln}\rho[x,\hat{x}])$. Consequently, going to generalized spherical coordinates,
\beq
\la \left(x_i(t)-x_i(t')\right)^2 \ra_{Z_N}=\restriction{\frac{\d \SS_{\rm IG}[h]}{\d h(t,t')}}{h=0}\underset{d\to\io}{=}-\restriction{\frac{\d \Phi(\bm Q^{\rm sp},h)}{\d h(t,t')}}{h=0}=\restriction{\frac{\d }{\d h(t,t')}\int\dd u\dd u'\DE(u,u')h(u,u')}{h=0}=\DE(t,t')
\eeq
We deduce that the adimensional rescaled correlation functions 
\beq
\frac{2 d}{\s^2N}\sum_i  x_i(t) \cdot x_i(t') \underset{d\to\io}{\sim} C(t,t')\,,\hskip15pt    \frac{2 d}{\s^2N} \sum_{i,\mu}  \frac{\delta  x_i^\mu(t) }{\delta \bar h_i^\mu(t')} \underset{d\to\io}{\sim} R(t,t') \ , \hskip15pt\frac{ d}{\s^2N}\sum_i  |x_i(t) -x_i(t')|^2 \underset{d\to\io}{\sim} \Delta(t,t')
\eeq
(with external fields $\bar h_i$) are non-fluctuating, imposed by their saddle-point value.

\subsubsection{Force-force correlation and its relation to the memory kernel}\label{sub:ffcorr}

Here we establish a connection between a microscopic force-force correlation $\la \sum_{i<j}F_{ij}(t)\cdot F_{ij}(t') \ra_{Z_N}$, where $F_{ij}=-\nabla V(x_i-x_j)$,  and the memory kernel $M$. 
To generate such terms we will use a random shift similar to the MK model~\cite{MK11}, and once again resort to the SUSY notation for compactness. We will thus consider a shift of vector $A_{ij}\bm g(a)$ 
for each pair of particles, where $A_{ij}$ are Gaussian centered random vectors in $d$ dimensions, of variance $\Sigma_A^2$, independent and identically distributed. 
We will note $\DD A=\prod_{\m=1}^d\dd A^\m e^{-(A^\m)^2/2\Sigma_A^2}/\sqrt{2\p \Sigma_A^2}$ their common measure. 
$\bm g$ is a scalar time-dependent external field that will be sent to zero in the end, in order to recover the original model. Again for compactness, we will note averages over the $A_{ij}$ 
by an overbar, \ie $\overline{A_{ij}^\m}=0$ and $\overline{A_{ij}^\m A_{kl}^\n}=\Si_A^2\d_{ik}\d_{jl}\d^{\m\n}$. The dynamical action becomes
\beq
\AA[\{\bm x_i\},\{A_{ij}\},\bm g]=\sum_{i=1}^N \Phi(\bm x_i)+\sum_{i<j}^{1,N}\int\dd a\, V(\bm x_i(a)-\bm x_j(a)+A_{ij}\bm g(a))
\eeq
We still note $\bm F_{ij}^\m(a)=-\nabla^\m V(\bm x_i(a)-\bm x_j(a)+A_{ij}\bm g(a))$, knowing that we recover the previously defined force by $F^\m_{ij}(t)=\restriction{\bm F_{ij}^\m(a)}{0}$ where the 0 stands 
for $\bm g$ and all Grassmann variables being sent to zero.
Let us compute the second derivative of the generating dynamic functional $Z_N[\bm g]$:
\beq
\begin{split}
\frac{\d Z_N[\bm g]}{\d \bm g(a)}&=\int \prod_{i=1}^N\mathrm{D}\bm x_i\, e^{-\AA[\{\bm x_i\},\{A_{ij}\},\bm g]}\sum_{i<j}A_{ij}^\m \bm F_{ij}^\m(a)\\
\frac{\d^2 Z_N[\bm g]}{\d \bm g(a)\bm g(b)}&=\int \prod_{i=1}^N\mathrm{D}\bm x_i\, e^{-\AA[\{\bm x_i\},\{A_{ij}\},\bm g]}\left[\sum_{\substack{i<j\\k<l}}A_{ij}^\m A_{kl}^\n\bm F_{ij}^\m(a)\bm F_{kl}^\n(b)+\bm\d(a,b)
\sum_{i<j}A_{ij}^\m A_{ij}^\n\nabla^\n\bm F_{ij}^\m(a) \right]
\end{split}
\eeq
where repeated Greek indices are summed over. Sending the external field to zero and averaging over the random shifts directly give $\d \overline{Z_N}[\bm g]/\d \bm g(a)=\bm 0$ and
\beq\label{eq:d2ZN}
\restriction{\frac{\d^2 \overline{Z_N}[\bm g]}{\d \bm g(a)\bm g(b)}}{0}=\Si_A^2\la\sum_{i<j}F_{ij}(t)\cdot F_{ij}(t') \ra_{Z_N}
\eeq
which is the original force-force correlation looked for. As in~\cite{MK11}, one can compute the average $\overline{Z_N}[\bm g]$ introducing an averaged Mayer function
\beq
\overline{f_{ij}}[\bm g]=\int\DD A\, f(\bm x_i(a)-\bm x_j(a)+A\bm g(a))=\int\DD A\,\left[e^{-\int\dd a\, V(\bm x_i(a)-\bm x_j(a)+A\bm g(a))}-1\right]
\eeq
For a non-zero $\bm g(a)$ and large enough $\Si_A$, we still have the same crucial fact that $\overline{f_{ij}}[\bm g]=1+ O(\VV_d(\G)/\VV)$ where $\G$ is a typical length of a trajectory (for finite times), due to the requirement that two 
trajectories overlap to feel the effect of the potential. As a consequence, we can repeat the MK computation and obtain in the thermodynamic limit $\overline{Z_N}[\bm g]=e^{N\SS[\bm g]} $ with the action
\beq
\SS =-\int \mathrm{D}\bm{x}\, \rho(\bm{x})(\mathrm{ln}\rho(\bm{x})+\Phi(\bm{x}))+\frac{N}{2}\int \mathrm{D}[\bm{x},\bm{y}] \,\rho(\bm{x})\rho(\bm{y})\int\DD A\,f(\bm{x}-\bm{y}+A\bm g(a))
\eeq
The difference here is that we cannot simplify further by translation invariance since the shift is time-dependent. Nevertheless, we can still compute derivatives of $\overline{Z_N}[\bm g]$, reminding that due to probability conservation $\overline{Z_N}[\bm g]=1$:
\beq
\begin{split}
 \restriction{\frac{\d \overline{Z_N}}{\d \bm g(a)}}{0}&=N\restriction{\frac{\d \SS}{\d \bm g(a)}}{0}=0\\
 \restriction{\frac{\d^2 \overline{Z_N}}{\d \bm g(a)\bm g(b)}}{0}&=N\restriction{\frac{\d^2 \SS}{\d \bm g(a)\bm g(b)}}{0}+N^2\restriction{\frac{\d \SS}{\d \bm g(a)}}{0}\restriction{\frac{\d \SS}{\d \bm g(b)}}{0}
 =N\restriction{\frac{\d^2 \SS}{\d \bm g(a)\bm g(b)}}{0} 
\end{split}
\eeq
We will use, as in~\ref{sub:int}, the $d\to\io$ limit to compute $\SS$. We know in this limit that the trajectory density $\r(\bm x)$ is determined, at leading order, by a saddle-point equation implying only the Jacobian of the change to generalized spherical coordinates, hence it is independent of the external field at this order. Thus we can overlook this dependence and compute derivatives of the Mayer function, leading to, after averaging:
\beq
\restriction{\frac{\d^2 \SS}{\d \bm g(a)\bm g(b)}}{\bm g=0}=\Si_A^2\frac N2 \int \mathrm{D}[\bm{x},\bm{y}] \,\rho(\bm{x})\rho(\bm{y})e^{-\int\dd a\, V(\bm x-\bm y)(a)} \left[\bm F(\bm x-\bm y)(a)\cdot\bm F(\bm x-\bm y)(b)+\bm\d(a,b)\nabla\cdot\bm F(\bm x-\bm y)(a)\right]
\eeq
We are interested in the ``boson-boson'' part, so we can focus on the first term only, which reads 
\beq\label{eq:FFVV}
\restriction{\bm F(\bm x-\bm y)(a)\cdot\bm F(\bm x-\bm y)(b)}{0}=\frac{(x-y)(t)\cdot (x-y)(t')}{|x-y|(t)|x-y|(t')}V'\left(|x-y|(t)\right)V'\left(|x-y|(t')\right)
\eeq
The content of subsection~\ref{sub:scalings} is that the trajectory of $(x-y)(t) \sim X$ plus a correction that is in $1/d$. 
For any finite time, the vector $X$ can be considered to be constant, with $|X| = \s$ and on average $(X^\m)^2 \sim \s^2/d$.
This is because particles do not move by $ O(1)$ in a finite time.
The correction term has zero average and $|(x-y)(t)| = \s (1 + h(t)/d)$.
Recalling the definitions $\bar V(h) = V(\s(1+h/d))$ and $F(h)=-\bar V'(h)$, we have $F(h) = -V'(\s(1+h/d)) \s/d$.
Using this and the same analysis\footnote{The normalization, giving a factor proportional to the packing fraction as in~\ref{sub:norm2}, follows from a similar analysis since, though the Mayer function $f(\bm x-\bm y)$ constraining the trajectories to be close is not there anymore (it is replaced by $1+f$), $\bm F(\bm x-\bm y)$ has the same role.} of this two-body term as in~\ref{sub:int}, we get, at leading order in $d\to\io$:
\beq
\restriction{\frac{\d^2 \overline{Z_N}[\bm g]}{\d \bm g(a)\bm g(b)}}{0}=\Si_A^2\frac{Nd^3\wh\varphi}{2\s^2}\int\dd h_0\,e^{-\b w(h_0)}\la F\left(h(t)\right)F\left(h(t')\right)\ra=\Si_A^2\frac{Nd^3}{\s^2}M(t,t')
\eeq
We conclude from~\eqref{eq:d2ZN}
\beq\label{eq:linkMFF}
\boxed{M(t,t')=\frac{\s^2}{Nd^3}\sum_{i<j}\la F_{ij}(t)\cdot F_{ij}(t') \ra_{Z_N}}
\eeq

We make a final comment about two-body correlations that have the same structure as the interaction term. The truncated virial expansion in $d\to\io$
tells us that the two-body density of trajectories, which is the average of $\tilde\r^{(2)}$ defined in~\eqref{eq:rho2}, is simply given by\footnote{The $N^2$ factor only comes from the choice of a different normalization of $\r[x,\hat x]$ its 
definition~\eqref{eq:defrho}, with respect to the one used in liquid theory.}~\cite{MH61,Sa58}
\beq
\r^{(2)}[x,y,\hat x,\hat y]=\la\tilde\r^{(2)}[x,y,\hat x,\hat y]\ra_{Z_N}=N^2\r[x,\hat x] \r[y, \hat y] \left(1+f[x - y, \hat x-\hat y]\right)=N^2\r(\bm x) \r(\bm y)e^{-\int\dd a\,V(\bm x-\bm y)(a)}
\eeq
We obtain the relation, for a function $\OO$,
\beq
\begin{split}
\frac1N \sum_{i\neq j} \la \OO\left(x_{ij}(t)\right) \OO\left(x_{ij}(0)\right) \ra
&= \frac1N \int {\rm D}[x,\hat x] {\rm D}[y,\hat y] \r^{(2)}[x,y,\hat x,\hat y] \OO\left(x(t) - y(t)\right) \OO\left(x(0)-y(0)\right) \\
&\sim N \int {\rm D}[x,\hat x] {\rm D}[y,\hat y] \r[x,\hat x] \r[y, \hat y]\left(1+f[x - y, \hat x-\hat y]\right)  \OO\left(x(t) - y(t)\right)  \OO\left(x(0)-y(0)\right)
\end{split}
\eeq
To this kind of function we can apply the same reasoning as in subsection~\ref{sub:int}, with the replacement\\ $f \to (1+f) \OO(t) 
\OO(0)$, if $\OO$ rejects trajectories that do not get close enough\footnote{This is important for \eg the normalization as in~\ref{sub:norm2}.}, 
as was the Mayer function's role, now replaced by $1+f=e^{-W}$ (which is 1 for most trajectories), giving an average over the effective dynamics with potential.
This argument gives more directly Eq.~\eqref{eq:linkMFF}. However one has to be careful for correlations of a higher number of particles, as in the next subsection.

\subsubsection{Stress-stress correlation and its relation to the memory kernel}

Here we repeat the former procedure to obtain the link between the correlation of the off-diagonal ($\m \neq \n$) components of the stress tensor, which reads~\cite{hansen,Yo12}
\beq
\begin{split}
\sigma_{\m\n} = \sum_{i<j} (x_i-x_j)^\m \nabla^\nu V (x_i-x_j)
\end{split}\eeq
Once again, we use external random fields to generate the correlation function. This amounts to turn the dynamical action into
\beq
\begin{split}
\AA[\{\bm x_i\},\{A_{ij},B_{ij}\},\bm g_1,\bm g_2,\bm h_1,\bm h_2]=&\sum_{i=1}^N \Phi(\bm x_i)-\sum_{i<j}^{1,N}\int\dd a\,\left[A_{ij}\bm g_1(a)+B_{ij}\bm g_2(a)\right]\cdot (\bm x_i-\bm x_j)(a)\\
&+\frac12\sum_{i<j}^{1,N}\int\dd a\,\left[ V\left(\bm x_i-\bm x_j+A_{ij}\bm h_1\right)(a)+ V\left(\bm x_i-\bm x_j+B_{ij}\bm h_2\right)(a)\right]
\end{split}
\eeq
where $\bm g_1,\bm g_2,\bm h_1,\bm h_2$ are external $d$-dimensional fields and $\{A_{ij},B_{ij}\}$ are one-dimensional independent identically distributed centered Gaussian random variables 
of variance $\Si_A^2$ and $\Si_B^2$, respectively. Note that when the external fields are zero, we recover the original model. Using the shorthand notation 
$\partial_{1234}\equiv\d^4 /\d \bm g_1^\m(a)\bm g_2^\m(b)\bm h_1^\n(a)\bm h_2^\n(b)$, we have, microscopically,
\beq\label{eq:stressln}
\begin{split}
     \restriction{\partial_{1234}\overline{Z_N}}{\bm g_i=\bm h_i=\bm 0}&=\int \prod_{i=1}^N\mathrm{D}\bm x_i\, e^{-\AA[\{\bm x_i\}]}\sum_{\substack{i<j\\k<l\\m<n\\p<q}}\frac14\overline{A_{ij}(\bm x_i-\bm x_j)^\m(a) B_{kl}(\bm x_k-\bm x_l)^\n(b)A_{mn}\bm F_{mn}^\m(a)B_{pq}\bm F_{pq}^\n(b)}\\
     &=\frac{\Si_A^2\Si_B^2}{4}\la\sum_{\substack{i<j\\k<l}}(\bm x_i-\bm x_j)^\m(a)(\bm x_k-\bm x_l)^\n(b)\bm F_{ij}^\m(a)\bm F_{kl}^\n(b)\ra=\frac{\Si_A^2\Si_B^2}{4}\la \bm\sigma_{\m\n}(a)\bm\sigma_{\m\n}(b)\ra_{Z_N}
\end{split}
\eeq
$\ln\overline{Z_N}$ generates the connected correlation functions (cumulants), so that, using another shorthand notation $\partial^n$ relative to any $n^{\rm th}$ derivative with respect to the fields involved
in $\partial_{1234}$:
\beq\label{eq:cumulants}
\begin{split}
 \partial_{1234}\ln\overline{Z_N}=\partial^4\ln\overline{Z_N}=&\;\frac{1}{\overline{Z_N}}\underbrace{\partial^4\overline{Z_N}}_{1~ \rm term}\;-\;\frac{1}{\overline{Z_N}^2}\underbrace{\partial^1\overline{Z_N}\partial^3\overline{Z_N}}_{4~ \rm terms}\;-\;\frac{1}{\overline{Z_N}^2}\underbrace{\partial^2\overline{Z_N}\partial^2\overline{Z_N}}_{3~ \rm terms}\;+\;\frac{2}{\overline{Z_N}^3}\underbrace{\partial^1\overline{Z_N}\partial^1\overline{Z_N}\partial^2\overline{Z_N}}_{6~ \rm terms}\\
 &-\;\frac{6}{\overline{Z_N}^4}\underbrace{\partial^1\overline{Z_N}\partial^1\overline{Z_N}\partial^1\overline{Z_N}\partial^1\overline{Z_N}}_{1~ \rm term}
\end{split}
\eeq
By isotropy (or average over the disorder), when evaluated at zero external field, all terms containing a first derivative are zero. Furthermore, second derivative terms are also zero, either due to $\overline{AB}=0$ for terms 
involving different times, or by isotropy for terms involving different indices\footnote{These are averages of an expression proportional to $(\bm x-\bm y)^\m(\bm x-\bm y)^\n$ which is its own 
opposite when rotating by an angle $\p/2$ in the $(\m,\n)$ plane.} $\m\neq\n$. We are thus left with only one term, which is  $\partial_{1234}\overline{Z_N}$ since $ \overline{Z_N}=1$.
Once again as in~\cite{MK11}, one can compute $\ln\overline{Z_N}[\bm g_1,\bm g_2,\bm h_1,\bm h_2]$ introducing an averaged Mayer function
\beq
\begin{split}
\overline{f_{ij}}[\bm g_1,\bm g_2,\bm h_1,\bm h_2]&=\int\DD A\DD B\, f(\bm x_i(a)-\bm x_j(a),\bm g_1,\bm g_2,\bm h_1,\bm h_2)\\
&=\int\DD A\DD B\,\left[e^{\int\dd a\,\left[\left(A\bm g_1(a)+B\bm g_2(a)\right)\cdot (\bm x_i-\bm x_j)(a)-\frac12 V\left(\bm x_i-\bm x_j+A\bm h_1\right)(a)-\frac12 V\left(\bm x_i-\bm x_j+B\bm h_2\right)(a)\right]}-1\right]
\end{split}
\eeq
Note that, for finite times, $e^{-\frac12 \int \dd a\,V\left(\bm x_i-\bm x_j+A\bm h_1\right)(a)}-1$ is zero except if $h_1$ is for some time approximately in the same direction as $x_i-x_j$, 
and in that case it is of order $ O(\G/L)$ where once again $\G$ is the typical length of a trajectory and $L^d\sim \VV$, for large enough $\Si_A$. This implies, for fixed non-zero $\bm h_1$ and small 
$\bm g_1=\varepsilon\bm{\tilde g}_1/\Si_A$ with $\bm{\tilde g}_1$ of order one,
\beq
\begin{split}
 \int\DD A\, e^{\int\dd a\,  A\bm g_1(a)\cdot (\bm x_i-\bm x_j)(a)-\frac12 \int \dd a\,V\left(\bm x_i-\bm x_j+A\bm h_1\right)(a)}&\simeq\int\DD A\, e^{\int\dd a\,  A\bm g_1(a)\cdot (\bm x_i-\bm x_j)(a)}+
  O(\G/L)\\
 &=e^{\Si_A^2\left[\int\dd a\,  A\bm g_1(a)\cdot (\bm x_i-\bm x_j)(a)\right]^2/2}+ O(\G/L)\\&=1+ O(\varepsilon^2)+ O(\G/L)
\end{split}
\eeq
One must be careful with the order of limits, the reasoning is the following here:
\begin{enumerate}
 \item we fix $\bm h_1$ (non-zero)
 \item we fix $\bm g_1=\varepsilon\bm{\tilde g}_1/\Si_A$ with $\bm{\tilde g}_1$ of order one
 \item we take large $\Si_A$ so that $A h_1$ covers the whole line spanned by $h_1$, and $\DD A\sim \dd A/L$
 \item we then take $\ee\to0$. In the end we will set all external fields to zero so defining them only in the neighborhood of 0 is enough\footnote{We can choose $\varepsilon=1/\Si_A$ if do not wish to introduce an additional parameter.}.
\end{enumerate}
The same holds for the terms involving $\bm g_2$ and $\bm h_2$, with $B$ instead of $A$. This way, $\overline{f_{ij}}= O(\varepsilon^2)+ O(\G/L)$ is small and we may repeat the same steps as in~\cite{MK11}, so that $\ln\overline{Z_N}=N\SS$, where 
\beq
\SS =-\int \mathrm{D}\bm{x}\, \rho(\bm{x})(\mathrm{ln}\rho(\bm{x})+\Phi(\bm{x}))+\frac{N}{2}\int \mathrm{D}[\bm{x},\bm{y}] \,\rho(\bm{x})\rho(\bm{y})\int\DD A\DD B\,f(\bm{x}-\bm{y},\bm g_1,\bm g_2,\bm h_1,\bm h_2)
\eeq
We only need to compute, to leading order\footnote{As in~\ref{sub:ffcorr}, for $d\to\io$ we can ignore the external fields dependence of the trajectory density $\r$.}
\beq
\begin{split}
\restriction{\partial_{1234}\SS}{\bm g_i=\bm h_i=\bm 0}&=\frac{N}{2}\int \mathrm{D}[\bm{x},\bm{y}] \,\rho(\bm{x})\rho(\bm{y})e^{-\int\dd a V(\bm x-\bm y)(a)}
\frac14\overline{ A(\bm x-\bm y)^\m(a)B(\bm x-\bm y)^\m(b)A\bm F^\n(\bm x-\bm y)(a)B\bm F^\n(\bm x-\bm y)(b)}\\
&=\frac{N}{8}\Si_A^2\Si_B^2\int \mathrm{D}[\bm{x},\bm{y}] \,\rho(\bm{x})\rho(\bm{y})e^{-\int\dd a V(\bm x-\bm y)(a)}(\bm x-\bm y)^\m(a)(\bm x-\bm y)^\m(b)\bm F^\n(\bm x-\bm y)(a)\bm F^\n(\bm x-\bm y)(b)
\end{split}
\eeq
Likewise~\eqref{eq:FFVV}, we have for $d\to\io$
\beq
\begin{split}
\restriction{(\bm x-\bm y)^\m(a)(\bm x-\bm y)^\m(b)\bm F^\n(\bm x-\bm y)(a)\bm F^\n(\bm x-\bm y)(b)}{0}&=\frac{\left[(x-y)^\m\right]^2(t)\left[(x-y)^\m\right]^2(t')}{|x-y|(t)|x-y|(t')}V'\left(|x-y|(t)\right)V'\left(|x-y|(t')\right)\\
&\sim \frac{(X^\m)^2(X^\n)^2}{|X|^2}V'\left(|x-y|(t)\right)V'\left(|x-y|(t')\right)\\
&\sim F\left(h(t)\right)F\left(h(t')\right)
\end{split}
\eeq
We deduce, as in the last subsection,
\beq
\restriction{\frac{\d^4 \ln\overline{Z_N}}{\d \bm g_1^\m(a)\bm g_2^\m(b)\bm h_1^\n(a)\bm h_2^\n(b)}}{0}=\Si_A^2\Si_B^2\frac{Nd\wh\varphi}{8}\int\dd h_0\,e^{-\b w(h_0)}\la F\left(h(t)\right)F\left(h(t')\right)\ra=\frac{\Si_A^2\Si_B^2}{4}dNM(t,t')
\eeq
and from~\eqref{eq:stressln} and~\eqref{eq:cumulants}
\beq
\boxed{M(t,t')=\frac{1}{Nd}\la \sigma_{\m\n}(t)\sigma_{\m\n}(t') \ra_{Z_N}}
\eeq
We conclude that the memory function coincides with the force-force
and stress-stress correlations.

\clearpage

\end{widetext}

\bibliographystyle{mioaps}
\bibliography{HS}

\end{document}